\documentclass[12pt]{article}
\usepackage{a4wide}
\usepackage{amsmath,amssymb,amscd,amsthm,epsfig}
\usepackage{graphics}
\usepackage{verbatim}

\def\eps{\epsilon}

\def\tr{{\rm tr}}

\def\mi{{\rm i}}
\def\e{\rm e}
\def\defi{\stackrel{\rm def}{=}}
\def\hn{{\cal H}_{N,\theta}}
\def\shbar{\sqrt{\hbar}}
\def\C{\mathbb{C}}
\def\N{\mathbb{N}}
\def\cO{\mathcal{O}}
\def\P{\mathcal{P}}
\def\R{\mathbb{R}}
\def\T{\mathbb{T}}
\def\Z{\mathbb{Z}}
\def\Sp{\mathcal{S}_+}

\newtheorem{prop}{Proposition}
\newtheorem{theo}{Theorem}
\newtheorem{lem}{Lemma}

\begin{document}

\title{Scarred eigenstates for quantum cat maps of minimal periods}

\author{\it Fr\'ed\'eric Faure\thanks{
Laboratoire de Physique et Mod\'elisation des Milieux Condens\'es (LPM2C)
(Maison des Magist\`eres Jean Perrin, CNRS), BP 166,
38042 Grenoble C\'edex 9, France. e-mail:frederic.faure@ujf-grenoble.fr}\\
\it St\'ephane Nonnenmacher\thanks{ Service de Physique Th\'eorique,
CEA/DSM/PhT Unit\'e de recherche associ\'ee au CNRS CEA/Saclay 91191
Gif-sur-Yvette c\'edex, France. e-mail:nonnen@spht.saclay.cea.fr}
\\
\it Stephan De Bi\`evre\thanks{UFR de Math.--UMR AGAT, Universit\'e des
Sciences et Technologies de Lille,
59655 Villeneuve d'Ascq, France. email: debievre@agat.univ-lille1.fr} }

\maketitle

\begin{abstract}
In this paper we construct a sequence of eigenfunctions of the ``quantum
Arnold's cat map'' that, in the semiclassical limit, show a strong scarring
phenomenon on the periodic orbits of the dynamics. More precisely, those
states have a semiclassical limit measure that is the sum of $1/2$ the
normalized Lebesgue measure on the torus plus $1/2$ the normalized Dirac
measure concentrated on any a priori given periodic orbit of the dynamics. It
is known (the Schnirelman theorem) that ``most'' sequences of eigenfunctions
equidistribute on the torus. The sequences we construct therefore provide an
example of an exception to this general rule. Our method of construction and
proof exploits the existence of special values of $\hbar$ for which the
quantum period of the map is relatively ``short'', and a sharp control on the
evolution of coherent states up to this time scale. We also
provide a pointwise description of these states in phase space,
which uncovers their ``hyperbolic'' structure in the vicinity of the fixed
points and yields more precise localization estimates.

\end{abstract}



\newpage

\section{Introduction}
One of the main problems in quantum chaos is the understanding of the
semiclassical behaviour of the eigenfunctions of quantum dynamical systems
having a chaotic classical limit. The main theorem in this context is the
Schnirelman theorem \cite{sc, cdv, z1, hmr, bodb2}. It roughly states that
``most'' eigenfunctions equidistribute on the available phase space in the
classical limit. This leaves open the question of the existence of exceptional
sequences of eigenfunctions with a different limit. In the case of ``hard
chaos'' (uniformly hyperbolic systems), numerical computations have shown the
presence of ``scars'' on certain eigenfunctions \cite{he}, {\em i.e.} a visual
enhancement of the  wavefunction on an unstable periodic orbit. Up to now all
theories of this phenomenon have required some kind of averaging over a
(semiclassically large) set of eigenfunctions \cite{bog,berry89,he,
heller-kaplan}. In addition, scarring is often described in the physics
literature as a weak type of localization, compatible with Schnirelman's
(measure-theoretic) equidistribution, as opposed to ``strong scarring''
\cite{rudnick-sarnak}, which implies that the limiting measure has a component
supported on a periodic orbit and therefore does not equidistribute.  We show
in this paper that, for the quantized ``Arnold's cat map'', strongly scarred
sequences do indeed  exist for any periodic orbit (more generally, for any
finite union of periodic orbits). This is, to the best of our knowledge, the
first example of this kind in hyperbolic systems. A construction of
exceptional sequences of eigenfunctions not equidistributing in the
semiclassical limit was recently announced \cite{schubert} for the
quantization of certain ergodic piecewise affine
transformations on the torus, but these do not correspond to ``scars'' since
the systems in question have no periodic orbits.

Our construction is based on intuitively clear ideas that we now briefly
sketch. For unfamiliar notation, we refer to Sections
\ref{sec:linearcldyn}--\ref{s:cohstat}.  Precise statements of our results will
be given below.

Let $M\in$ SL$(2,\Z)$ be a hyperbolic automorphism of the 2-dimensional torus
$\T$ and $\hat M$ its quantization on the $N$-dimensional quantum Hilbert
space ${\mathcal H}_{N,\theta}$, where $2\pi\hbar N =1$. We will construct
strongly scarred quasimodes of $\hat M$ that, for certain values of $N$, will
be shown to be eigenfunctions. For that purpose we will use three ingredients.
First, the time-energy uncertainty relation  in the following simple form
($T\in \N, \phi\in\R$):
\begin{equation}\label{eq:uncrel}
\parallel (\hat M- \e^{\mi\phi}\hat I)\sum_{t=-T}^{T-1} \e^{-\mi\phi t} \hat M^t\parallel=
\parallel \e^{-2\mi\phi T}\hat M^{2T} - \hat I\parallel \leq 2.
\end{equation}
Second, precise estimates on intuitively clear phase space localization
properties of coherent states. Third, a remark on the quantum period of $\hat
M$ \cite{bondb1} (Section \ref{s:min quantum period}).

Let $x_0, x_1=Mx_0,\dots, x_\tau=M^\tau x_0=x_0$ be a periodic orbit of period
$\tau$ of  $M$. Let $\vert x_0, \tilde c_0, \theta\rangle$ be a ``squeezed''
coherent state in $\hn$ centered on the point $x_0$ and consider $\hat
M^t\vert x_0, \tilde c_0, \theta\rangle$ for $t\in\Z$. Note first that this
state is still a squeezed coherent state and that,  for small enough $t$, it
is localized around $x_t$. In fact, the support of the Husimi function of this
state is an ellipse stretched along the unstable direction of the dynamics
through the point $x_t$, with its major axis roughly of size $\sqrt \hbar
\e^{\lambda t}$, where $\lambda$ is the (positive) Lyapounov exponent of the
dynamics (Section \ref{s:cohstat}). Introducing the Ehrenfest time
$T=\frac{|\ln\hbar|}{\lambda}$, the support is therefore microscopic as long
as $t\leq (1-\epsilon)T/2$. For longer times, between $T/2$ and $T$, the
support of the Husimi function of   $\hat M^t \vert x_0, \tilde c_0,
\theta\rangle$ starts to wrap around the torus and it was shown in
\cite{bondb1} that it equidistributes on that time scale.

We shall consider the ``discrete time quasimode''
\begin{equation}\label{eq:quasimode}
|\Phi^{\rm disc}_\phi\rangle =\sum_{t=-T}^{T-1} \e^{-\mi\phi t}\hat M^t \vert
x_0,
 \tilde c_0, \theta\rangle=\sum_{j=1}^4 |\Phi_{j,\phi}^{\rm disc}\rangle
\end{equation}
and its ``components''
\begin{equation}\label{e:definition |Phi_j>}
|\Phi_{j,\phi}^{\rm disc}\rangle =
\sum_{t=-T+(j-1)\frac{T}{2}}^{-T+j\frac{T}{2}-1}
 \e^{-\mi\phi t}\hat M^t \vert x_0, \tilde c_0, \theta\rangle.
\end{equation}
We note that similar states were considered before in the study of scars, see
for instance \cite{borondo,heller-kaplan} and references therein.
\begin{figure}[htbp]
{\par\centering
\resizebox*{0.7\columnwidth}{!}{\includegraphics{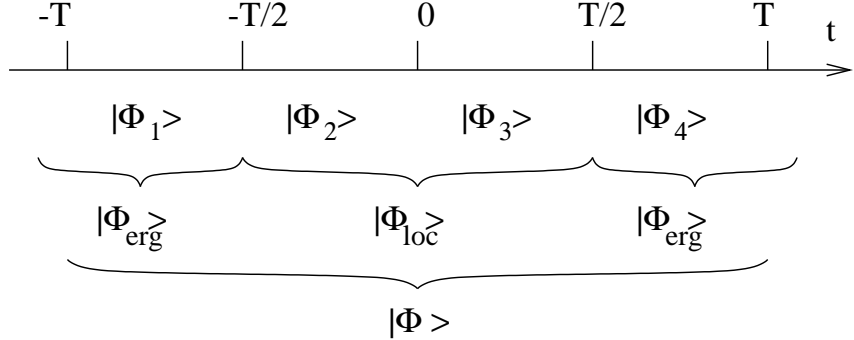}} \par}
\caption{\label{f:time-scale}Partition of the time interval $[-T,T]$ into four
equal parts, and of the quasimode $|\Phi_\phi\rangle$ into corresponding
components.}
\end{figure}
We shall introduce a ``continuous time'' version $|\Phi^{\rm
cont}_\phi\rangle$ of those quasimodes later. We will write
$|\Phi_\phi\rangle$ in statements true for both the discrete and continuous
time quasimodes.

Let us for simplicity concentrate on the case where $x_0=0, \tau=1$. Our
crucial technical estimate (Section \ref{s:cohstat}--Proposition
\ref{prop:crucial}) says that there exists $C>0$ so that
\begin{equation}\label{eq:crucial}
 \langle \tilde c_0, \theta \mid \hat M^t\mid  \tilde c_0, \theta\rangle
= \frac{1}{\sqrt{\cosh(\lambda t)}}+ I(t),\; {\rm with}\: |I(t)|\leq
C\e^{-\lambda(T-\frac{|t|}{2})}.
\end{equation}
This implies rather easily (Proposition \ref{prop:ortho}) the existence of a
smooth, strictly positive function $S_1(\phi,\lambda)$ so that
$$
\langle \Phi_\phi\mid \Phi_\phi\rangle\sim 2 S_1(\phi,\lambda) T.
$$
Using (\ref{eq:uncrel}) one concludes readily that
\begin{equation}\label{e:quasi property}
\parallel (\hat{M}-\e^{\mi\phi}\hat I)|\Phi_\phi\rangle_n \parallel \leq
\sqrt{\frac{2}{{S_1(\phi,\lambda)T}}}
\left(1+\frac{\mathcal{O}(1)}{S_1(\phi,\lambda)T}\right)\,,
\end{equation}
justifying the name ``quasimode''. Here we used the notation $|\psi\rangle_n=
|\psi\rangle/\sqrt{\langle \psi|\psi\rangle}$ for any non-zero
$|\psi\rangle\in {\mathcal H}_{N,\theta}$.

\begin{figure}[htbp]
{\par\centering
\resizebox*{0.9\columnwidth}{!}{\includegraphics{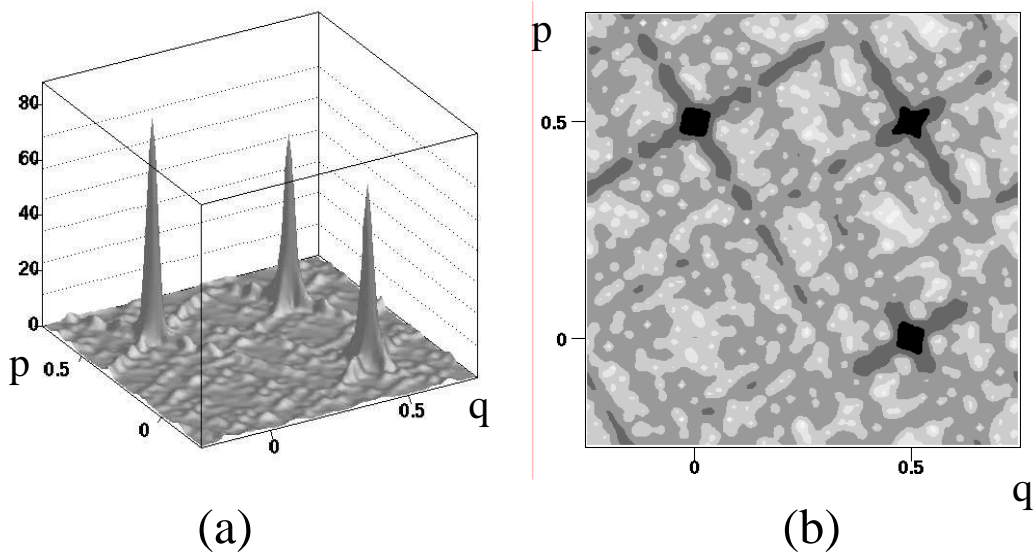}}
\par}
\caption{\label{f:quasimode} Husimi distribution of the state
$|\Phi_\phi\rangle_n$, constructed for the cat map \eqref{e:cat map} on the
orbit of period $3$ starting from $x_0=(0,0.5)$. The quantum parameters read
$N=1/(2\pi \hbar )=500$,  $\phi=0$. (a): 3D plot on a linear scale. (b): 2D
plot in logarithmic scale (darker = higher values).}
\end{figure}

To analyze the phase space properties of the above quasimodes, we first show
as a further consequence of (\ref{eq:crucial}) that the four states
$|\Phi_{j,\phi}\rangle$ have the same norm, asymptotically proportional to
$\sqrt T$ as $\hbar$ goes to $0$ and that they are asymptotically orthogonal
in the semiclassical limit. In fact, this is easily understood intuitively by
noting for example that the Husimi function of $|\Phi_{1,\phi}\rangle$ is
supported along the stable manifold of the periodic orbit, and that of
$|\Phi_{4,\phi}\rangle$ along the unstable one, so that they have essentially
disjoint supports, which is at the origin of their orthogonality. To put it
differently, since the unstable and stable manifolds intersect at  homoclinic
points, our results show that the contribution of these intersections in the
phase space integral expressing the overlap $\langle
\Phi_{1,\phi}|\Phi_{4,\phi}\rangle$ is small for small $\hbar$. Note that
although the homoclinic interferences do not contribute significantly to the
above integral, they are nevertheless clearly visible on the pointwise
behaviour of the Husimi distribution of $|\Phi_\phi\rangle$, which is
represented in Figure~\ref{f:quasimode} and that will be further studied in
Section \ref{s:pointwise} (for ``continuous time'' quasimodes). 
The pointwise estimates obtained there will show that the
Husimi density concentrates along 
``classical hyperbolas'' asymptotic to the stable and unstable manifolds; 
they will
at the same time provide estimates on the rate of convergence to the limit
measure, as well as other localization indicators (namely, $L^s$ norms
of the Husimi density).

It is furthermore clear from the previous discussion on the phase space
localization properties of the evolved coherent states that
$|\Phi_{1,\phi}\rangle$ and $|\Phi_{4,\phi}\rangle$ are sums of states that
equidistribute on the torus, whereas  $|\Phi_{2,\phi}\rangle$ and
$|\Phi_{3,\phi}\rangle$ are sums of states that localize on the periodic
orbit. One therefore expects (and we shall prove in 
Sections~\ref{s:quasimodes}--\ref{s:other_orbit}) that
$$
\lim_{\hbar\to0} \ _n\!\langle \Phi_{j,\phi}| \hat f |\Phi_{j,\phi}\rangle_n =
\int_{\T} f(x) dx\qquad{\rm if}\ j=1,4,
$$
and that
$$
\lim_{\hbar\to0} \ _n\!\langle \Phi_{j,\phi}| \hat f |\Phi_{j,\phi}\rangle_n
=\frac{1}{\tau}\sum_{i=0}^{\tau-1} f(x_i)\qquad{\rm if}\ j=2,3.
$$
Here $\hat f$ is either the Weyl or anti-Wick quantization of $f\in
C^\infty(\T)$. In other words, the Wigner and hence also the Husimi function
of $|\Phi_{2,3,\phi}\rangle$ converge (weakly) to the Dirac measure on the
periodic orbit, whereas the ones of $|\Phi_{1,4,\phi}\rangle$ equidistribute,
{\em i.e.} converge to the Lebesgue measure.  This suggests grouping these 
states
two by two, defining:
\begin{equation}\label{eq:locerg}
|\Phi_{{\rm erg},\phi}\rangle =|\Phi_{1,\phi} \rangle +
|\Phi_{4,\phi}\rangle\;\ {\rm and }\;\ |\Phi _{{\rm loc},\phi}\rangle
=|\Phi_{2,\phi} \rangle + |\Phi_{3,\phi}\rangle.
\end{equation}

Using the above information we shall finally prove (Propositions
\ref{prop:semiclassical properties of Phi_tot} and \ref{prop:qmell}) that, for any $\phi\in
[-\pi,\pi]$,
\begin{equation}\label{e:equidistributeall}
\lim_{\hbar\to0} \ _n\!\langle \Phi_\phi| \hat f |\Phi_\phi\rangle_n
=\frac{1}{2} \int_{\T^2} f(x) dx + \frac{1}{2}
\left[\frac{1}{\tau}\sum_{j=0}^{\tau-1} f(x_j)\right].
\end{equation}
In other words, the semiclassical limit measure of the sequence of quasimodes
$|\Phi_\phi\rangle_n$ is the measure
$$
\frac{1}{2} dx + \frac{1}{2} \left[\frac{1}{\tau}\sum_{j=0}^{\tau-1}
\delta_{x_j}\right].
$$
This shows that the quasimodes $|\Phi_\phi\rangle_n$ are strongly scarred.

We then conclude using a particular property of the quantum period of $\hat
M$. We recall that the quantum cat map $\hat M$ has an $\hbar$ dependent
``quantum period'' $P$, {\em i.e.} $\hat M^P=\e^{-\mi\varphi}\hat I$ for
some $\varphi\in [0,2\pi[$. The eigenvalues of $\hat M$ on $\hn$ are 
therefore all of
the form $\e^{-\mi\phi_j}$, with $\phi_j=\varphi/P + 2\pi j/P$, $j=1,\dots,
P$. Note that $P$ plays the role here of the Heisenberg time of the system,
since $\Delta \phi_j \sim 1/P$. Since, for general $\hbar$, the quantum period
$P$ is of order $\hbar^{-1}$ \cite{ke}, it is considerably longer than the
Ehrenfest time $T$, which grows only logarithmically in $\hbar^{-1}$.
Nevertheless, developing an argument in \cite{bondb1}, we will show that,
for any hyperbolic matrix in SL$(2,\Z)$
 there exists a subsequence $(\hbar_k)_{k\in\N}$ of
values of $\hbar$ tending to zero for which $P=2T+\cO (1)$ 
(see also \cite{kuru2}).
For those values
the Heisenberg and Ehrenfest times of the system coincide and the
$|\Phi_\phi\rangle_n$ therefore constitute a sequence of eigenfunctions of
$\hat M$ that strongly scar, provided $\phi=\phi_j$ for some $j\in\{1\dots
P\}$. It should be noted that, for the values of $\hbar$ considered, the
number of distinct eigenvalues $\phi_j$ is of order $|\ln\hbar|$, so that
 the eigenvalue degeneracy is very large, namely of order
$(\hbar|\ln\hbar|)^{-1}$.

Our main result can finally be summarized as follows:

\begin{theo} \label{th:betascar} Let $M$ and $(\hbar_k)_{k\in\N}$ be as above.
Let $0\leq \beta\leq 1/2$ and let $\mathcal P=\{x_0,\ldots,x_{\tau-1}\}$ 
be a periodic orbit of $M$.
Then there exists  a sequence  $(\psi_{j_k})_{k\in\N}$ of  eigenfunctions 
of $\hat M$ on $\mathcal{H}_{N_k,\theta}$ with the property  that, for all 
$f\in C^\infty(\T^2)$,
\begin{equation}\label{eq:scarbeta}
\lim_{k\to\infty}\ _n\!\langle\psi_k|\hat f|\psi_k\rangle_n = \beta
\frac{1}{\tau}\sum_{j=0}^{\tau-1} f(x_j) + (1-\beta) \int_{\T^2}\ f(x)\,dx.
\end{equation}
\end{theo}

Our result helps to complete the picture of the semiclassical eigenfunction
behaviour of quantized toral automorphisms known to date. Indeed, beyond the
general Schnirelman theorem for these models \cite{bodb2} the following
results are known. First, suppose $M$ is of ``checkerboard form'', meaning 
$AB\equiv 0\equiv CD\bmod 2$. Then all eigenfunctions of 
$\hat M$ semiclassically
equidistribute, provided one takes the limit along a density one subsequence
of values of $N$ \cite{kuru2}, for which the quantum period is larger than 
$\sqrt N$. Note that this sequence excludes  the values $N_k$ for
which the period is very short. Second, it is shown in \cite{kuru1,mezz2} 
that for such $M$ there
exists a basis of eigenfunctions that equidistribute as $N$ tends to infinity,
without restrictions on $N$. This basis is constructed as a common eigenbasis
for $\hat M$ and its ``quantum symmetries'', which are shown in \cite{kuru1}
to be sufficiently numerous to drastically reduce (if
not to lift) the degeneracies of the eigenvalues. Finally, one may wonder if
it would be possible to construct a sequence of eigenfunctions of $\hat M$
that has as a limit measure
$$
\beta \frac{1}{\tau}\sum_{j=0}^{\tau-1} \delta_{x_j} + (1-\beta) dx,
$$
with $\beta>1/2$. 
It is proven in \cite{fred-steph-maximal} that
this is impossible, so that the above quasimodes are in a sense
maximally localized (the bound $\beta>(\sqrt 5-1)/2\cong 0.62$ had been previously
obtained by \cite{bondb2}).


\section{Linear dynamics on the plane}\label{sec:linearcldyn}
In this section we recall some known results we will need in the sequel. For
details not given here we refer to \cite{f}.

\subsection{Classical linear flow}
The most general quadratic Hamiltonian on $\R ^{2}$ is ($\alpha,
\beta,\gamma\in\R)$:
\begin{equation}\label{eq:hamilt}
H(q,p)=\frac{1}{2}\alpha q^{2}+\frac{1}{2}\beta p^{2}+\gamma qp.
\end{equation}
Assuming $\gamma ^{2}>\alpha \beta $, $H$ generates a hyperbolic flow
$x(t)=(q(t), p(t))$ on $\R^2$, given by $x(t)=M{(t)}x(0)\ (t\in\R)$, where for
each $t\not=0$, $M(t)$ is a hyperbolic matrix in SL$(2,\R)$. Explicitly, for
$t=1$
\begin{equation}\label{eq:M}
M\defi M(1)=\left( \begin{array}{cc} A & B\\ C & D
\end{array}\right) \in {\rm SL}\left( 2,\R \right) ,
\end{equation}
{\it i.e.} $AD-BC=1$, and
\begin{equation}\label{eq:hyperb}
 \left\{\begin{array}{ll} A=\cosh \lambda +\frac{\gamma }{\lambda
}\sinh \lambda  & B=\frac{\beta }{\lambda }\sinh \lambda\\
C=-\frac{\alpha }{\lambda }\sinh \lambda & D=\cosh \lambda-\frac{\gamma
}{\lambda }\sinh \lambda
\end{array}\right.
\end{equation}
where $\lambda=\sqrt{\gamma ^{2}-\alpha \beta }>0$ is the Lyapounov exponent.
Note that $M$ has two real eigenvalues $\e^{\pm\lambda}$ and hence two real
eigenvectors corresponding to an unstable and a stable direction for the
dynamics. They have respective slopes $s_+=\tan\psi _{+}, s_-=\tan\psi _{-}$.
Clearly, any hyperbolic matrix $M\in$ SL$(2,\R)$ with Tr$M>2$ is of the above
form for a unique $\alpha, \beta, \gamma$ (the case Tr$M<-2$ is treated by
using the map $-M$). The  expressions in
(\ref{eq:M})--(\ref{eq:hyperb}) still make sense in the elliptic case, when
$\gamma ^{2}<\alpha \beta$ and $-2<$Tr$M<2$. In terms of the complex 
coordinate $z=\frac{1}{\sqrt{2}}\left( q+\mi p\right)$, the Hamiltonian in 
(\ref{eq:hamilt}) reads
\begin{equation}\label{eq:compl}
H=\frac{c}{2}z^{2}+\frac{\overline{c}}{2}\overline{z}^{2}+bz\overline{z}, \
{\rm with}\  b=\frac{1}{2}\left( \alpha +\beta \right) \in \R, \
c=\frac{1}{2}\left( \alpha -\beta \right) -\mi \gamma \in \C .
\end{equation}
and $ \lambda =\sqrt{\left| c\right| ^{2}-b^{2}}.  $ We shall write
$M_{(c,b)}$ for the matrix $M$ constructed via
(\ref{eq:M})--(\ref{eq:compl}), whenever $b^2\not=|c|^2$.

\begin{figure}[htbp] {\par\centering
\resizebox*{0.8\columnwidth}{!}{\includegraphics{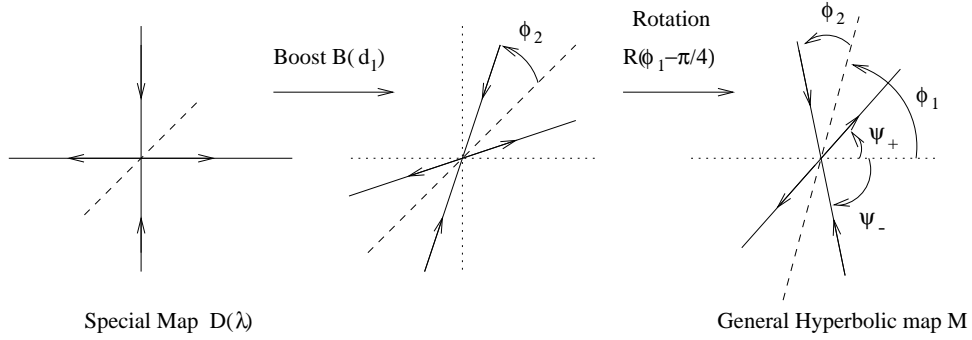}}
\par}
\caption{\label{f:decomposition}Decomposition of the general linear hyperbolic
map $M_{(c,b)}$ as in (\ref{e:M is QDQ}). }
\end{figure}

We will make use of the following convenient decomposition of a general
hyperbolic matrix $M$ (Tr$M>2$). We first introduce some notation. For
$\mu\in\R_+$ we define:

$$
D(\mu )\defi M_{(c=-\mi \mu ,b=0)},\quad B(\mu )\defi 
M_{\left( c=-\mu ,b=0\right)},\quad R(\mu)\defi M_{(c=0,b=-\mu)}.
$$
Clearly, $D(\mu)$ is hyperbolic, with the $q$ and $p$ axes as unstable and
stable axes. $B(\mu)$ is also hyperbolic, with eigenaxes forming angles
$\psi_+=\frac{1}{2}\arg(-\mi\bar c)=\frac{\pi}{4}=-\psi_-$ with the horizontal.
$R(\mu)$, on the other hand, is just a rotation of angle $\mu$ and hence
elliptic.  Any hyperbolic matrix $M_{(c,b)}$ as in (\ref{eq:M}) can be
decomposed as:
\begin{equation}\label{e:M is QDQ}
M_{(c,b)}=Q D(\lambda ) Q^{-1},   \ {\rm with}\ Q=R(b_1) B(b_{2}),
\end{equation}
where $b_1\in \left[ -\frac{\pi }{2},\frac{\pi }{2}\right] $, $ b_{2}\in \R $
are defined as follows. We denote by $\phi _{1}\in \left[ -\frac{\pi
}{2},\frac{\pi }{2}\right] $ the angle between the $q$ axis and the bisector
between the stable and unstable axes of $M_{(c,b)}$, and by $\phi _{2}\in
\left] 0,\frac{\pi }{4}\right] $ the angle between the bisector and the stable
axis of $M_{(c,b)}$ (Figure~\ref{f:decomposition}). In terms of those, one
has:
\begin{equation}\label{eq:parameterchange}
\sinh \left( 2b_{2}\right) =\frac{1}{\tan \left( 2\phi _{2}\right) }, \qquad
b_1=\phi _{1}-\frac{\pi }{4}.
\end{equation}

This last decomposition has the following interpretation. The general
hyperbolic map $M_{(c,b)}$ is obtained from the special case $D(\lambda )$
($\lambda=\sqrt{|c|^2-b^2}>0$) by a change of coordinates  $Q$ yielding a
transformation from the $(q,p)$ frame into the unstable-stable frame. The
unstable (respectively stable) direction is given by the vectors $ {v
}_{+}=Q\,  {e}_{q}$, $ {v }_{-}=Q\,  {e}_{p}$ (which are, in general, not
normalized). Above, we decomposed $Q$ into the transformation $B(b_{2})$ which
changes the angle between the stable and unstable axis, and the rotation
$R(b_1)$ which rotates the whole frame (Figure~\ref{f:decomposition}).

We finally remark, for later purposes, that there exists another
decomposition: given $M\in\mathrm{SL}(2,\R)$, 
$\exists!\tilde c\in\C$, $\mu\in]-\pi, \pi]$ so that
\begin{equation}\label{e:squeeze!}
M=M_{(\tilde c,0)} R(\mu).
\end{equation}


\subsection{Linear quantum dynamics}

In terms of  the usual annihilation and number operators $
a=\frac{1}{\sqrt{2\hbar}}\left( \hat{q}+i\hat{p}\right) $, and $
\hat{n}=\frac{1}{2}\left( a^{\dagger}a+aa^{\dagger}\right) , $ the Weyl (or
canonical) quantization of $H$ in (\ref{eq:hamilt}) is defined as the
self-adjoint operator $\hat H$ on $L^{2}(\R )$ given by:
\begin{equation}\label{e:Hamiltonian}
\hat{H}=\frac{1}{2}\alpha \hat{q}^{2}+\frac{1}{2}\beta \hat{p}^{2}+\gamma
\frac{1}{2}\left( \hat{q}\hat{p}+\hat{p}\hat{q}\right)   =\hbar \left(
\frac{c}{2}a^{2}+\frac{\overline{c}}{2}a^{\dagger 2}+b\hat{n}\right).
\end{equation}
The quantum evolution operator for time $t=1$ which corresponds to $M_{(c,b)}$
is then:
\begin{equation}
\label{e:operator M} \hat{M}_{(c,b)}=\exp\left\{-\mi \frac{\hat{H}}{\hbar
}\right\}.
\end{equation}
The quantization of the matrix $-M_{(c,b)}=M_{(c,b)}R(\pi)$ can be defined 
as $\hat M_{(c,b)}\hat P=\hat P\hat M_{(c,b)}$ 
where $\hat P=-\mi\hat R(\pi)$ is the parity operator.
The unitary operators $\hat{M}_{(c,b)}$, $\hat M_{(c,b)}\hat P$ yield a 
projective representation of
SL$(2,\R)$ (which resembles the metaplectic
representation). We  will in most of the paper omit to indicate the
$\hbar$-dependence of the operators $\hat H$ and $\hat M_{(c,b)}$.

Let $v= v_1  e_q + v_2  e_p \in \R^2$ and let $T_v:\R^2\rightarrow \R^2$
denote the translation on classical phase space by $v$.  The corresponding
quantum translation operator is defined by:
\begin{equation}
\label{e:quantum translation} \hat{T}_v=\exp \left(- \frac{\mi }{\hbar }
\left(v_1\hat{p}-v_2\hat{q}\right)\right) .
\end{equation}
These quantum translations satisfy the algebraic identity
\begin{equation}
\label{e:Heisenberg} \hat{T}_v\, \hat{T}_{v'}=\e^{\mi S}\, \hat{T}_{v+v'},
\end{equation}
with $S=\frac{1}{2\hbar }\left( v_2 v_1'-v_1 v_2'\right)
=-\frac{1}{2\hbar}v\wedge v'$, so they generate an (irreducible) unitary
representation of the Heisenberg group.   For any matrix $M\in {\rm SL}(2,\R
)$, one trivially has $ M\, T_v M^{-1}=T_{Mv}$. This intertwining persists at
the quantum level:
\begin{equation}
\label{e: M and T} \hat{M}\, \hat{T}_v\, \hat{M}^{-1}=\hat{T}_{Mv}.
\end{equation}


\section{Classical and quantum automorphisms of the torus}\label{s:dyntor}

\subsection{Classical automorphisms and their invariant manifolds}
\label{s:classaut}

Consider the torus $\T=\R ^{2}/\Z ^{2}$ as a symplectic manifold with the
two-form $dq\wedge dp$. Then any $M\in$ SL$(2,\Z)$ defines a
(discrete) symplectic dynamics on $\T$ in the obvious way. We are interested
in the case where $M$ is hyperbolic: 
the corresponding dynamical system is then an Anosov
system \cite{arnoldavez}. The stable and unstable manifolds of any point
$x\in\T$ are obtained by wrapping the lines with slopes $s_{\pm}$ passing
through $x$ around the torus.  We present here some properties of these
manifolds that we will need in subsequent sections.

\begin{figure}[htbp]
{\par\centering
\resizebox*{0.4\columnwidth}{!}{\includegraphics{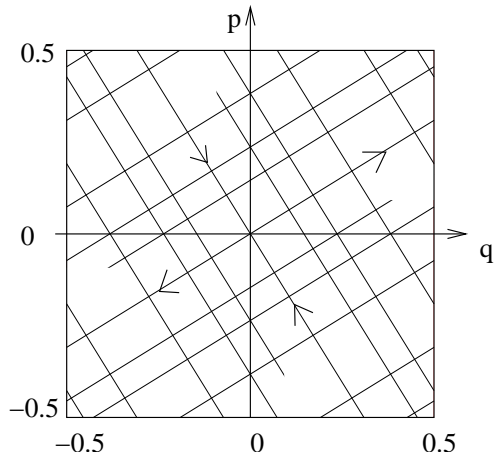}} \par}

\caption{\label{f:varietes}The stable and unstable axes through $0$ of the map
$M_{\rm Arnold}$ wrap around the torus at infinity. We have only represented
the first six occurrences.}
\end{figure}

A simple example we will use for numerical illustrations is the so called
``Arnold's cat map'' \cite{arnoldavez}
\begin{equation}
\label{e:cat map} M_{\rm Arnold}=\left( \begin{array}{cc} 2 & 1\\ 1 & 1
\end{array}\right) .
\end{equation}
Its Lyapounov coefficient is $ \lambda _{0}=\log \left(
\frac{3+\sqrt{5}}{2}\right) \approx 0.9624$. The stable and unstable manifolds
of the fixed point $x=0$ are depicted in Figure~\ref{f:varietes}.

For any hyperbolic matrix $M$, the slopes $s_+$ and $s_-$ of the unstable 
and stable
directions are quadratic irrationals ({\it i.e.} the solutions of a quadratic
equation with integer coefficients). It is well known \cite{kh} that any
quadratic irrational $s$ satisfies the following diophantine inequality:
$$
\exists C(s)>0,\ \forall k\in\Z,\ \forall l\in \N^*,\quad
\left|s-\frac{k}{l}\right|\geq C(s)\frac{1}{l^{2}}\Longleftrightarrow
\left|ls-k\right|\geq C(s)\frac{1}{l}.
$$
This means that quadratic irrationals are poorly approximated by rationals, in
the sense that, to get an approximation with an error $\epsilon $, you need a
rational with a denominator of order at least $\epsilon ^{-1/2}$.

This inequality will be used in the following manner. Consider the
eigenvectors $v_\pm$ of $M_{(c,b)}$ defined as $v_+=Qe_q$ , $v_-=Qe_p$ (with
$Q$ the matrix defined in Eq.~\eqref{e:M is QDQ}). As usual, their dual basis
$u_\pm $ (defined as $v_+\cdot u_+=1,\: v_+\cdot u_-=0$, \emph{etc.}) can be
used to express the coordinates of a point $x$ in the basis $v_\pm$:
\begin{equation}\label{e:decompo-n}
 x = q'(x) v_+ + p'(x)  v_-,\quad {\rm with}\quad q'(x)=
 x\cdot u_+,\ \ p'(x)= x\cdot u_-.
\end{equation}
We call d$(x,\Z^2_*)$ the distance between a point $x\in\R^2$ and 
$\Z^2_*=\Z^2\setminus\{0\}$, and we will estimate it for $x$ on the (un)stable 
axis: 
\begin{equation}\label{eq:distance}
\exists C>0,\ \forall x\in\R v_{\pm},\quad {\rm d}(x, \Z^2_*) 
\geq \frac{C}{\|x\| +1}.
\end{equation}
To prove this, first note that, for any $n\in\Z$ s.t. $n_q\neq 0$, we have
\begin{equation}
\label{eq:dioph1} |p'(n)|=|n\cdot u_-|=|u_{-,p}|\,
 \left|n_q\frac{u_{-,q}}{u_{-,p}}+n_p\right|\geq
 \frac{C(s_+)|u_{-,p}|}{|n_q|},
\end{equation}
where we have used the fact that $u_{-,q}/u_{-,p}=s_+$ is a quadratic
irrational. Interchanging $n_q$ and $n_p$, we obtain a first set of
inequalities:
\begin{lem}\label{lem:diophantine}
There is a constant $C$ (depending on $M$) such that, for any integer lattice
point $n\neq 0$,
$$
|p'(n)|\geq \frac{C}{\|n\|}\quad\mbox{and}\quad |q'(n)|\geq \frac{C}{\|n\|}.
$$
\end{lem}
We can now prove (\ref{eq:distance}) as follows. For each $x\in\R v_{+}$,
there exists an $n\in\Z^2_*$ so that
$$
{\rm d}(x, \Z^2_*)=\| n-x\| \geq \frac{|n\cdot u_-|}{\|u_-\|}\geq
\frac{C_{\pm}}{\|u_-\| \|n\|}.
$$
Since, obviously, $\|n-x\|\leq 1/\sqrt{2}$, (\ref{eq:distance}) follows easily.

We will in addition need a slightly refined statement. If the lattice point
$n\neq 0$ is in a sufficiently thin strip around the unstable axis, it
satisfies $\|p'(n)v_-\|\leq 1/2\leq \|n\|/2$, which implies the lower bound
$|q'(n)|\geq \frac{\|n\|}{2\|v_+\|}$. Together with the above lemma, this
entails $|p'(n)|\geq\frac{C'_1}{|q'(n)|}$ for a certain $C'_1$. Interchanging
$p'\leftrightarrow q'$, we see that the same inequality holds for points in a
sufficiently thin strip around the stable axis. Outside the union of these
strips, this inequality can be violated by at most a finite set of lattice
points; therefore, upon reducing the constant $C'_1$ we obtain the main
technical result of this section:
\begin{lem}
There exists a constant $C_o>0$ (depending on $M$) such that, for any integer
points $n\neq m$ of the plane, their coordinates along the (un)stable
directions satisfy:
\begin{equation}
\label{eq:dioph3} |q'(n)-q'(m)|\geq \frac{C_o}{|p'(n)-p'(m)|}.
\end{equation}
\end{lem}
These inequalities precisely control the sparseness of the lattice points
inside a strip around the unstable axis: the narrower the strip, the farther
successive lattice points have to be from each other.

%

\subsection{Quantum mechanics on the torus}\label{s:quantum torus}
We recall as briefly as possible the basic setting for the quantum mechanics
of a system with $\T$ as phase space, as well as the quantization of the
automorphism $M$, referring to \cite{hb,de,bodb2} and references therein for
further details. In order to define the  Hilbert space associated to $\T$, we
first consider the translation operators $\hat{T}_{1}=\hat{T}_{(1,0)},\quad
\hat{T}_{2}=\hat{T}_{(0,1)}$, which satisfy  $\hat{T}_{1}\hat{T}_{2}=\e^{-\mi
/\hbar }\hat{T}_{2}\hat{T}_{1}$ as a result of (\ref{e:Heisenberg}). So for
the values of $\hbar $ defined as:\begin{equation} \label{e:N} N=\frac{1}{2\pi
\hbar }\in \N ^{*},
\end{equation}
one has the property $\left[ \hat{T}_{1},\hat{T}_{2}\right] =0.$ The Hilbert
space $L^{2}(\R )$ may then be decomposed as a direct integral of the joint
eigenspaces of $\hat{T}_{1}$ and $\hat{T}_{2}$:
\begin{equation}\label{e:decomposition}
L^{2}(\R )  =  \int^{\oplus} \hn\,
\frac{d^2\theta}{(2\pi)^{2}},\ \ \ 
\hn =  \left\{ |\psi\rangle
\in {\cal S}'(\R) \ \big\vert\  \hat{T}_1 |\psi \rangle 
=\e^{\mi\theta_1}|\psi\rangle,  
\hat{T}_2|\psi\rangle =\e^{\mi\theta_2}|\psi\rangle \right\} .
\end{equation}
The `angle' $\theta =(\theta_{1},\theta_{2})\in [0,2\pi [^{2}$ thus describes
the periodicity properties of the wave function under translations by an
elementary cell. $\mathcal{H}_{N,\theta }$ is $N$-dimensional.

We can define a projector $\hat{P}_{\theta }$ from ${\cal S}(\R) $ onto the
space $\mathcal{H}_{N,\theta }$:
\begin{equation}
\label{e:Projector} 
\hat{P}_\theta=\sum _{(n_1,n_2)\in \Z ^{2}}
\e ^{-\mi n_1\theta_1 -\mi n_{2}\theta_2}\: \hat{T}_{1}^{n_{1}}\:
\hat{T}_2^{n_2}=\sum_{n\in \Z^2}\e^{-\mi\theta\cdot n+\mi\delta_n}\:\hat{T}_n.
\end{equation}
The phase $\delta_n=-n_{1}n_{2}N\pi $ comes from the decomposition
$\hat{T}_n=\e^{-\mi\delta_n}\, \hat{T}_{1}^{n_1}\hat{T}_{2}^{n_2}$.

The Weyl quantization of a function 
$f(x)=\sum_{k\in\Z^2} f_k \e^{2\mi\pi (x\wedge k)}$ is
  an operator on $\mathcal{H}_{N,\theta }$ defined by
\begin{equation}\label{eq:weyl}
\hat f=\sum_{k\in\Z^2} f_k\: \hat T_{k/N}.
\end{equation}
For $|\psi\rangle\in \mathcal{H}_{N,\theta }$, its ``Wigner function''
$W_\psi(x)$ is the distribution implicitly defined via
\begin{equation}\label{e:wigner}
\langle \psi| \hat f | \psi\rangle =\int_{\T} f(x)\ W_\psi(x)\ dx,\ \ 
\mbox{ so that }\ \tilde W_\psi(k) = \langle \psi| \hat T_{k/N}|
\psi\rangle
\end{equation}
where the $\tilde W_\psi(k)=\int_{\T} \e^{2\mi\pi(x\wedge k)}\;W_\psi(x)\,dx$ 
are the Fourier coefficients of $W_\psi$.

Let now $M\in {\rm SL}(2,\Z)$, so that  $A,B,C,D$ (see Eq.~\eqref{eq:M})
are integers. One then easily deduces from (\ref{e: M and T}) and
(\ref{e:Projector}) that the quantum map $\hat M$ satisfies:

\begin{equation}
\hat{M}\, \hat{P}_{\theta }=\hat{P}_{\theta '}\,\hat{M},
\quad\mbox{with}\quad\theta '=\theta M^{-1} +2\pi \frac{N}{2}\left(CD,AB\right).
\label{e:dual map}
\end{equation}
The constant shift on the right hand side (RHS) 
is due to the phases $\delta_n$ appearing
in \eqref{e:Projector}. $\hat{M}$ will define an endomorphism in
$\mathcal{H}_{N,\theta }$ provided $\theta'\equiv\theta\bmod 2\pi$, {\it i.e.}
provided $\theta $ is a fixed point of the dual map
defined in \eqref{e:dual map}.  Given a hyperbolic
matrix $M$, such a fixed point exists for any $N$ \cite{de}. 
In particular, for any matrix $M$
the angle $\theta=(0,0)$ (periodic wavefunctions) is 
a fixed point if $N$ is even, while $\theta=(\pi,\pi)$ 
(antiperiodic wave wavefunctions) is a fixed point for $N$ odd. We will always
make this choice for our numerical examples.

From now on, we will assume that $M=\pm M_{(c,b)}\in {\rm SL}(2,\Z )$ 
is a fixed hyperbolic matrix defining a dynamics on the plane and on 
the torus. We will
therefore no longer indicate its dependence on $(c,b)$. We will also assume
that $\hbar $ is such that \eqref{e:N} holds, and for this $\hbar $ we select
an angle such that $\theta'\equiv\theta $. In general, $\theta$ can depend
on $\hbar$, but we will not indicate this dependence.


\section{Coherent states and their evolution}\label{s:cohstat}

\subsection{Standard and squeezed coherent states}
With the normalized state $|0\rangle $ defined by $a|0\rangle =0$, a
``standard'' coherent state is
\begin{equation}
\label{e:|z>} |x\rangle =\hat{T}_{x}|0\rangle ,\quad x=(q,p)\in \R ^{2}.
\end{equation}
More generally, we define for each $\tilde{c}\in \C^*$ the ``squeezed''
coherent states $|x,\tilde{c}\rangle$ by
\begin{equation}
\label{e:Rz} |\tilde{c}\rangle =|0,\tilde{c}\rangle
=\hat{M}_{(\tilde{c},0)}|0\rangle , \quad
|x,\tilde{c}\rangle=\hat{T}_{x}|\tilde{c}\rangle,
\end{equation}
where the ``squeezing operator'' $\hat{{M}}_{(\tilde{c},0)}$  is defined by
(\ref{e:operator M}), with $\tilde{b}=0$. Note that, in view of
(\ref{e:squeeze!}), given $M\in\mathrm{SL}(2,\R), \exists! \tilde c\in \C,
\sigma\in [0,2\pi[$ such that
\begin{equation}\label{e:squeeze!!}
\hat M|0\rangle = \e^{\mi \sigma} |\tilde c\rangle.
\end{equation}
For more details on coherent states, we refer to \cite{gil, pel}.

To avoid confusion, we will use a tilde for the parameters of the squeezing
operator $\hat{{M}}_{(\tilde c,0)}$, and keep untilde notations for the
parameters of the dynamics defined by the matrix $M\defi \pm M_{(c,b)}$ 
that are
at any rate kept fixed throughout the further discussion. 
In the $L^{2}(\R)$ representation, the state $|x,\tilde{c}\rangle $ is a
Gaussian wave packet with mean position $q$. Its Fourier transform is centered
around the mean momentum $p$.
For any state $|\psi \rangle \in L^{2}\left( \R \right) $, we define its
Bargmann function as $x\mapsto \langle x,\tilde c|\psi\rangle$, and its Husimi
function to be the positive function $\mathcal{H}_{\tilde c, \psi }$ defined
on phase space $\R ^{2}$ by:
\begin{equation}\label{eq:husimi}
\mathcal{H}_{\tilde c,\psi }\left( x\right) =\frac{\left| \langle
x,\tilde{c}|\psi \rangle \right| ^{2}}{2\pi\hbar}, \quad \textrm{which
satisfies}\quad\int_{\R^2} \mathcal{H}_{\tilde c,\psi }\left( x\right) dx = \|
\psi \|^2_{L^2(\R)}.
\end{equation}
Note that for given $|\psi\rangle$, the Bargmann and Husimi functions depend
on the choice of $\tilde c$. Also, the function 
$x\mapsto \langle x,\tilde c|\psi\rangle$ is the product of a Gaussian 
factor with a function
holomorphic with respect to a $\tilde c$-dependent holomorphic structure. The
term Bargmann function is usually reserved for the holomorphic factor, but we
find it convenient to adopt here a slightly different convention.

\begin{figure}[htbp]
{\par\centering
\resizebox*{0.4\columnwidth}{!}{\includegraphics{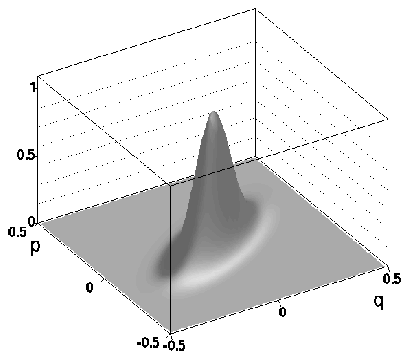}}
\resizebox*{0.4\columnwidth}{!}{\includegraphics{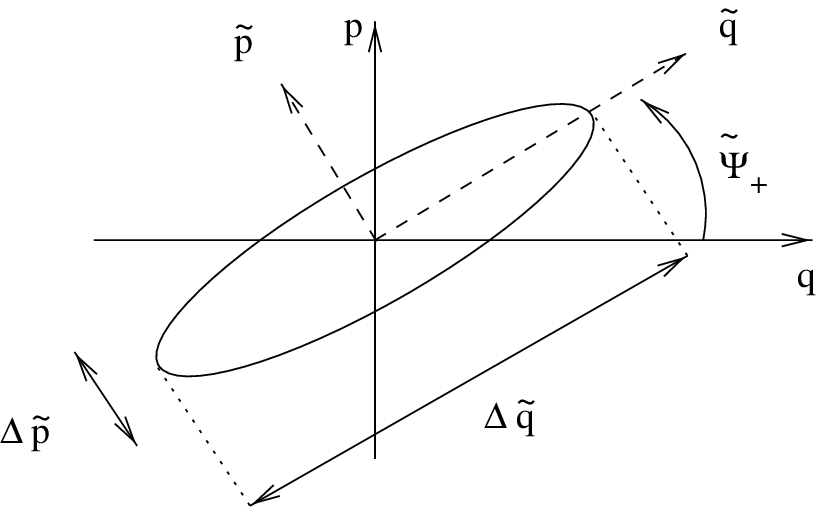}}
\par}
\caption{\label{f:etat squeezed} Modulus square of the Bargmann function of a
squeezed coherent state $|\tilde{c}\rangle $, as given in \eqref{e:Hus of
|c>}. The inverse Planck's constant $N=1/h=40$, and the squeezing parameter
$\tilde c=-\mi|\tilde c|\e^{-2\mi\tilde\psi_+}$ with $|\tilde c|=0.962$,
$\tilde\psi_+=32^\circ$ (this corresponds to $\tilde c_1$ for the
map $\hat M_{\rm Arnold}$). (a) Three dimensional picture. (b) Typical size and
orientation of the distribution: ellipse ``supporting'' the distribution.}
\end{figure}

We will need the explicit expression of the (standard) Bargmann and Husimi
functions of the squeezed coherent state $|\tilde{c}\rangle$:
\begin{equation}\label{e:Hus of |c>}
\langle x,0|\tilde c\rangle=\frac{1}{\sqrt{\cosh |\tilde{c}|}} \exp\left\{-\mi
\frac{\tilde{q}\tilde{p}\tanh |\tilde{c}|}{2\hbar }\right\}
\exp\left\{-\frac{1}{2}\left(\frac{\tilde{q}^{2}}{\Delta \tilde{q}^{2}}
+\frac{\tilde{p}^{2}}{\Delta \tilde{p}^{2}}\right)\right\}.
\end{equation}
Here the unstable-stable frame $(\tilde{q},\tilde{p})$ of the symmetric matrix
$M_{(\tilde c, 0)}$ is easily seen from the formulas in Section~
\ref{sec:linearcldyn} to be obtained from $(q,p)$ by a rotation of angle
$\tilde{\psi}_+$ (Figure~\ref{f:etat squeezed}), and the widths are given by
\begin{equation}
\Delta \tilde{q}^{2}=  \frac{2\hbar }{\left( 1-\tanh \left| \tilde{c}\right|
\right) }, \quad \Delta \tilde{p}^{2}=  \frac{2\hbar} {\left( 1+\tanh \left|
\tilde{c}\right|\right) }.
\end{equation}

Standard and squeezed coherent states on the torus are defined to be the images
of the previous coherent states by the projector $\hat{P}_\theta$. We use the
notation:
\begin{eqnarray}
|x,\tilde c, \theta \rangle =\hat{P}_{\theta }|x,\tilde c\rangle \in
 {\mathcal H}_{N,\theta}.
\end{eqnarray}
These states are asymptotically normalized:
$$
 \langle  x, \tilde c, \theta| x,\tilde c,\theta\rangle=
1+ \cO (\e^{-C(\tilde c)/\hbar})
$$
and satisfy a resolution of the identity on the Hilbert spaces ${\cal
H}_{N,\theta}$ \cite{bodb2}:
\begin{equation}
\int_{\T} \frac{dqdp}{2\pi\hbar}\ |x,\tilde{c},\theta\rangle\langle
x,\tilde{c},\theta| = \hat I_{{\cal H}_{N,\theta}}.
\end{equation}
Similarly as above, one defines for any $|\psi\rangle\in{\mathcal
H}_{N,\theta}$ its Bargmann ``function'' $x\mapsto \langle x,\tilde c,\theta
|\psi\rangle$ (which is actually a {\it section} of a suitable line bundle
over $\T$, i.e. a quasiperiodic function on $\R^2$, but this shall not
interest us here), and Husimi function $ \mathcal{H}_{\tilde c,\psi,\theta }(
x) =N\left| \langle x,\tilde{c},\theta|\psi \rangle \right|^{2}, $ a {\it bona
fide} function on the torus (of which we omit to indicate the $N$-dependence).


\subsection{The evolution of coherent states}\label{s:evolution}
Before turning to  quasimodes, we need to study in detail the  quantum
evolution of the squeezed coherent state $|\tilde{c},\theta\rangle$  which is
given by $ |t;\tilde c,
\theta\rangle\defi\hat{M}^{t}|\tilde{c},\theta\rangle$, $t\in\Z$. 
 We will
extend this notation to any {\it real} time, by $|t;\tilde c,
\theta\rangle\defi\hat{P}_\theta \e^{-\mi\hat{H}t/\hbar}|\tilde c\rangle$. Due
to (\ref{e:squeeze!!}), the states $|t;\tilde c,
\theta\rangle$ are again squeezed coherent states (up to a global phase), so
this evolution defines a time flow $\tilde{c}(t)$ on the family of squeezed
coherent states centered at the origin. All squeezed states at the origin
have even parity: 
$\hat P|\tilde c\rangle=|\tilde c\rangle$, so that the 
evolution of $|\tilde c\rangle$ through the map 
$\hat M\hat P$ is the same as through $\hat M$ (yet, these two maps might 
require different values for $\theta$, see Eq.~\eqref{e:dual map}).

It will turn out that $|t;\tilde c, \theta\rangle$  will be most simply
described if the initial squeezed state $|\tilde{c}_{0},\theta\rangle$ at time
$t=0$ is well chosen in terms of the decomposition \eqref{e:M is QDQ}.
Defining, with the notations of \eqref{e:M is QDQ}--\eqref{eq:parameterchange},
$\tilde{c}_{0}=-b_2 \e^{-2\mi b_1}$, it is  easy to check that
$|\tilde{c}_{0}\rangle=\e^{-\mi b_1/2} \hat{Q}|0\rangle$ since $\hat M_{(\tilde
c_0,0)}=\hat R(b_1)\hat B(b_2)\hat R(-b_1)$ and $\hat{R}(-b_1)|0\rangle
=\e^{-\mi b_1/2}|0\rangle$. Then, with $\hat M = \hat Q \hat D(\lambda)\hat
Q^{-1}$,
\begin{equation}\label{e:positivity}
\hat{M}^{t}|\tilde{c}_{0}\rangle =\e^{-\mi b_1 /2}\hat{Q}\, \hat{D}(\lambda
t)|0\rangle, \qquad\langle \tilde{c}_{0}|\hat{M}^{t}|\tilde{c}_{0}\rangle
=\langle 0|\hat{D}(\lambda t)|0\rangle =\frac{1}{\sqrt{\cosh (\lambda t)}}\in
\R ^{+},
\end{equation}
so the overlap $\langle \tilde{c}_{0}|\hat{M}^{t}|\tilde{c}_{0}\rangle$
is real positive for all times. 

For later purposes we note that, defining, for $s\in\R$, $\tilde c_s\in\C,
\sigma_s\in [0,2\pi[$ by
\begin{equation}\label{e:defcs}
\e^{-\mi\frac{\hat H}{\hbar}s}|\tilde c_0\rangle = \e^{\mi \sigma_s} |\tilde
c_s\rangle,
\end{equation}
(see \eqref{e:squeeze!!}), it is clear that
$\langle \tilde{c}_s|\e^{-\mi\hat{H}t/\hbar}|\tilde{c}_s\rangle$ is real 
positive for all
$t$. In fact, it can be shown that the $\tilde c_s$ are the only values 
of $\tilde c$ with
this property. Among all $s$, $s=0$ maximizes $\left| \langle
0|\tilde{c}_s\rangle \right| ^{2}$, so $|\tilde{c}_{0}\rangle$ is in a 
sense the
most localized state among all $|\tilde{c}_s\rangle$.

In this paper, we will almost exclusively build quasimodes from coherent states with
``squeezing'' $\tilde c_0$; this choice is made for pure convenience, and our
main semiclassical results apply to more general squeezings as well 
(see Section~\ref{s:robustness} and Appendix~\ref{s:changesqueeze}).

Before turning to $|t;\tilde c, \theta\rangle\in\hn$, we first  describe the
evolved state $|t;\tilde c_0\rangle\defi\e^{-\mi\hat{H}t/\hbar}|\tilde
c_0\rangle \in L^2(\R)$, by studying its Husimi function on the plane, as
defined in \eqref{eq:husimi}. It will be convenient (but again not absolutely
necessary for our results, see Section~\ref{s:changesqueeze}) to adapt the
choice of $\tilde c$ in the definition of this Husimi function to the dynamics
$M$ by putting $\tilde c=\tilde{c}_{0} $.  One then computes
\begin{equation}\label{e:Husimi of evolved plane c.s.}
{\mathcal H}_{\tilde c_0, t}(x)\defi \frac{\left| \langle
\tilde{c}_{0}|\hat{T}^{\dagger }_{x}|t;\tilde{c}_{0}\rangle \right|
^{2}}{2\pi\hbar}=\frac{\left| \langle 0|\hat{Q}^{\dagger } \hat{T}^{\dagger
}_{x} \hat{Q}\hat{D}\left( \lambda t\right) \hat{Q}^{\dagger }
\hat{Q}|0\rangle \right| ^{2}}{2\pi\hbar}= \frac{\left| \langle
0|\hat{T}^{\dagger }_{Q^{-1}x }\hat{D}\left( \lambda t\right) |0\rangle
\right| ^{2}}{2\pi\hbar}.
\end{equation}
It is now natural to use the coordinates  $\left( q',p'\right) =Q^{-1}\left(
q,p\right) \in \R ^{2}$ attached to the unstable-stable basis $({v }_{+,\,
}{v }_{-})$ (see Eq.~\eqref{e:decompo-n}). In terms of these, the Husimi
function is a Gaussian drawn on the unstable and stable axes:
\begin{equation}
\label{e:<R|TM|R>} \mathcal{H}_{\tilde c_0, t}(x) =\frac{1}{2\pi\hbar\cosh
(\lambda t)}\exp \left( -\frac{q^{\prime 2}}{\Delta q^{\prime
2}}-\frac{p^{\prime 2}}{\Delta p^{\prime 2}}\right) ,
\end{equation}
with
\begin{equation}\label{e:widths}
\Delta q^{\prime 2}=\frac{2\hbar }{1-\tanh (\lambda t)}
\stackrel{t\to\infty}\sim\hbar\, \e ^{2\lambda t},
 \qquad \Delta
p^{\prime 2}=\frac{2\hbar }{1+\tanh (\lambda t)}  =  \e ^{-2\lambda t}\Delta
q^{\prime 2}\stackrel{t\to\infty}\rightarrow \hbar .
\end{equation}
The Husimi distribution of the evolved state $|t;\tilde
c_0\rangle$ therefore spreads
exponentially (with rate $\lambda $) in the unstable direction of the map, and
has a finite transverse width $\sqrt{\hbar }$. It ``lives'' in an
elliptic region of phase space centered on the origin and of area $\Delta q'
\Delta p'\sim \hbar \e^{\lambda t}$. Due to conservation of the total
probability, the height of the distribution decreases exponentially.

\begin{figure}[htbp]
{\par\centering \resizebox*{!}{3.6cm}{\includegraphics{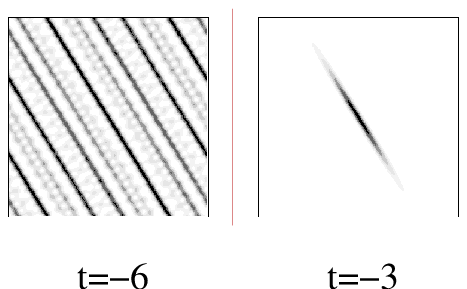}}
\resizebox*{!}{3.6cm}{\includegraphics{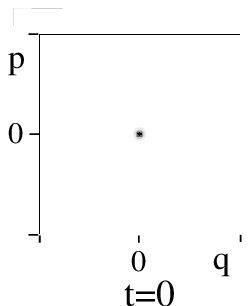}}
\resizebox*{!}{3.6cm}{\includegraphics{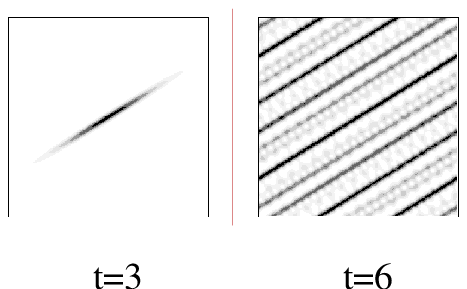}} \par}

\caption{\label{f:evol}Husimi function of the state
$|t;\tilde{c}_{0},\theta \rangle $ for the dynamics \eqref{e:cat map} 
and $N=1/(2\pi \hbar )=500$. One has $T\approx 8.37$.}
\end{figure}

We now turn to $|t;\tilde c_0,\theta\rangle
=\hat M^t|\tilde c_0,\theta\rangle$, $t\geq 0$ and its 
Husimi function
$$
{\mathcal H}_{\tilde c_0, t, \theta}(x) = N|\langle x,\tilde c_0 | \hat
P_\theta |t; \tilde c_0 \rangle |^2.
$$
It is clear from (\ref{e:Projector}) that $\langle x,\tilde c_0| \hat P_\theta
|t; \tilde c_0 \rangle$ is obtained by summing (up to some phases) the
translates of  the function $\langle x, \tilde c_0 |t; \tilde
c_0\rangle$ into the different phase space cells of size $1$ centered on the
points of  $\Z^2$ (the cell around $0$ will be called the fundamental cell
$\mathcal{F}$).  Consequently, it follows from
(\ref{e:<R|TM|R>})--(\ref{e:widths}) that this function is non-negligible at a
point $x\in\mathcal{F}$ only if $x$ lies within a distance $\sqrt \hbar$ from
a stretch of length $\Delta q'\approx\sqrt \hbar \e^{\lambda t}=\e ^{\lambda
(t-T/2)}$ of the unstable manifold through $0$ (Figure~\ref{f:evol}). Here we
introduced   the \textbf{Ehrenfest time} as
\begin{equation}
\label{e:Ehrenfest time} T\defi \frac{|\log \hbar |}{\lambda }.
\end{equation}

Since at time $|\log \hbar |/(2\lambda )=T/2$, $\Delta q'$ reaches the size
$1$ ({\em i.e.} the size of the torus), it is clear that for shorter times the
Husimi function ${\mathcal H}_{\tilde c_0, t, \theta}$ lives in an elliptic
region of shrinking diameter $\sqrt \hbar \e^{\lambda t}$ around $0$.

For times larger than $T/2$, this Husimi function starts to wrap itself around
the torus along the unstable axis or, equivalently, the support of some of the
translates $\langle x+n, \tilde c_0|t;\tilde c_0\rangle $  
start to enter into the fundamental
cell. The diophantine properties guarantee that the branches of the piece of
length $\Delta q'$ of the unstable manifold passing through the origin are
roughly at a distance $1/\Delta q'=\e^{-\lambda(t-T/2)}$ from each other (Fig.~
\ref{f:varietes}). Consequently, as long as $\Delta p'<<
\e^{-\lambda(t-T/2)}$, {\em i.e.} as long as $t\leq (1-\epsilon)T$, the main
contribution to 
$\langle x,\tilde c_0|\hat P_\theta|t;\tilde c_0\rangle $
and hence to the Husimi function $\mathcal{H}_{\tilde c_0, t,\theta}$ comes
from a single term $\langle x+n,\tilde c_0|t;\tilde c_0\rangle $ for
most $x\in\mathcal{F}$. We say there are no interference effects. The regime
$(1+\epsilon)T/2\leq t\leq (1-\epsilon)T$ was studied in \cite{bondb1} where
it was proven that on that time scale the Husimi function equidistributes on
the torus.

For longer times $t\geq (1+\epsilon)T$, when the area $\Delta p'\Delta q'$ 
occupied by the
support of $\mathcal{H}_{\tilde c_0,t}$  becomes larger than the area of the
torus itself, several terms may contribute equally to $\langle x,\tilde
c_0|\hat P_\theta |t;\tilde c_0\rangle$. In the next subsection  we 
give a detailed control on the onset of this ``interference regime'' up to
time $2T$ for the Husimi function of $|t;\tilde c_0,\theta\rangle$
evaluated at the origin $x=0$; we shall show that the interferences remain
``small'' up to the time $2T$.

As a last remark, we point out that the above discussion is symmetric with
respect to time reversal. For negative times, $\mathcal{H}_{\tilde c_0, t,
\theta}$  spreads along the stable direction, reaches the boundary of
$\mathcal{F}$ around $-T/2$, and will interfere with itself for $t\leq -T$.


\subsection{Estimating the interference effects}\label{s:interference bound}
As explained in the introduction, our crucial technical estimate concerns the
{\it autocorrelation function} for the state $|\tilde c_0,\theta\rangle $,
given by $\langle \tilde c_0,\theta |\hat{M}^t|\tilde c_0,\theta\rangle$. More
generally, we will need control on
\begin{equation}
\label{e:def de I(t)} \langle \tilde c_s,\theta |\hat{M}^t|\tilde c_s,\theta
\rangle = \langle \tilde c_s|\hat{P}_{\theta}\hat{M}^t|\tilde c_s\rangle
=\langle \tilde c_s|\hat{M}^t|\tilde c_s\rangle +I(t, s),
\end{equation}
where we separated the contribution of the term $n=(0,0)$ (the
``plane overlap''), from the remaining terms:
\begin{equation}\label{definition I(t)}
I(t,s)\defi \sum _{n\in \Z^{2}_*}\e ^{-\mi n\cdot\theta +\mi
\delta_n}\: \langle \tilde c_s|\hat{T}_n\,\hat M^t|\tilde c_s\rangle.
\end{equation}
This remainder represents the interference of the evolved plane coherent state
with the lattice-translated initial state. We will show that these
contributions tend to $0$ as $N\to\infty$, uniformly for all times $
|t|\leq 2(1-\epsilon)T$, for any fixed $\epsilon>0$.

A trivial upper bound is
\begin{equation}
\label{e:L(t)} \left|I(t,s)\right| \leq \sum _{n\in \Z^{2}_*}
\left|\langle n,\tilde{c}_{s}|\e^{-\frac{i}{\hbar}\hat H t}|\tilde
c_s\rangle\right|\defi J_0(t,s),
\end{equation}
and we shall estimate the RHS. 
Note that we extended $I(t,s)$ in
the natural way to real times $t$. The detailed proofs of the estimates below
are given in Appendix~\ref{a:interference term}; here we limit ourselves to
explaining the underlying ideas and to an instructive comparison with a
numerical example. For simplicity, we will concentrate on the case $s=0$.

\begin{figure}[htbp]
{\par\centering \resizebox*{0.70\columnwidth}{!}{\includegraphics{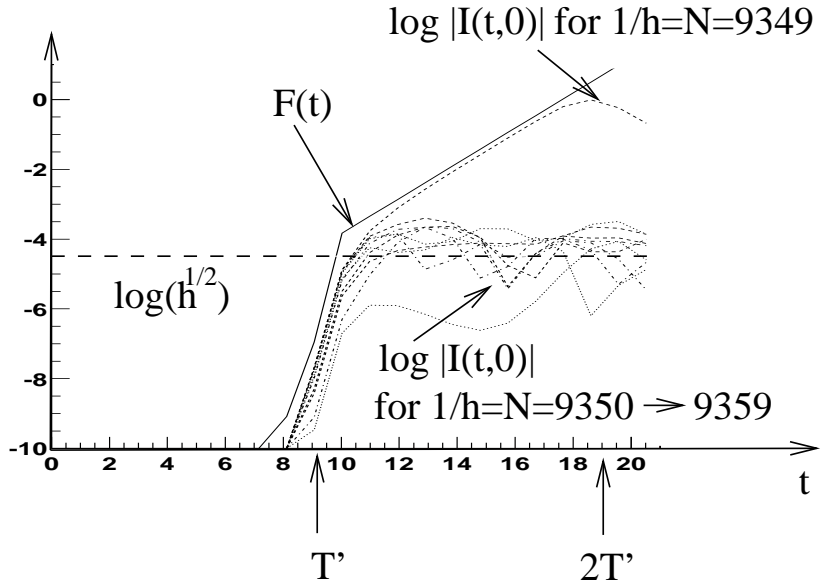}}
\par} \caption{\label{f:interf} Numerical calculations of $\log|I(t,0)|$, for
the map \eqref{e:cat map}. The heuristic upper bound $F(t)$ (solid line) is
defined in terms of the shifted Ehrenfest time $T'=\log(N)/\lambda$:
$F(t)=-\frac{\lambda t}{2}-\e^{\lambda \left(T'-t\right)}-1$ for $0<t<T'$ and
$F(t)=\frac{\lambda }{2}\left( t-2T'\right)+0.5$ for $T'<t<2T'$. The
horizontal dashed line at $\log(h^{1/2})$ gives the order of magnitude of the
plateau for $t>T'$.}
\end{figure}

We define a time-dependent metric on the plane adapted to the Gaussian in
\eqref{e:<R|TM|R>}:
$$
\|\, x\,\|^{2}_{t}\ \defi\  \frac{1}{2}\left( \frac{q'(x)}{\Delta
q'(t)}\right) ^{2}+\frac{1}{2}\left( \frac{p'(x)}{\Delta p'(t)}\right)^{2}.
$$
The RHS of \eqref{e:L(t)} is simply the sum of this Gaussian of height
$H_t=(\cosh\lambda t)^{-1/2}$ evaluated at all
nonzero integer lattice points. The diophantine properties proven in Section~
\ref{s:classaut} provide information on the position of the integer lattice
with respect to the ellipse $\{\|x\|^2_t=\nolinebreak 1\}$ and allow us to 
prove the 
following
estimates:

\begin{itemize}
\item for relatively short times (meaning $|t|\leq (1-\epsilon)T$), 
all lattice points $n\neq 0$
are far outside the support of the Gaussian so that $\|n \|_t$ is large. In
fact, the distance $\| n\|_t$ reaches its minimum for a single point $n_o$ 
(more precisely a finite number $\mathcal{N}$ of points), with $\|
n_o\|^2_t>c\,\frac{\e ^{-\lambda |t|}}{\hbar }>> 1$. Note that, here and in the
following, we write $f(\hbar)<<g(\hbar)$ when
$\lim_{\hbar\to0}f(\hbar)/g(\hbar)=0$. $J_0(t,0)$ is 
dominated by the contribution of this finite set of points, given by
$\mathcal{N}\,H_t\exp\left\{-\| n_o\|^2_t\right\}$, the contributions of
farther points being much smaller.  The precise bound proven in the appendix
reads:
\begin{equation}\label{e:boundshort}
|t|\leq T\Longrightarrow |I(t,0)|\leq 2\sqrt{2}\, \e ^{-\lambda |t|/2}\,
\exp\left\{-C_{o}\frac{\e ^{-\lambda |t|}}{2\hbar}\right\} 
 \left[ 1+C\e^{\lambda (|t|-T)/2}\right],
\end{equation}
where the constant $C_{o}$ is the parameter of the diophantine equation
(\ref{eq:dioph3}), and $C$ can be computed explicitly (it depends only on
$M$).

\item For times $|t|\geq T$, a large number of lattice points
($\mathcal{N}_t= \Delta q'(t)\Delta p'(t)\sim \e^{\lambda (|t|-T)}$) are
contained in the ellipse ({\it i.e.} satisfy $\|n\|_t\leq 1$), and their
collective contribution dominates the RHS of \eqref{e:L(t)}:
$|I(t)|\lesssim \mathcal{N}_t H_t\sim\e^{\lambda (-T+|t|/2)}$. This is indeed
essentially what we prove:
\begin{equation}\label{e:boundlong}
T\leq |t|\Longrightarrow |I(t,0)|\leq \frac{2\pi \sqrt{2}}{C_{o}}\, 
\e^{\lambda(-T+ |t|/2)}\, \left[ 1+C'\e ^{\lambda (T-|t|)/2}\right] ,
\end{equation}
where $C'$ can be computed explicitly in terms of $M$. This upper bound
becomes of order unity for $|t|\simeq 2T$.
\item From the definition (\ref{e:def de I(t)}), we have trivially for
any time
\begin{equation*}
|I(t,0)|\leq  \langle\tilde c_0,\theta|\tilde c_0,\theta\rangle +\langle
\tilde c_0|\hat M^t|\tilde c_0\rangle \leq 1+\cO \left(\e^{-C(\tilde
c_0)/\hbar}\right) +\frac{1}{\sqrt{\cosh (\lambda |t|)}}.
\end{equation*}
\end{itemize}
Combining these estimates (generalized to $s\neq 0$), 
one obtains the following proposition:
\begin{prop}\label{prop:crucial}
There exist positive constants $C$, $C'$, $C''$ such that for all times $t\in
\R$, and for all $s$ in a bounded interval
\begin{equation}
\label{e:upper bound for I(t)} \vert I(t,s)\vert \leq J_0(t,s)\leq \min \left( C\hbar \e
^{\lambda |t|/2},\: 1+\sqrt{2}\e ^{-\lambda |t|/2}+C'\e^{-C''/\hbar}\right) .
\end{equation}
\end{prop}
This shows that the interferences remain small until times of
order $2T$. The existence of ``short quantum periods'' for certain values of
$\hbar$ (see the introduction and Section~\ref{s:min quantum period}) implies
that $I(t,0)$ is of order $1$ at $t=P\simeq 2T$ for these values of $\hbar$.
This is further illustrated in Figure~\ref{f:interf}.

Figure~\ref{f:interf} shows numerical calculations of $\log| I(t,0)|$ for
values of Planck's ``constant'' $N=9349\rightarrow 9359$ and compares them
to $F(t)$, which is essentially given by the upper bounds
\eqref{e:boundshort}--\eqref{e:boundlong}. We observe that, whereas
\eqref{e:boundshort} is close to optimal, the same is not true for
(\ref{e:boundlong}) {\em for most values of} $N$: there is a 
``plateau'' $\log|I(t,0)|\simeq \log(\hbar^{1/2})$ for $t>T'$, where
$T'=\log(N)/\lambda$ is a shifted Ehrenfest time. This plateau can be
explained by assuming that the phases which multiply the different terms in
$I(t,0)$ are uncorrelated, like independent random phases. For $t>>T$, the RHS
of \eqref{definition I(t)} could then be replaced by a sum of many ($\simeq
\mathcal{N}_t$) terms with identical moduli $H_t$ but random uncorrelated
phases, similar to a 2-dimensional random walk.  The modulus of the sum ({\it
i.e.} the length of the random walk) has a typical value $|I(t,0)|\sim
\sqrt{\mathcal{N}_t}\,H_t\sim \hbar^{1/2}$, independent of time: this is
indeed what we see numerically. 

However, for the value of $N=9349$, corresponding to a ``short quantum
period'' $P=19$, as discussed in Section \ref{s:min quantum period},
$\log|I(t,0)|$ is close to the upper bounds
(\ref{e:boundshort})--(\ref{e:boundlong}) up to time $P\simeq 2T'$. In such
exceptional cases ---crucial in this paper--- there appears strong
correlations between the phases in the sum $I(t,0)$: the random walk somehow
becomes ``rigid'', which makes its total length of the same order as the sum
of individual lengths, $|I(t,0)|\sim J_0(t,0)\sim\mathcal{N}_t H_t$. 
This rigidity
can actually by analyzed directly from the explicit expression for the phases
\cite{fred-steph-phases}: one first
finds that for these special values of Planck's constant $N=N_k$ 
and $t$ in the interval $T<t<2T$, 
the phases corresponding to the relevant $\sim\mathcal{N}_t$ lattice points 
are all close to $2d$-th roots of unity, where $d=(\tr M)^2-4$ (in
the example $M=M_{\rm Arnold}$ and $N_k=9349$, the relevant phases are all close to unity). 
Then, the sum of these $\sim\mathcal{N}_t$ phases behaves like $G(M,N_k)\mathcal{N}_t$ 
for $\mathcal{N}_t>>1$, and one can check that the prefactor $G(M,N_k)$ (a Gauss sum) is 
bounded away from zero uniformly ({\it e.g.} $G(M_{\rm Arnold},9349)=1$). This
explains the behaviour $|I(t,0)|\sim J_0(t,0)$. This situation drastically
differs from the case of a  ``generic'' $N$, where the relevant phases are more or less
equidistributed over the circle.


\section{Quasimodes at the origin}\label{s:quasimodes}

\subsection{Continuous time versus discrete time quasimodes}
\label{s:continuous}

We are now ready to  study the quasimodes \eqref{eq:quasimode} and
\eqref{eq:locerg} ``associated'' with the periodic orbits of the dynamics
generated by $M$, as discussed in the introduction. To alleviate the
notations, we start with the case where the orbit is simply the fixed point
$(0,0)\in \T$. The rather straightforward generalization to arbitrary orbits
is given in Section~\ref{s:other_orbit}. Note that the Ehrenfest time
$T=\frac{|\ln\hbar|}{\lambda}$ is in general not an integer: whenever $T$ or
$T/2$ appears in a sum boundary, they should therefore be replaced  by the
nearest integer.

It will be convenient to also consider slightly modified quasimodes, for which
the initial state is not the squeezed coherent state $|\tilde
c_0,\theta\rangle$ as in \eqref{eq:quasimode}, but rather the following
superposition of squeezed coherent states:
\begin{equation}\label{e:def de |psi_0,phi>}
\hat{P}_{\theta }\: \int ^{1}_{0}dt\,\e ^{-\mi\phi t}\e^{-\frac{\mi}{\hbar}\hat
H t}|\tilde c_0\rangle.
\end{equation}
The ``continuous time'' version of the quasimodes defined in
\eqref{eq:quasimode} then reads:
\begin{align}\label{e:cont quasimode}
|\Phi^{\rm cont}_\phi\rangle&\defi
\hat{\mathcal{P}}_{-T,T,\phi}\:\hat{P}_{\theta }\: \int ^{1}_{0}dt\,
\e ^{-\mi\phi t}\e^{-\frac{\mi}{\hbar}\hat H t}|\tilde c_0\rangle\\
&=\hat{P}_{\theta }\:\int_{-T}^T dt\, \e ^{-\mi \phi t}
\e^{-\frac{\mi}{\hbar}\hat H t}|\tilde c_0\rangle.\label{e:plane torus}
\end{align}
Here we introduced, for any $\phi\in\R, t_0<t_1\in\Z$, the operator
\begin{equation}\label{eq:proj}
\hat{\mathcal P}_{t_0, t_1,\phi} = \sum_{t=t_0}^{t_1-1} \e^{-\mi t\phi}\hat
M^t,
\end{equation}
and the equality \eqref{e:plane torus} follows from a trivial computation.

These quasimodes can also be decomposed into 4 parts $|\Phi^{\rm
cont}_{j,\phi}\rangle$, obtained by integrating in $t$ over time intervals of
length $T/2$, then projecting the obtained state in $\hn$. A remarkable and
useful property (derived from Poisson's formula) is that we can recover the
``discrete time'' quasimodes $|\Phi^{\rm
disc}_\phi\rangle$ defined in \eqref{eq:quasimode} from the ``continuous
time'' ones:
$$
|\Phi^{\rm disc}_\phi\rangle=\sum_{k\in\Z}|\Phi^{\rm cont}_{\phi+2\pi
k}\rangle.
$$
Notice that the state in (\ref{e:def de |psi_0,phi>}) is not $2\pi$-periodic
with respect to $\phi$ so that the quasimodes $|\Phi^{\rm cont}_{\phi}\rangle$
depend on the ``quasienergy'' $\phi\in\R$.

The main reason for considering continuous time quasimodes is that 
they are easily
connected with generalized eigenstates of the Hamiltonian $\hat H$, which
allows to \emph{pointwise} describe their Husimi densities, a task we turn to
in  Section~\ref{s:pointwise}.

In the next subsection, we start our study of the above quasimodes. We will
use from now on the notation $|\Phi_\phi\rangle$  in statements that are valid
both for $|\Phi^{\rm disc}_\phi\rangle$ and $|\Phi^{\rm cont}_{\phi}\rangle$
(and similarly for $|\Phi_{j,\phi}\rangle$).

\subsection{Orthogonality of the states $|\Phi_{j,\phi}\rangle_n$ at fixed
$\phi$\label{s:orthogonality}}

\begin{prop}\label{prop:ortho}
(i) The states $|\Phi_{j,\phi}\rangle$, $j=1,2,3,4$ and $|\Phi_\phi\rangle$
satisfy, as $\hbar\to0$
\begin{equation}\label{eq:norms}
\langle \Phi_{j,\phi}|\Phi_{j,\phi}\rangle=\frac{T}{2}S_1(\lambda,\phi)+\cO
(1), \qquad\langle \Phi_{\phi}|\Phi_{\phi}\rangle= 2T S_1(\lambda,\phi)+\cO
(1),
\end{equation}
where the smooth function $S_1(\lambda,\phi)$ is strictly positive for all
$\phi\in\R$ and $\mathcal{O}(1)$ is uniformly bounded in $\phi$. In particular
these states do not vanish for small enough $\hbar$ and the normalized
quasimodes $|\Phi_\phi\rangle_n$ satisfy \eqref{e:quasi property}.

(ii) Furthermore, for all $\phi\in\R$, the $|\Phi_{j,\phi}\rangle_n$ become
mutually orthogonal in the semiclassical limit: for all $j\not=
k\in\{1,\ldots,4\}$
\begin{equation}\label{e:asymptotic norm}
\lim_{\hbar\to0} \ _n\!\langle \Phi_{j,\phi}|\Phi_{k,\phi}\rangle_n=0.
\end{equation}
The limit is uniform for all $\phi$ in a bounded interval.

(iii) Consequently, for all $\phi\in\R$,
$$
_n\!\langle \Phi_\phi|\Phi_{j,\phi}\rangle_n\rightarrow 1/2,\quad _n\!\langle
\Phi_\phi |\Phi _{{\rm erg},\phi}\rangle_n\rightarrow 1/\sqrt{2}\quad
\textrm{and}\quad _n\!\langle \Phi_\phi|\Phi_{{\rm loc},\phi}
\rangle_n\rightarrow 1/\sqrt{2}.
$$
\end{prop}

\begin{proof} 
(i) We first give a detailed proof for the ``continuous time'' quasimodes.
Writing $k=j-i\in\{0,1,2,3\}$, a simple computation yields (see
(\ref{e:defcs}))
\begin{equation*}
\langle \Phi_{i,\phi}^{\rm cont}|\Phi_{j,\phi}^{\rm cont}\rangle
=\sum_{t=0}^{\frac{T}{2}-1}\sum_{t'=0}^{\frac{T}{2}-1}\int_0^1 ds\ \int_0^1
ds' \,\e^{-\mi (t-t'+s-s'+kT/2)\phi } \,\langle \tilde{c}_{0} |
\hat{P}_{\theta }\e^{-\frac{i}{\hbar}\hat H(t-t'+s-s'+kT/2)} |\tilde
c_{0}\rangle .
\end{equation*}
Using (\ref{e:def de I(t)}) and (\ref{e:L(t)}) this becomes:
\begin{equation}\label{e:<Phi_i|Phi_j>}
\langle \Phi_{i,\phi}^{\rm cont}|\Phi_{j,\phi}^{\rm cont}\rangle
=\int^{T/2}_{-T/2}ds\; \left(\frac{T}{2} -|s|\right) \: \e^{-\mi (s+kT/2)\phi} 
\langle
\tilde{c}_{0}| \e^{-\frac{i}{\hbar}\hat H(s+kT/2)}|\tilde c_{0}\rangle
+ \mathrm{error}
\end{equation}
where
$$
\mathrm{error}\leq \int_0^1 ds\ \int_0^1 ds'
\sum_{t=-\frac{T}{2}}^{\frac{T}{2}-1} \big(\frac{T}{2}-|s|\big)\ 
J_0(t + k\frac{T}{2} + s-s', s').
$$
Using the bound (\ref{e:upper bound for I(t)}), one readily finds that the
second term is $\cO (\hbar^{\frac{3-k}{4}})$.

To estimate the norm of $|\Phi_j^{\rm cont}\rangle$, there remains to compute
the integral in (\ref{e:<Phi_i|Phi_j>}) in the case $i=j$, that is $k=0$:
\begin{align*}
\int^{T/2}_{-T/2}ds\; (T/2 -|s|) \, \e^{-\mi s\phi } \langle
\tilde{c}_{0}|\e^{-\frac{i}{\hbar}\hat H s}|\tilde c_0\rangle &=\int
^{T/2}_{-T/2}ds\,
\frac{(T/2-| s|) \e^{-\mi s\phi }}{\sqrt{\cosh ( \lambda s) }}\\
&=\frac{T}{2} \, S_{1}^{\rm cont}(\lambda ,\phi ,T/2)-S_{2}^{\rm cont}(\lambda
,\phi ,T/2),
\end{align*}
where the (real) functions $S_1^{\rm cont}$, $S_2^{\rm cont}$ are defined as
follows:
\begin{equation}\label{e:S1}
S_{1}^{\rm cont}(\lambda ,\phi ,\tau)\defi \int ^{\tau}_{-\tau}dt\,
\frac{\e^{-\mi t\phi }}{\sqrt{\cosh \left( \lambda t\right) }},  \quad
S_{2}^{\rm cont}(\lambda ,\phi ,\tau)\defi \int ^{\tau}_{-\tau}dt\,
\frac{\left| t\right| \e^{-\mi t\phi }}{\sqrt{\cosh \left( \lambda t\right) }}.
\end{equation}
The limits of $S_{i}^{\rm cont}(\lambda,\phi,\tau)$ as $\tau\rightarrow\infty$
clearly exist. We only give the value for $S_{1}^{\rm cont}$, the most
relevant one for our purposes \cite{Ba}:
\begin{equation}
S_{1}^{\rm cont}(\lambda ,\phi)\defi \lim_{\tau\to\infty}S_{1}^{\rm
cont}(\lambda ,\phi ,\tau ) =\frac{1}{\lambda \sqrt{2\pi }} \left|\Gamma
\left(\frac{1}{4}+\mi\frac{\phi }{2\lambda }\right)\right| ^{2}.
\end{equation}
For fixed $\lambda$, this function is maximal for $\phi=0$ 
(with value $\approx 5.244/\lambda$), and decreases
as $\sqrt{\frac{4\pi}{\lambda|\phi|}}\e^{-\pi|\phi|/2\lambda}$ for 
$|\phi|\to\infty$. A crucial property is the \emph{strict positivity} of this 
function, for all values $\lambda >0$, $\phi \in \R$.  

The computation of $\langle \Phi_\phi|\Phi_\phi\rangle$ is similar.

(ii) We now estimate the overlaps $\langle\Phi_{i,\phi}|\Phi_{j,\phi}\rangle$
for $j\neq i$, by estimating the first integral of
\eqref{e:<Phi_i|Phi_j>} in the cases $3\geq k\geq 1$:
$$
\left| \int ^{T/2}_{-T/2 }ds\; \left(T/2 -|s|\right) \frac{\e^{-\mi (s+kT/2
)\phi }}{\sqrt{\cosh (\lambda (s+kT/2 ))}}\right| \leq
\frac{4\sqrt{2}}{\lambda ^{2}}\e ^{-\frac{\lambda
(k-1)T}{4}}=\cO(\hbar^{\frac{k-1}{4}}).
$$
Taking into account the estimate of the error in
(\ref{e:<Phi_i|Phi_j>}), we see that for any $i\neq j$, the overlap $\langle
\Phi_{i,\phi}^{\rm cont}|\Phi_{j,\phi}^{\rm cont}\rangle $ is bounded by a
constant (even by $\cO (\hbar^{1/4})$ for $|i-j|=2)$. As a result,
\begin{equation}
\label{e:asymptotic orthogonality} \forall i\neq j,\quad_n\!\langle
\Phi_{i,\phi}^{\rm cont}|\Phi_{j,\phi}^{\rm cont}\rangle_n =\frac{\langle
\Phi_{i,\phi}^{\rm cont}|\Phi_{j,\phi}^{\rm cont}\rangle } {\langle
\Phi_{i,\phi}^{\rm cont}|\Phi_{i,\phi}^{\rm cont}\rangle }\leq \frac{C}{T}.
\end{equation}
This proves (ii). Part (iii) is now obvious.

To treat the case of the discrete quasimodes, the integrals over  time have to
be replaced by sums over integers. For instance, the expressions defined in
(\ref{e:S1}) are replaced by
$$
S^{\rm disc}_{1}(\lambda ,\phi ,\tau)\defi \sum _{|t|\leq \tau} \frac{\e^{-\mi
t\phi }}{\sqrt{\cosh \left( \lambda t\right) }},
$$
and similarly for $S_2$. The sum $S_1^{\rm disc}(\lambda,\phi)=
\lim_{\tau\to\infty}S^{\rm disc}_{1}(\lambda ,\phi,\tau)$ is also nonnegative for
all $\lambda>0$, $\phi\in[-\pi,\pi]$. Indeed, Poisson's formula induces the
identity
$$
S^{\rm disc}_1(\lambda,\phi)= \sum_{k\in \Z } S_1^{\rm cont}(\lambda ,\phi
+2k\pi).
$$
The norms of the discrete quasimodes therefore satisfy an estimate similar to
\eqref{e:asymptotic norm}, upon replacing $S_1^{\rm cont}$ by $S_1^{\rm
disc}$. The other estimates are identical as for the continuous version.
\end{proof}


\subsection{Quasimodes of different quasienergies}
\label{s:Delta phi-orthogonality} We now compare quasimodes
$|\Phi_\phi\rangle$ of different quasienergies and show:

\begin{prop}\label{prop:quasi-basis}
Let $\phi_0$ be an arbitrary angle in $[0,2\pi[$, and
$$
\phi_k=\phi_0+\frac{\pi }{T}k,\quad k=1,\ldots ,2T.
$$
The $2T$ quasimodes $|\Phi_{\phi_k}\rangle_n$ become
mutually orthogonal in the semiclassical limit: $\forall k'\neq k$,
$_n\!\langle \Phi_{\phi_{k'}}|\Phi_{\phi_k}\rangle_n=\cO(1/T)$.
\end{prop}
This is an immediate consequence of the following finer estimate:
\begin{prop} Let $I\subset\R$ be a fixed bounded interval.
There exists a constant $C>0$ such that, given any semiclassically vanishing
function  $\theta(\hbar)$ and $n\in\Z^*$, if $\phi,\phi'\in I$, and if the
phase shift $\Delta\phi=\phi'-\phi$ satisfies $|\Delta \phi -\frac{n\pi
}{T}|\leq \theta (\hbar )\frac{|n|}{T}$, then we have, for small enough
$\hbar$, $_{n}\!\langle \Phi_{\phi '}|\Phi_{\phi }\rangle_{n}\leq
C\left(\theta (\hbar) +\frac{1}{T}\right)$.
\end{prop}
\begin{proof}
As before, we write the proof for the continuous time quasimodes. The overlap
$\langle \Phi_{\phi'}^{\rm cont}|\Phi_{\phi }^{\rm cont}\rangle$ is given by
an expression similar to (\ref{e:<Phi_i|Phi_j>}). Using the estimate
(\ref{e:upper bound for I(t)}) for $I(t,s)$, we obtain
\begin{align*}
\langle \Phi ^{\rm cont}_{\phi '}|\Phi ^{\rm cont}_{\phi }\rangle &= \int
^{T}_{-T}dt\int^{T}_{-T}dt'\:
\frac{\e ^{\mi (t'\phi'-t\phi) }}{\sqrt{\cosh \lambda (t-t')}}+\cO (1)\\
&=\int ^{2T}_{-2T}ds\: \frac{\e^{\mi s\bar{\phi}}}{\sqrt{\cosh (\lambda s)}}\:
\frac{\sin \left\{ \Delta \phi \left(T -|s|/2\right) \right\} }{\Delta \phi
/2}+\cO (1)
\end{align*}
where we introduced $\bar{\phi }\defi \frac{\phi '+\phi }{2}$. This integral
is bounded above by $\frac{2S_1(\lambda ,0)}{|\Delta \phi |}$, so that for a
phase difference bounded away from zero ({\em i.e.} $|\Delta \phi |\geq c>0$),
the scalar product of the normalized states is $_n\!\langle \Phi_{\phi
'}|\Phi_\phi\rangle_n=\cO (T^{-1})$.
We are however more interested in the case where $\Delta\phi$ is
$\hbar$-dependent and semiclassically small: $\Delta \phi \rightarrow 0$.
Inserting $|\sin\{\Delta\phi(T-|s|/2)\}-\sin\{\Delta\phi T\}|\leq
|s|\Delta\phi/2$ in the integral and using (\ref{eq:norms}), we get for
$\hbar\to 0$, $\Delta\phi\to 0$:
$$
_n\!\langle \Phi_{\phi '}^{\rm cont}|\Phi_{\phi }^{\rm cont}\rangle_n =
\frac{S^{\rm cont}(\bar\phi)}{\sqrt{S^{\rm cont}(\phi)S^{\rm cont}(\phi')}}
\frac{\sin \left( T\Delta\phi \right) }{T\Delta \phi }\, +\cO (1/T).
$$
The first term can be as large as $1$, for $\Delta \phi << T^{-1}$. 
It will also be large for values $\Delta\phi =\frac{\pi (n+1/2)}{T}$ 
with $n$ an integer, $|n|<< T$, where it takes
the value $\pm 1/T\Delta \phi $. At the opposite extreme, the term vanishes for
$\Delta \phi =\frac{n\pi }{T}$, $n$ a nonzero integer, and close to this value
it behaves like $(-1)^{n}\frac{T}{\pi n}(\Delta \phi -\frac{n\pi }{T})$. 
\end{proof}

We are now set to analyze, in the next subsections, the phase space
distributions of the quasimodes $|\Phi_\phi\rangle_n$ and of their components.


\subsection{Localization of $|\Phi_{{\rm loc},\phi}\rangle_n $ near the origin
\label{s:localization}} Recall that $|\Phi_{{\rm loc},\phi}\rangle
=|\Phi_{2,\phi}\rangle +|\Phi_{3,\phi}\rangle$. We will show the following:

\begin{prop}\label{prop:localization}
Let $\phi\in\R$. Then, for any $f\in C^\infty(\T^2)$,
\begin{equation}\label{eq:loc1}
\lim_{\hbar\to0}\ _n\!\langle\Phi_{{\rm loc},\phi }|\hat f|\Phi_{{\rm
loc},\phi }\rangle_n=f(0)\quad {and}\quad\lim_{\hbar\to0} \int_{\T}
f(x)\,\mathcal{H}_{\tilde c_0,{\rm loc},\phi,\theta}(x)\, dx=f(0).
\end{equation}
where  $\mathcal{H}_{\tilde c_0, {\rm loc},\phi, \theta}(x)=N|\langle x,\tilde
c_0,\theta|\Phi_{{\rm loc},\phi}\rangle_n|^2$ is the Husimi
function of $|\Phi_{{\rm loc},\phi}\rangle_n$. It follows that the
semiclassical measures $\mathcal{H}_{\tilde c_0,{\rm loc},\phi,\theta}(x) dx$
and the Wigner distribution converge to the delta measure at the origin. 
All limits are uniform for $\phi$ in a bounded interval.
\end{prop}

Using a more physical terminology, one can say that the quasimodes
$|\Phi_{{\rm loc},\phi }\rangle_{n}$  strongly scar (or localize) 
on the fixed point
$0\in\T^2$ of the map $M$.

\begin{proof}
 As before, we write the proof for 
$|\Phi^{\rm cont}_{{\rm loc},\phi}\rangle$, given by
$$
|\Phi^{\rm cont}_{{\rm loc},\phi}\rangle =\hat{P}_\theta \int_{-T/2}^{T/2}dt\,
\e^{-\mi\phi t}|t;\tilde c_0\rangle.
$$
This is a sum of evolved coherent states for times $|t|\leq T/2$. At this
maximal time, the length $\Delta q'$ of the Husimi function of
$\e^{-\frac{i}{\hbar}\hat H t}|\tilde c_0\rangle$ reaches the size of the
torus. To control the contribution of the nonlocalized states at $t\approx
T/2$, we first select a function $\Theta (\hbar )$ such that in the
small-$\hbar $ limit $1<<\Theta (\hbar )<<T$.  We then split
$|\Phi^{\rm cont}_{{\rm loc},\phi}\rangle$ in two pieces:
\begin{eqnarray*}
|\Phi^{\rm cont}_{{\rm loc},\phi}\rangle &=&\hat{P}_\theta \int_{|t|\leq
\tau_*} dt\, \e^{-\mi\phi t} |t;\tilde c_0\rangle +
\hat{P}_\theta \int_{\tau_*\leq |t|\leq \frac{T}{2}}dt\, \e^{-\mi\phi t}
|t;\tilde c_0\rangle\\
&=&|\Phi'\rangle + |\Phi''\rangle,
\end{eqnarray*}
where $\tau_*\defi [T/2-\Theta (\hbar )/\lambda ]$. From the proof of
Proposition \ref{prop:ortho}  it is clear that
\begin{equation}\label{eq:help1}
\langle\Phi' |\Phi'\rangle\sim 2\tau_* S_{1}^{\rm cont}(\lambda ,\phi ) \sim
TS_{1}^{\rm cont}(\lambda,\phi )\sim \langle\Phi _{{\rm loc},\phi} |\Phi_{{\rm
loc},\phi}\rangle\quad\mbox{when}\ \hbar\to 0.
\end{equation}
The norm of the remainder $|\Phi''\rangle$ is estimated similarly:
\begin{equation}\label{eq:help2}
\langle\Phi''|\Phi''\rangle \sim \frac{\Theta}{\lambda} 
S_1^{\rm cont}(\lambda,\phi)
\leq C \Theta(\hbar) = o(T).
\end{equation}
In the interval $|t|\leq \tau_*$, the ellipses supporting the states
$|t;\tilde c_0\rangle $ have lengths 
$\shbar\e^{\lambda t}\leq\e^{-\Theta(\hbar)}\to 0$. Considering the disk
$D_{\Theta}$ centered at the origin and of radius $\e ^{-\Theta (\hbar )/2}$,
the Husimi functions of these states are therefore semiclassically
concentrated inside $D_{\Theta }$. We will show below that $|\Phi^{\rm
cont}_{{\rm loc},\phi}\rangle_n $ is also concentrated inside this disk.

Using (\ref{eq:help1}) and (\ref{eq:help2}), together with the obvious
$|a+b|^2\leq 2(|a|^2+|b|^2)$, one finds
\begin{eqnarray}
\int_{\T\setminus D_\Theta} N|\langle x, \tilde c_0,\theta|\Phi_{{\rm
loc},\phi}^{\rm cont}\rangle_n|^2\, dx &\leq& \frac{C}{T}
 \int_{\T\setminus D_\Theta} N\left(|\langle x,
\tilde c_0,\theta|\Phi'\rangle|^2 +
|\langle x, \tilde c_0,\theta|\Phi''\rangle|^2\right)dx\nonumber\\
&\leq& \frac{C}{T} \left(\int_{\T\setminus D_\Theta}
N|\langle x, \tilde c_0,\theta|\Phi'\rangle|^2 \,dx
+ \langle \Phi''|\Phi''\rangle\right).\label{eq:help3}\\
&\leq&C\frac{\Theta(\hbar)}{T},\label{eq:help4}
\end{eqnarray}
The last inequality comes from the observation that the Bargmann function
$\langle x,\tilde c_0,\theta|\Phi'\rangle $ is a sum of Gaussians of widths
smaller than $\e ^{-\Theta (\hbar )}$ so simple analysis shows that the
integral in (\ref{eq:help3}) is $\mathcal{O}\left(N\exp(-c\e^{\Theta(\hbar)})\right)$.
 Consequently, (\ref{eq:help4}) holds and yields the proposition provided we
choose $\log\log N <<\Theta<< \log N$.
For discrete quasimodes, we only need to replace $S_1^{\rm cont}$ by $S_1^{\rm
disc}$ in the above estimate.\end{proof}

For later purpose, we notice that the previous proof can be applied
to the states
\begin{equation}\label{e:locstates}
|\Phi^{\rm cont}_{t_1,t_2}\rangle=
\hat P_\theta\int_{t_1}^{t_2} \e^{i\phi t}|t;\tilde c_0\rangle,\qquad
\mathrm{with}\qquad -\frac{T}{2}\leq t_1\leq0\leq t_2\leq \frac{T}{2}.
\end{equation}
These states indeed localize at the origin in the sense of equations
(\ref{eq:loc1}) and (\ref{eq:help4}). The same is obviously true for the
discrete analogues of these states: 
$|\Phi^{\rm disc}_{t_1,t_2}\rangle=\hat{\mathcal P}_{t_1,t_2}
|\tilde c_0,\theta\rangle$. 
Note that $|\Phi_{2,\phi}\rangle$ and $|\Phi_{3,\phi}\rangle$ are of this
type.

\subsection{Equidistribution of $|\Phi_{{\rm erg},\phi}\rangle_n $}\label{s:phi erg}
Recalling that $|\Phi_{\rm erg, \phi}\rangle=|\Phi_{1,\phi}\rangle
+|\Phi_{4,\phi}\rangle$ we have
\begin{prop}\label{prop:ergodicity of |Phi_erg>}
Let $\phi\in\R$. Then, for any $f\in C^{\infty}(\T^2)$
\begin{equation}\label{e:equidistribute Phi erg}
\lim_{\hbar\to0}\,_n\!\langle\Phi_{{\rm erg},\phi}|\hat f|\Phi_{{\rm
erg},\phi}\rangle_n =\int_{\T} f(x) dx = \lim_{\hbar\to0}\int_{\T} f(x)
\mathcal{H}_{\tilde c_0,{\rm erg},\phi,\theta}(x)\, dx,
\end{equation}
where $\mathcal{H}_{\tilde c_0, {\rm erg},\phi, \theta}(x)=N|\langle x, \tilde
c_0,\theta|\Phi_{{\rm loc}, \phi}\rangle_n|^2$ is the Husimi
function of $|\Phi_{{\rm erg}, \phi}\rangle_n$.
It follows that the Husimi measure $\mathcal{H}_{{\rm erg},\phi}(x)\, dx$
and the Wigner distribution converge to the Liouville measure on the torus. 
The limits are uniform for $\phi$ in a bounded interval.
\end{prop}
The states $|\Phi_{\rm erg,\phi}\rangle$ are said to semiclassically 
equidistribute on the torus.

\begin{proof}
We will use the algebraic structure of the
quantized automorphisms in the proof. We will drop the index $\phi$ from the
notations. It is clearly enough to show that, for each  $k\in\Z^2_*$, we have
$$
\lim_{\hbar\to0}\langle \Phi_{\rm erg}|\hat{T}_{k/N}|\Phi_{\rm erg}\rangle=0.
$$
For that purpose, we write
\begin{equation}\label{e:decompo overlap}
\langle \Phi_{\rm erg}|\hat{T}_{k/N}|\Phi_{\rm erg}\rangle =\langle \Phi
_{1}|\hat{T}_{k/N}|\Phi _{1}\rangle +\langle \Phi _{4}|\hat{T}_{k/N}|\Phi
_{4}\rangle +\langle \Phi _{1}|\hat{T}_{k/N}|\Phi _{4}\rangle +\langle \Phi
_{4}|\hat{T}_{k/N}|\Phi _{1}\rangle .
\end{equation}
We first estimate the  two diagonal terms of the RHS. Using $|\Phi_{1}\rangle
=\e ^{\mi \phi T}\hat{M}^{-T}|\Phi_{3}\rangle$, $|\Phi_{4}\rangle=\e^{-\mi\phi
T}\hat{M}^{T}|\Phi_{2}\rangle$ and the intertwining property (\ref{e: M and
T}), we get
$$
\langle \Phi _{1}|\hat{T}_{k/N}|\Phi _{1}\rangle +\langle \Phi
_{4}|\hat{T}_{k/N}|\Phi _{4}\rangle=\langle\Phi_{3}|\hat{T}_{k_+}|\Phi
_{3}\rangle +\langle\Phi_{2}|\hat{T}_{k_-}|\Phi _{2}\rangle.
$$
Here $k_\pm \defi M^{\pm T}k/N\in(\Z/N)^2$ are of order $1$ (see below),
so that we transformed the
``microscopic'' translation by ${k/N}$ (of order $\hbar$) into
``macroscopic'' ones.
Each term is therefore the overlap between the state $|\Phi_{2}\rangle$ 
or $|\Phi_{3}\rangle$ 
localized in a small disc $D_{\Theta}$ centered at the origin of the
torus (cf. Eqs.~(\ref{eq:help3},\ref{e:locstates})), and a translated state
localized in the disc $D_{\Theta
,\pm }\defi D_{\Theta }+ k_\pm$ centered at the point $k_\pm $ mod $\Z^2$. 
This overlap will consequently be small provided $k_\pm$ is
sufficiently far away from the integer lattice.  We prove this fact
using (\ref{e:decompo-n}):
\begin{align*}
k_+ &=q'(k)\,\e^{T \lambda }/N\, {v}_{+}+p'(k)\,\e ^{-T \lambda}/N\,
 v_- = C(N)q'(k)v_+ + \cO (\hbar^2) v_-,\\
k_- &=q'(k)\,\e ^{-T \lambda }/N\,  v_+ +p'(k)\,\e^{T \lambda }/N\,  v_- =C(N)
p'(k) v_- +\cO (\hbar^2)v_+,
\end{align*}
where $2\pi\e^{-\lambda}\leq C(N) \leq 2\pi \e^{\lambda}$ since $T/2$ is the 
closest integer to $|\ln\hbar|/2\lambda$. 
Now, $q'(k)\not=0\not= p'(k)$ since the slopes of $v_{\pm}$ are irrational.
Consequently, $k_{\pm}$ are at a finite distance from $\Z^2$ for small
enough $\hbar$, and the disks $D_{\Theta}$ and $D_{\Theta,\pm}$ do not
intersect each other. We can thus
estimate the overlap:
\begin{align*}
_n\!\langle \Phi_3|\hat{T}_{k_+}|\Phi_3\rangle_n&= \int_{\T}
\: _n\!\langle \Phi_{3}|x,\tilde c_0,\theta\rangle \, \langle x,\tilde
c_0,\theta|\hat{T}_{k_+}|\Phi_{3}\rangle\: N\,dx
_{n} \\
&=\left\{\int_{\T\setminus D_\Theta} +\int_{D_\Theta}\right\}\: _n\!\langle
\Phi_{3}|x,\tilde c_0,\theta\rangle \, \langle x,\tilde
c_0,\theta|\hat{T}_{k_+}|\Phi_{3}\rangle_{n}\, N\, dx.
\end{align*}
Using the Cauchy-Schwarz inequality, the first integral is bounded as
\begin{multline*}
\Big |\int_{\T\setminus D_\Theta} \:_n\!\langle \Phi_{3}|x,\tilde
c_0,\theta\rangle \, \langle x,\tilde
c_0,\theta|\hat{T}_{k_+}|\Phi_{3}\rangle_{n}\ N\, dx\Big|
\leq\\
\sqrt{\int _{\T\setminus D_{\Theta}} \mathcal{H}_{\tilde
c_0,3,\theta}(x)\,dx } \sqrt{\int_{\T\setminus  D_{\Theta }}
\mathcal{H}_{\tilde c_0,3,\theta}(x-k_+)}\,dx\, \leq C \sqrt{\frac{\Theta}{T}}
\end{multline*}
where we used (\ref{eq:help3}) applied to $|\Phi_{3}\rangle$. The integral
over $D_{\Theta }$ is treated similarly, exchanging the roles of both
factors: now the second factor semiclassically converges to zero due to the
inclusion $D_{\Theta }\subset (\T\setminus D_{\Theta ,+})$.
In the end, we get for $\log\log N << \Theta(\hbar) << \log N$
\begin{equation}
\label{e:quantum ergod.} \ _n\!\langle \Phi _{1}|\hat{T}_{k/N}|\Phi
_{1}\rangle_n= _{n}\!\langle \Phi_3|\hat M^T\hat{T}_{k/N}\hat
M^{-T}|\Phi_{3}\rangle_{n} = \cO \left(\sqrt{\frac{\Theta(\hbar)}{T}}\right)
\end{equation}
uniformly for $\phi$ in a finite interval. The
proof goes through unaltered for the second overlap $_{n}\langle \Phi
_{4}|\hat{T}_{k/N}|\Phi _{4}\rangle _{n}$ and in fact for any $\hat
M^T|\Phi_{t_1, t_2}\rangle_n$ as in (\ref{e:locstates}), leading to:

\begin{lem}\label{prop:ergodicity}
Consider a semiclassically diverging function $\log|\log \hbar| 
<< \Theta (\hbar)<< |\log \hbar|$ and $k\in \Z^2_*$. 
Given a bounded interval, there exists a constant $C$
so that for all $\phi$ in the interval
$$
\big|\, _n\!\langle \Phi_{t_1, t_2}|\hat{M}^{-T}\hat{T}_{k/N}\hat{M}^{T}
|\Phi_{t_1,t_2}\rangle_n\big| \leq C\sqrt{\frac{\Theta (\hbar )}{T}}.
$$
\end{lem}
As a result, the states $\hat{M}^{T}|\Phi_{t_1, t_2}\rangle_n$
equidistribute as $\hbar\to 0$, which implies that the integral of
their Husimi function over a {\it
fixed} domain of area $\mathcal{A}$ converges to $\mathcal{A}$. We now use
this information to finish the proof of  Proposition~\ref{prop:ergodicity of
|Phi_erg>}.

We enlarge the Figure~\ref{f:time-scale} and define the additional state
$|\Phi_{5}\rangle =\e ^{-\mi \phi T}\hat{M}^{T}|\Phi_{3}\rangle $, which,
according to Lemma \ref{prop:ergodicity}, equidistributes. Now, using the
same intertwining property as above, we rewrite the nondiagonal terms in the
RHS of (\ref{e:decompo overlap}) as
$$
\langle \Phi_{2}|\hat{M}^{T/2}\hat{T}_{k/N}\hat{M}^{-T/2}|\Phi_{5}\rangle
+\langle\Phi_{5}|\hat{M}^{T/2}\hat{T}_{k/N}\hat{M}^{-T/2}|\Phi_{2}\rangle
=\langle \Phi_{2}|\hat{T}_{k'}|\Phi_{5}\rangle +\langle
\Phi_{5}|\hat{T}_{k'}|\Phi_{2}\rangle,
$$
with the vector $k'\defi M^{T/2}k/N$. Each term is the overlap
between a state localized near
the origin ({\it e.g.}  $\langle\Phi_{2}|$) and an equidistributed one
({\it e.g.}  $\hat{T}_{k'}|\Phi_{5}\rangle$). It is natural 
to expect that they are
asymptotically orthogonal.

To prove this fact, we  proceed as above:
\begin{multline}\label{eq:decoupe}
\big|\,_n\!\langle \Phi_{2}|\hat{T}_{k'}|\Phi_{5}\rangle_{n}\big|\leq 
\sqrt{\int_{\T\setminus  D(r)}dx\,\mathcal{H}_{\tilde c_0,2,\theta}(x)} 
\sqrt{\int_{\T\setminus D(r)}dx\,\mathcal{H}_{\tilde c_0,5,\theta}(x-k')}
\\
+
\sqrt{\int_{D(r)}dx\, \mathcal{H}_{\tilde c_0,2,\theta}(x)},
\sqrt{\int_{D(r)}dx\, \mathcal{H}_{\tilde c_0,5,\theta}(x-k')},
\end{multline}
where $D(r)$ is the disc of radius $r$ centered at $0$. Using the
semiclassical localization of $|\Phi_2\rangle_n$  at the origin 
and the equidistribution of $|\Phi_5\rangle_n$, we find
$$
\limsup_{\hbar\to0}
\big|_n\!\langle \Phi_{2}|\hat{T}_{k'}|\Phi_{5}\rangle_{n}\big|\leq
\sqrt{\pi}r.
$$
Since this is true for any $r>0$, $\lim_{\hbar\to0}\big|_n\!\langle
\Phi_{2}|\hat{T}_{k'}|\Phi_{5}\rangle_{n}\big|=0$. 
We now control all the terms
of (\ref{e:decompo overlap}) and after taking care of the normalizations we
obtain Proposition \ref{prop:ergodicity of |Phi_erg>}.\end{proof}


\subsection{Semiclassical properties of
$|\Phi_{\phi}\rangle =|\Phi_{{\rm loc},\phi}\rangle +|\Phi_{{\rm
erg},\phi}\rangle $}\label{s:phi tot}

We now finally consider the ``full'' quasimode $|\Phi_{\phi}\rangle$. It is
the sum of two states, one localized, the second equidistributed.

\begin{prop}\label{prop:semiclassical properties of Phi_tot}
For any $\phi\in\R$,
(\ref{e:equidistributeall}) holds with $\tau=1, x_0=0$. The limit is uniform 
for $\phi$ belonging to
a bounded interval.
\end{prop}

\begin{proof} 
It is again enough to study 
$_n\!\langle \Phi|\hat T_{k/N}|\Phi\rangle_n$ and
to show
$$
\lim_{\hbar\to0} \ _n\!\langle \Phi|\hat
T_{k/N}|\Phi\rangle_n=\frac{1}{2}(1+\delta_{k,0}).
$$
The results of the previous subsections imply immediately that this reduces to
showing
$$
\lim_{\hbar\to0} \ _n\!\langle \Phi_{\rm loc}|\hat T_{k/N}|\Phi_{\rm
erg}\rangle_n=0.
$$
This in turn is proven as in the previous subsection through the use of the
Cauchy-Schwarz inequality and cutting the integral over $\T$ into the
integral over a small disc around the origin and an integral over 
the complement (see (\ref{eq:decoupe})).
\end{proof}

To conclude this section, let us remark that the semiclassical properties of
the various quasimodes we introduced are
not altered if we replace $T$ in the sum or integration boundaries
by an integer that differs from it by a finite amount, bounded
as $\hbar$ goes to zero. This will occasionally be useful in the sequel.


\section{Pointwise description of the quasimodes\label{s:pointwise}}
In the last sections, we showed that the Husimi and Wigner 
functions of the quasimodes $|\Phi_\phi\rangle_n$ converge to the measure
$\frac{1+\delta_0}{2}$ in the semiclassical limit.
The crucial tools of the proof were, on the one hand, 
precise estimates of the overlaps 
$\langle \tilde c_0,\theta|\hat M^t|\tilde c_0,\theta\rangle$ 
(obtained using the diophantine properties of the invariant axes), on the other
hand the algebraic intertwining between $\hat M$ and the quantum 
translations. 

Still, it would be interesting to know the speed at which this
convergence takes place, or to compute more refined ``indicators'' of 
the localization of the quasimodes. 

In this section, we will use a more ``direct'' yet slightly more
cumbersome route which will yield more precise
information on the phase space distribution of the ``continuous time'' 
quasimodes. The main step of this route is the 
\emph{pointwise} description of the Bargmann and Husimi functions
of $|\Phi^{\rm cont}_\phi\rangle$. This description will then
provide an estimate of the speed of convergence to the limit semiclassical
measure; at the same time, it will allow us to compute alternative 
localization indicators, like
the $L^s$-norms of the Husimi functions. The pointwise estimates will
also uncover the ``hyperbolic''
structure of the Husimi functions
near the origin, a structure
already emphasized by several 
authors for finite-time quasimodes \cite{heller-kaplan,wisniacki} and 
for spectral Wigner and Husimi functions \cite{rivas}.


\subsection{Plane quasimodes}
Our final objective is to estimate the Bargmann function 
$\langle x,\tilde c_0,\theta|\Phi^{\rm cont}_\phi\rangle$ for 
$x\in\mathcal{F}$ the 
fundamental domain. For this purpose, we start from
quasimodes of the Hamiltonian $\hat H$:
\begin{equation}\label{e:plane projector} 
|\Psi_{\phi,t}\rangle \defi 
\int_{-t}^{t}ds\, \e^{-\mi \phi s} 
\e ^{-\mi \hat{H}s/\hbar }|\tilde{c}_{0}\rangle.
\end{equation}
The torus quasimode $|\Phi^{\rm cont}_\phi\rangle$ is obtained
by projecting $|\Psi_{\phi,T}\rangle$ onto $\hn$ 
(cf. Eq.~\eqref{e:plane torus}). 
In this subsection, we will
study the Bargmann function of the plane quasimode $|\Psi_{\phi,T}\rangle$. 

Using the
rescaled variable $Z\defi \frac{q'-\mi p'}{2\shbar}$, this function
is given by the following integral:
\begin{equation}
\label{e:Bargmann for quasimode} \Psi_{\phi,T}(x)
\defi\langle x,\tilde{c}_0|\Psi_{\phi,T}\rangle
=\e^{-|Z|^2}
\int_{-T}^{T}ds\:\frac{\e ^{-\mi \phi s}}{\sqrt{\cosh \lambda s}}
\e^{Z^2\tanh\lambda s}.
\end{equation}
Through the  change of variables $U=Z^2(1-\tanh\lambda s)$, 
and using the parameter
$\mu\defi 1/4+\mi\phi/2\lambda$, this integral may be rewritten as
\begin{equation}\label{e:integral for quasimode}
\Psi_{\phi,T}(x)
=\frac{\e^{Z^2-|Z|^2}}{\lambda 2^{\mu+1/2}Z^{2\mu}}
\int_{U_0}^{U_1}\frac{dU}{U}U^{\mu}\e^{-U}
\left(1-\frac{U}{2Z^2}\right)^{-\mu-1/2},
\end{equation}
with the boundaries $U_0=Z^2(1-\tanh\lambda T)\simeq 2Z^2\hbar^2$, 
$U_1=Z^2(1+\tanh\lambda T)\simeq 2Z^2(1-\hbar^2)$. 
This function satisfies the following symmetries 
(with obvious notations):
\begin{equation}\label{symmetries}
\Psi_{\phi,T}(Z)=\Psi_{\phi,T}(-Z)=\Psi_{-\phi,T}(\mi Z).
\end{equation}

\begin{figure}[htbp]
{\par\centering
\rotatebox{-90}{\resizebox*{0.50\columnwidth}{!}{\includegraphics{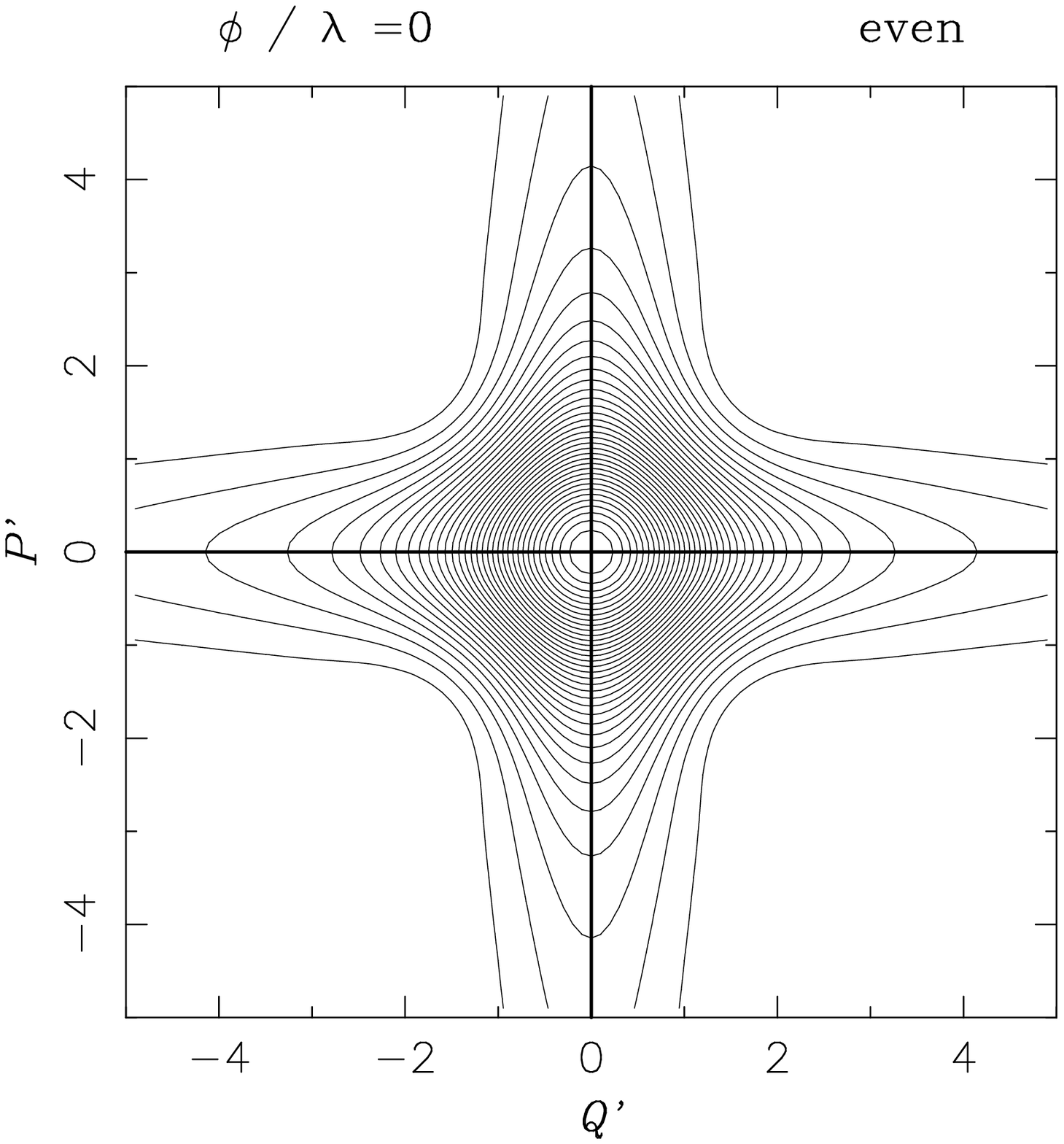}}}
\rotatebox{-90}{\resizebox*{0.50\columnwidth}{!}{\includegraphics{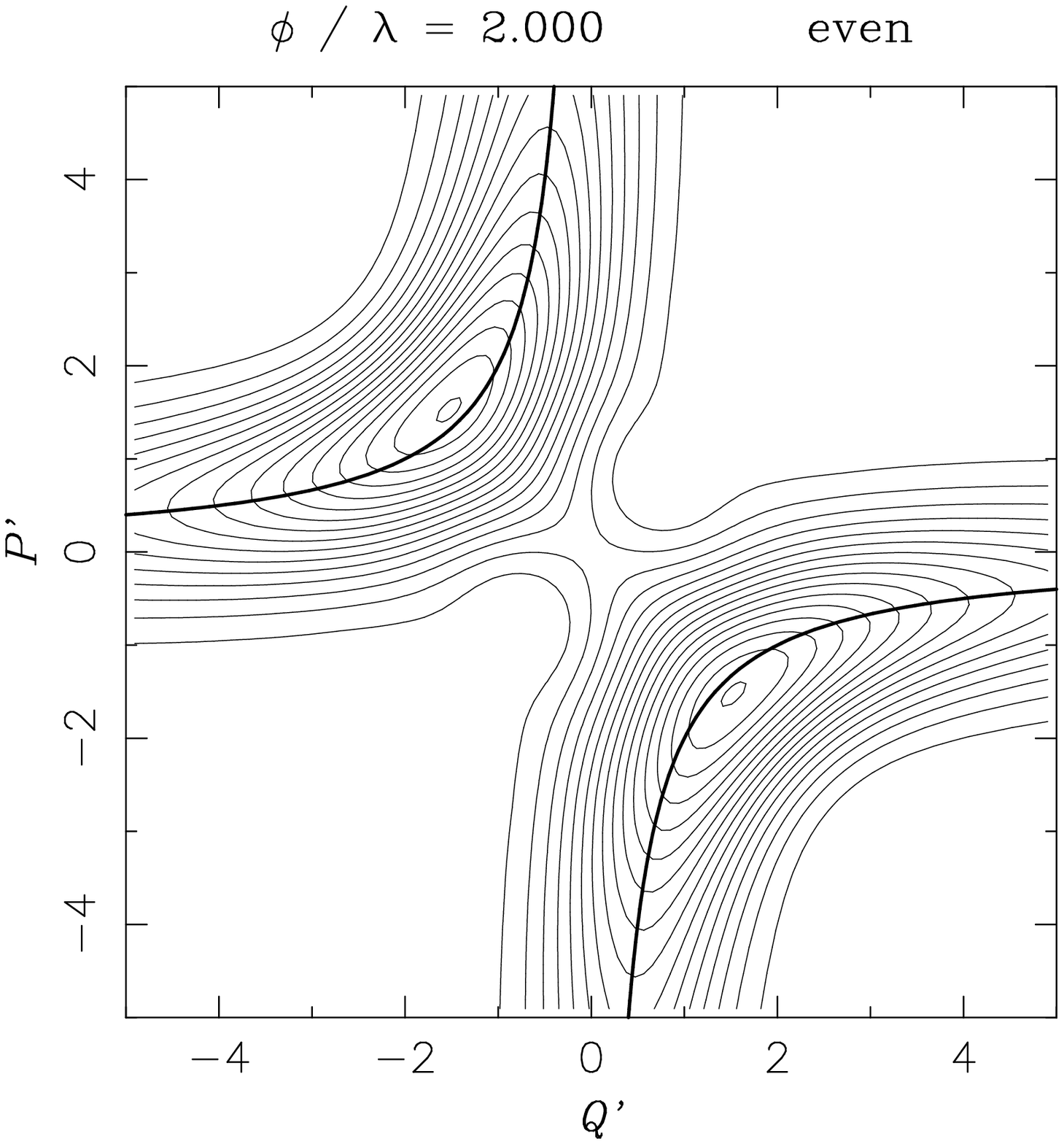}}}
\par}
\caption{\label{fig:plane eigenstates} Husimi functions of two
generalized eigenstates
$|\Psi^{(even)}_{\phi}\rangle$, in the coordinates $(Q',P')$.
The densities are plotted in linear scale, the contour step depending on the
plot. The classical energy hyperbolas are drawn in thick curves.}
\end{figure}

The hyperbolic Hamiltonian $\hat H$ admits no bound state in $L^2(\R)$, but
for any real energy $E=-\hbar\phi$, it has two independent generalized 
eigenstates, distinguished by their parity. 
In the limit $t\to\infty$, the quasimode
$|\Psi_{\phi,t}\rangle$ converges (in a sense explained below) to the even 
eigenstate, that we denote by $|\Psi_\phi^{(even)}\rangle$. 
From the identities
$H(x)=\lambda q'p'$, 
$\hat H=\lambda \hat Q\frac{\hat q\hat p+\hat p\hat q}{2}\hat Q^{-1}$, 
the Bargmann function of $|\Psi_\phi^{(even)}\rangle$
can be expressed in
terms of parabolic cylinder functions \cite{hyperbolic point,bateman}:
\begin{equation}
\label{e:parabolic cylinder functions} 
\langle x,\tilde c_0|\Psi^{(even)}_{\phi}\rangle
=C_\phi \e^{-|Z|^2}\: \left\{ D_{-1+2\mu}(2Z) 
+ D_{-1+2\mu}(-2Z)\right\}.
\end{equation}
The  normalization coefficient
$C_\phi=\pi\left(2^\mu \cosh(\pi\phi/\lambda)\Gamma(\mu+1/2)\right)^{-1}$ can
be computed from the value at $Z=0$.
For fixed $\phi$ and $\hbar$ small, this Bargmann function 
takes its largest values close to the origin 
(where it takes the value $S_1^{\rm cont}(\lambda,\phi)$), and is otherwise
concentrated
along the  hyperbola $\{q'p'=-\hbar\phi/\lambda\}$, which is the classical
energy surface $\{H(x)=-\hbar\phi\}$ (see below and Section~\ref{Ls norms} 
for more details). 
The Husimi functions
of two of these generalized eigenstates 
are displayed in Figure~\ref{fig:plane eigenstates}
in terms of the 
coordinates $(Q',P')=\frac{(q',p')}{\shbar}$.

\medskip 

From the integral expression \eqref{e:Bargmann for quasimode}, we see
that the Bargmann
functions of $|\Psi^{(even)}_{\phi}\rangle$ and $|\Psi_{\phi,T}\rangle$ are
semiclassically close to each other:
\begin{equation}\label{closeness}
\Psi_{\phi,T}(x) - 
\Psi^{(even)}_{\phi}(x) =\cO (\hbar^{1/2})
\quad \textrm{uniformly with respect to $x$ and $\phi$.}
\end{equation}

This equation together with 
\eqref{e:parabolic cylinder functions} yields a uniform approximation for
$\Psi_{\phi,T}(x)$. One cannot simplify this expression in the central region 
$\{x=\cO(\shbar)\}$.
On the other extreme, one can obtain asymptotic expansions for 
\eqref{e:integral for quasimode}
in the region $\{|x|>>\shbar\}$ ($\{|Z|>>1\}$). We will give formulas 
uniformly valid in the 
``positive sector''
$\Sp\defi\{Z\mid \arg(Z)\leq\frac{\pi}{4}(1-\eps)\}$, where
$\eps>0$ is fixed. The symmetries \eqref{symmetries} then allow to
fill the remaining three sectors (around the angles
$\pi/4+n\pi/2$, the function is exponentially small).

Expanding the last factor in the integral 
\eqref{e:integral for quasimode} into powers of $1/Z^2$, we get a sum of 
incomplete Gamma functions \cite[Chap. 9]{bateman}: 
$$\int_{U_0}^{U_1}\frac{dU}{U}U^{\mu}\e^{-U}
\left(1-\frac{U}{2Z^2}\right)^{-\mu-1/2}=\big(\gamma(\mu,U_0)-\gamma(\mu,U_1)\big)
+\frac{\mu+1/2}{2Z^2}\big(\gamma(\mu+1,U_0)-\gamma(\mu+1,U_1)\big)+\ldots
$$ 
These gamma functions have simple
asymptotics in two regimes:
\begin{itemize}
\item for $U_0<<1<<U_1$, that is, $x\in \Sp$, 
$\shbar<<|x|<<\frac{1}{\shbar}$,
they yield
\begin{align}\label{inner sector}
\Psi_{\phi,T}(x)
&=\frac{\Gamma(\mu)\e^{Z^2-|Z|^2}}{\lambda 2^{\mu+1/2}Z^{2\mu}}
\left(1+\cO(\frac{1}{|Z|^2}+\cO(\sqrt{\hbar Z})\right)\nonumber\\
&=\frac{\Gamma(\mu)}{\lambda 2^{1/2-\mu}}\ 
\frac{\hbar^{\mu}}{(q'-\mi p')^{2\mu}}
\e^{-\frac{p^{\prime 2} +\mi q'p'}{2\hbar}}\left(1+\cO\Big(\frac{\hbar}{|x|^2}\Big)+
\cO(\sqrt{\hbar^{1/2}|x|})\right).
\end{align}
This asymptotics also holds for the Bargmann function of
$|\Psi^{(even)}_\phi\rangle$ in the sector $|Z|>>1,\ Z\in \Sp$: it
indeed corresponds to known asymptotics of the parabolic cylinder 
functions $D_{-1+2\mu}$ 
\cite[Chap. 8]{bateman}. This gives for the Husimi function:
\begin{equation}\label{asymptotics for Husimi}
\frac{|\Psi_{\phi,T}(x)|^2}{2\pi\hbar}\sim 
\frac{S_1^{\rm cont}(\lambda,\phi)}{2\lambda\sqrt{\pi\hbar}}\ 
\frac{1}{|q'-\mi p'|}\ 
\e^{-\frac{p^{\prime 2}}{\hbar}
-2\frac{\phi}{\lambda}\frac{p'}{q'}}.
\end{equation}
For fixed $q'>>\shbar$, the $p'$-Gaussian of width $\sqrt\hbar$ 
is centered on the point 
$p'=-\hbar\phi/\lambda q'$, that is on the classical hyperbola. 
The function decreases as $\frac{1}{q'}$ along the ``crest''.

\item in the region $|x|>>\frac{1}{\shbar}$, $x\in \Sp$, 
the Bargmann function is
``dominated'' by the coherent state at time $T$:
\begin{equation}\label{outer sector}
\Psi_{\phi,T}(x)=\frac{\sqrt{2}}{\hbar^{1/2-\mi\phi/\lambda}(q'-\mi p')^2}
\e^{-\frac{p^{\prime 2}(1-\hbar^2)}{2\hbar}- \frac{q^{\prime 2}\hbar}{2}}
\e^{-\mi \frac{q'p'(1-2\hbar^2)}{2\hbar}}\left(1+
\cO\Big(\frac{1}{\hbar|x|^2}\Big)+\cO(\hbar^2)\right).
\end{equation}
\end{itemize}
The crossover between the $\frac{1}{\sqrt{q'}}$ decay and the 
$\e^{-\frac{q^{\prime 2}\hbar}{2}}$ decay 
is governed by the function $\gamma(\mu,U_0)$, with 
$U_0\sim \hbar q^{\prime 2}/2$ varying from small to
large values. 


\subsection{Pointwise description of the torus quasimodes}
\label{s:pointwise description}
Using the results in the last section, we will now derive semiclassical 
estimates
for the Bargmann function of the torus quasimode 
$|\Phi^{\rm cont}_\phi\rangle$:
\begin{align}\label{sum quasimode}
\Phi_\phi(x)&\defi \langle x,\tilde c_0,\theta|\Phi^{\rm cont}_\phi\rangle
=\langle x,\tilde c_0|\hat{P}_\theta|\Psi_{\phi,T}\rangle\nonumber\\
&=\sum_{n\in\Z^2}\e^{\mi\vartheta(x,n)}\:\Psi_{\phi,T}(x+n),\\
\textrm{with the phases}\quad
\vartheta(x,n)&=n\cdot \theta +\mi\delta_{n}-\mi \pi Nx\wedge n.\nonumber
\end{align}
From now on we restrict $x$ to the fundamental domain $\mathcal{F}$. 
We will split the above sum between 
a few ``dominant terms'' and a ``remainder'', which we then bound
from above by using similar methods as in Appendix~ 
\ref{a:interference term}. We will only provide a sketch of the proof.

From the last subsection, we know that the function $\Psi_{\phi,T}(x)$ 
is concentrated along the 
hyperbola $\{p'=-\hbar\phi/\lambda q'\}$, which is itself 
$\shbar$-close to the 
stable and unstable axes. We therefore define two strips 
$B_u$, $B_s$ around these axes:
$$
B_{u}=\left\{ x\in \R^2,\: |p'(x)|\leq 2\sqrt{\hbar T}\; \textrm{and}\;
|q'(x)|\leq \frac{C_o}{9\sqrt{\hbar T}}\right\},\quad B_s=\{q'\leftrightarrow 
p'\}.
$$
We call $B\defi B_u\cup B_s$ the union
of these strips, $Sq=B_u\cap B_s$ the ``central square'' 
and  $B_u^\T$, $B_s^\T$ and $B^\T$ their periodizations
on $\T$ or $\mathcal{F}$. 
The coefficient $C_o/9$ in the above definition is chosen such that 
$B_u$ (resp. $B_s$) 
does not intersect any of its integer translates 
(see Eq.~\eqref{eq:dioph3}). As a consequence, for any 
$x\in\mathcal{F}$ the intersection between the lattice $x+\Z^2$ and $B_u$
(resp. $B_s$)
is either empty, or it contains a single point noted $x+n_{u,x}$ (resp. 
noted $x+n_{s,x}$), with $n_{u/s,x}\in\Z^2$. 
These (possible) points define our ``dominant terms''
in \eqref{sum quasimode}. The remainder thus consists in the sum 
over $n\in\Z^2$ such that $(x+n)\not\in B$.
In order
to state the pointwise estimate, we define the
``modified characteristic functions'' $\chi_u(x)$, $\chi_s(x)$ on $\mathcal{F}$
as 
$$
\left\lbrace\begin{array}{l}
\chi_u(x)=\e^{\mi\vartheta(x,n_{u,x})}\mbox{ if $x\in B_u^\T$, $0$ otherwise}\\
\chi_s(x)=\e^{\mi\vartheta(x,n_{s,x})}\mbox{ if $x\in B_s^\T\setminus\! Sq$, $0$ 
otherwise.}\end{array}\right.
$$
(this definition is consistent: $n_{u,x}$ is well-defined iff
$x\in B_u^\T$). 
The slight asymmetry between $\chi_u$ and $\chi_s$ will prevent 
double counting for $x$ in the central square. 
\begin{prop}
The Bargmann functions of the quasimodes 
$|\Phi^{\rm cont}_\phi\rangle$ have the following expression, uniformly
for $x\in\mathcal{F}$ and $\phi$ in a bounded interval:
\begin{equation}
\label{e:Bargmann estimate} 
\langle x,\tilde{c}_{0},\theta|\Phi^{\rm cont}_\phi\rangle
=\chi_u(x)\,\langle x+n_{u,x},\tilde{c}_{0}|\Psi_{\phi,T}\rangle
+\chi_s(x)\,\langle x+n_{s,x},\tilde{c}_{0}|\Psi_{\phi,T}\rangle
+\cO(\hbar^{1/2}T^{1/4}).
\end{equation}
On the RHS, $|\Psi_{\phi,T}\rangle$ may be replaced by 
$|\Psi^{(even)}_{\phi}\rangle$.
\end{prop}
Notice that $\Psi_{\phi,T}(x)$ 
at the ``edge'' of $B_u$ or $B_s$ is of order $\cO(\hbar^{1/2}T^{1/4})$, so 
that the above estimate of the remainder is sharp.

\begin{figure}[htbp]
{\par\centering
\resizebox*{0.56\columnwidth}{!}{\includegraphics{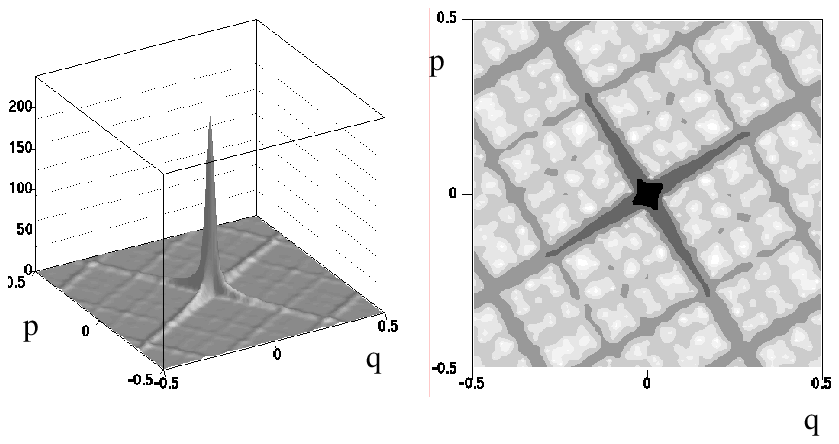}}
\resizebox*{0.30\columnwidth}{!}{\includegraphics{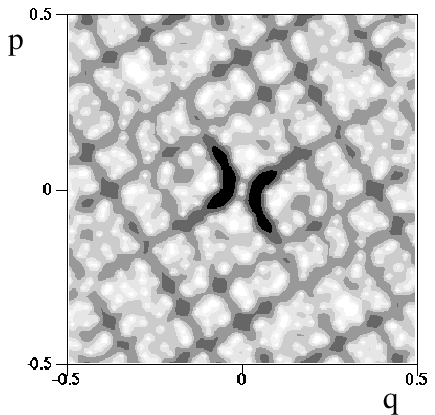}}\par}
\caption{\label{fig:continuous modes} Husimi functions of quasimodes
$|\Phi^{\rm cont}_{\phi=0}\rangle$ (left :3D linear scale; center: 
logarithmic scale)
and $|\Phi^{\rm cont}_{\phi=2\lambda_0}\rangle$ (right, logarithmic
scale) of the map $\hat M_{\rm Arnold}$ ($N=500$).}

\end{figure}

This equation gives a precise information for $x\in B^\T$, but also
a nontrivial upper bound in $\T\setminus B^\T$. 
It implies that the Bargmann (and Husimi)
function of $|\Phi^{\rm cont}_\phi\rangle$
is concentrated along (a portion of) the periodized classical hyperbola, 
itself asymptotically close 
to the invariant axes (see Fig.~\ref{fig:continuous modes} and 
compare with Fig.~\ref{fig:plane eigenstates}). 
These features were not visible
in the framework of Section~\ref{s:quasimodes}.

\medskip

\begin{proof}[Sketch of proof]
We have to find an upper bound for the sum
$
\sum_{n\in \Z^2,\: x+n\not\in B}|\Psi_{\phi,T}(x+n)|.
$
We first consider the points $x+n$ in the sector  $\Sp$; since they satisfy 
$|x+n|>>\shbar$, the Bargmann function is described by formulas 
\eqref{inner sector}--\eqref{outer sector}.  
As in Appendix~\ref{a:interference term}, we split the region
$\Sp\setminus B$ into a union of strips parallel to the unstable axis, of
width $\delta p'=\shbar$. 
The results of Section~\ref{s:classaut} imply that
two points $(x+n)$, $(x+m)$ in such a strip are separated by at least 
$|q'(n-m)|\geq C_o\hbar^{-1/2}$. Summing the estimates
(\ref{inner sector},\ref{outer sector}) in these strips, 
we obtain the ($x$-independent) upper bound 
$\cO(\shbar T^{1/4})$ for points in $\Sp$. From \eqref{symmetries}, 
the sum over the three other sectors leads to the same bound. 
\end{proof}



\subsection{Controlling the speed of convergence}\label{s:controlling}
Using the pointwise formula \eqref{e:Bargmann estimate}, we can now 
directly compute the Fourier coefficients of the Husimi function of 
$|\Phi^{\rm cont}_\phi\rangle$:
$$
\tilde{\mathcal{H}}_{\tilde c_0,\Phi^{\rm cont}_\phi}(k)
\defi\int_{\mathcal{F}}dx\,\e^{2\mi\pi x\wedge k}\;
N|\langle x,\tilde c_0,\theta|\Phi^{\rm cont}_{\phi}\rangle|^2, \quad k\in\Z^2.
$$
We will prove the following estimate:
\begin{prop}
The Fourier coefficients of the (non-normalized) Husimi function for the 
quasimode $|\Phi^{\rm cont}_\phi\rangle$ satisfy, 
uniformly for $\phi$ in a bounded interval and $k\in\Z^2$,
$|k|\leq \e^{\sqrt{T}}$:
\begin{equation}\label{e:Fourier integral}
\tilde{\mathcal{H}}_{\tilde c_0,\Phi^{\rm cont}_\phi}(k)=
S_1^{\rm cont}(\lambda,\phi)T\,(1+\delta_{k,0})
+\mathcal{O}(\sqrt{T}).
\end{equation}
\end{prop} 
This formula yields at the same time the norm of 
$|\Phi^{\rm cont}_\phi\rangle$, the convergence of the normalized quasimode
to the measure 
$\frac{1+\delta_0}{2}$, but also the remainder $\cO(T^{-1/2})$ in this
convergence (which we could not obtain with previous methods). 
We do not know whether
this estimate is sharp; in any case, we believe that the remainder 
cannot be smaller than
$\cO(T^{-1})$. Using the same methods, we can show that the remainder
in the convergence 
of $|\Phi^{\rm cont}_{\rm loc}\rangle_n$ to its limit measure $\delta_0$ 
behaves as $F(k)T^{-1}$, with a function $F(k)\not\equiv 0$.

\begin{proof}
From Eq.~\eqref{e:Bargmann estimate}, we split 
$\mathcal{H}_{\tilde c_0,\Phi_{\phi}}(x)$ 
into 3 components:
\begin{align}
\mathcal{H}^{\rm diag}(x)&=N\left(|\chi_u(x)\Psi_{\phi,T}(x+n_{u,x})|^2+
|\chi_s(x)\Psi_{\phi,T}(x+n_{s,x})|^2\right)\\
\mathcal{H}^{\rm interf}(x)&=N\left(\chi_u(x)\overline{\chi_s}(x)
\Psi_{\phi,T}(x+n_{u,x})\overline{\Psi_{\phi,T}}(x+n_{s,x})\ 
+\ \ c.c.\ \ \right)\\
\mathcal{H}^{\rm remain}(x)&=\cO(\hbar^{-1/2}T^{1/4})
\left(|\chi_u(x)\Psi_{\phi,T}(x+n_{u,x})|+|\chi_s(x)\Psi_{\phi,T}(x+n_{s,x})|\right)+\cO(\sqrt{T}).
\end{align}
We will show that the integrals over $\mathcal{F}$ of the ``remainder'' and 
the ``interference''
components are $\cO(T^{1/2})$, while the integral of 
$\e^{2\mi\pi x\wedge k}\mathcal{H}^{\rm diag}(x)$ 
yields the dominant contribution in \eqref{e:Fourier integral}. 

The integral of $\mathcal{H}^{\rm remain}$ on $\mathcal{F}$ is 
easy to treat.
It involves $\int_B dx\,|\Psi_{\phi,T}(x)|$, which we 
estimate by using
the asymptotics \eqref{inner sector} in the domain $x\in B$, 
$|x|>>\shbar$. 
This yields $\int_B dx |\Psi_{\phi,T}(x)|=\cO(\hbar^{1/2}T^{-1/4})$, 
so the integral of $\mathcal{H}^{\rm remain}$ is an $\cO(\sqrt{T})$.

\paragraph{Homoclinic intersections}

To understand the ``interference component'' $\mathcal{H}^{\rm interf}(x)$, we
have to describe a little bit the set
$(B_u^\T\cap B_s^\T)\setminus Sq$. It is
composed of a large number of small ``squares'' surrounding homoclinic 
intersections (some of them are clearly visible in 
Fig.~\ref{fig:continuous modes}). 
Each of these squares is indexed by a couple of (nonzero and nonequal) 
integer vectors
$(n_u,n_s)$ (finitely many such couples correspond to an actual
square in $B^\T$): 
$$
Sq_{n_u,n_s}\defi (B_u-n_u)\cap(B_s-n_s)=
\{|q'(x)+q'(n_s)|\leq 2\sqrt{\hbar T},\;
|p'(x)+p'(n_u)|\leq 2\sqrt{\hbar T}\}.
$$
Since we have excluded the central square, one can use the asymptotics 
\eqref{inner sector} for $\Psi_{\phi,T}(x+n_{u/s})$. The
integral of $|\mathcal{H}^{\rm interf}(x)|$ on $Sq_{n_u,n_s}$ is then
smaller than
$$
C\int_{Sq_{n_u,n_s}} 
dq'\,dp'\, \frac{\hbar^{-1/2}}{\sqrt{|q'(x+n_u)p'(x+n_s)|}}
\e^{-\frac{q^{\prime 2}(x+n_s)}{2\hbar}}
\e^{-\frac{p^{\prime 2}(x+n_u)}{2\hbar}},
$$
which admits the upper bound
$$\frac{C'\hbar^{1/2}}{\sqrt{|q'(n_u-n_s)p'(n_s-n_u)|}}
\leq C'\hbar^{1/2}\left(\frac{1}{|q'(n_u-n_s)|}+
\frac{1}{|p'(n_s-n_u)|}\right).
$$ 
We now want to sum the RHS over all homoclinic squares in $B^\T$. 
To compute the
sum over $1/q'$ (resp. $1/p'$), we consider the squares as
subsets of $B_u$ (resp. $B_s$), which orders them along the strip. 
Two successive squares do not overlap, so
their centers in $B_u$ (resp. in $B_s$) 
satisfy
$|\delta q'|\geq 4\sqrt{\hbar T}$. 
As a result, the total number of squares is less than
$\frac{C}{\hbar T}$, and summing their contributions we get
$$
\int_{\T}|\mathcal{H}^{\rm interf}(x)|\,dx\leq 
4C'\hbar^{1/2}\sum_{j=1}^{C/\hbar T} \frac{1}{j\,4\sqrt{\hbar T}}
=\cO\left(\frac{1}{\sqrt{T}}|\log(\hbar T)|\right)=\cO(\sqrt{T}).
$$
Notice that we ignored the phases present in 
$\mathcal{H}^{\rm interf}(x)$, as we had done in 
Section~\ref{s:interference bound} to estimate $I(t,s)$.

\paragraph{Diagonal contribution}
We now finish the proof by computing the integral
$$
\int_\mathcal{F}dx\;\e^{2\mi\pi x\wedge k}\mathcal{H}^{\rm diag}(x)
=N\int_B dx\; \e^{2\mi\pi x\wedge k}\;|\Psi_{\phi,T}(x)|^2.
$$
The wedge product $2\pi x\wedge k$ is
rewritten $k_u q'+k_s p'$ in the adapted coordinates. If $k\neq 0$, then
$k_u=2\pi v_+\wedge k$, $k_s=2\pi v_-\wedge k$ are bounded
away from zero (cf. Section~\ref{s:classaut}).

We give some details for the computation of the integral in the 
positive sector $\Sp$.
Let $\Theta(\hbar)$ be a semiclassically 
increasing function s.t. $1<<\Theta(\hbar)<<T^{1/4}$. 
The integral of $\mathcal{H}^{\rm diag}$ in the central region 
($|q'|<\Theta\shbar$) 
admits the obvious upper bound $\cO(\Theta^2)$.
 
In the region $\{x\in\Sp,\, q'>\Theta\shbar\}$, one can apply the asymptotics 
\eqref{asymptotics for Husimi}.  
After integrating over $p'$, we obtain 
$$\frac{S_1^{\rm cont}(\lambda,\phi)}{2\lambda}\int_{\Theta\shbar}^{\frac{C_o}{9\sqrt{\hbar T}}}dq'\;
\frac{\e^{\mi k_u q'}}{q'}\left(1+\cO(\e^{-4T})+
\cO\Big(\frac{\hbar}{q^{\prime 2}}\Big)+\cO(\hbar k_s^2)\right).
$$
This integral is easy to estimate:
\begin{itemize}
\item for $k=0$, it yields $$\frac{S_1^{\rm cont}(\lambda,\phi)}{2\lambda} 
\log\Big(\frac{C}{\hbar\sqrt{T}\Theta}\Big)+\cO(\Theta^{-2})
=\frac{S_1^{\rm cont}(\lambda,\phi)}{2}T +\cO\big(\log(\Theta \sqrt{T})\big).
$$
\item for $k\neq 0$, it
has the asymptotics \cite[Chap. 9]{bateman}
$$\frac{S_1^{\rm cont}(\lambda,\phi)}{2\lambda} |\log(\shbar\Theta k_u)|+\cO(1)=
\frac{S_1^{\rm cont}(\lambda,\phi)}{4}T+\cO(\log(\Theta k_u)). $$
\end{itemize}
Taking the 3 remaining sectors into account, we obtain the proposition.
\end{proof}


\subsection{Husimi function close to the origin and $L^s$ norms}
\label{Ls norms}
Besides providing the limit semiclassical measure, the pointwise formula 
\eqref{e:Bargmann estimate} allows us to compute 
different indicators of localization for the quasimode
$|\Phi_\phi^{\rm cont}\rangle_n$, namely the $L^s$ norms of its 
Husimi function \cite{prosen,NV2}:
$$
(s>0)\quad \|\mathcal{H}_{\Phi}\|_s\defi \left(\int_\T \left[\mathcal{H}_\Phi
(x)\right]^s\,dx\right)^{1/s}.
$$
For $s=2$, this defines a phase space analogy of the ``inverse
participation ratio'' used in condensed-matter physics; 
in the limit $s\to 1^+$, it
yields the Wehrl entropy of the state; for $s\to\infty$, this is
sup-norm of the Husimi density.

\begin{prop}
For any fixed $\infty\geq s>1$ and $\phi$ in a bounded interval, the $L^s$
norms of the quasimodes $|\Phi^{\rm cont}_\phi\rangle_n$ behave in the
semiclassical limit as
$$
\left\| \mathcal{H}_{\tilde c_0,|\Phi^{\rm cont}_\phi\rangle_n}\right\|_s\sim
\frac{C(s,\phi/\lambda)}{\hbar^{1-\frac{1}{s}}|\log\hbar|}.
$$
\end{prop}
By comparison, the $L^s$-norms of a coherent state 
$|\tilde c,\theta\rangle$  behave as 
$C'(s,\tilde c)\hbar^{-1+1/s}$ as $\hbar\to 0$, $\tilde c$ in a bounded set 
\cite{NV2}.
In the case of the sup-norm, we have a more precise statement 
(see Fig.~\ref{fig:plane eigenstates}):
\begin{prop}
For small enough $\hbar$, the maximum of 
$\mathcal{H}_{\tilde c_0,|\Phi^{\rm cont}_\phi\rangle_n}(x)$
is at the origin for $|\phi|/\lambda<0.5$, 
and $C(\infty,\phi/\lambda)=\frac{|\Gamma(1/4+\mi\phi/2\lambda)|^2}{2^{5/2}\pi^{3/2}}$. Conversely, for $|\phi|/\lambda>>1$, 
the maximum is close to the points $Q'=-P'=\pm \sqrt{\phi/\lambda}$ 
on the hyperbola, and 
$C(\infty,\phi/\lambda)\sim (2^{5/2}\sqrt{\pi|\phi|/\lambda})^{-1}$.
\end{prop}
\begin{proof}[Sketch of proof]
For any $s>1$, the decrease $\sim \frac{1}{|x|^s}$ of the Husimi function along
the hyperbola implies that most of the weight in the integral 
$\int_\mathcal{F} \mathcal{H}_{\Phi^{\rm cont}_\phi}^s$
is supported near the origin, so that this integral is 
close to $\int_{\R^2}\mathcal{H}_{\Psi^{(even)}_{\phi}}$. This yields 
the proposition, with the coefficients $C(s,\phi/\lambda)$ given as 
integrals of parabolic cylinder functions. 
The statements on the maxima derive from 
known results about parabolic cylinder functions. 
\end{proof}


\subsection{Odd-parity quasimodes}\label{odd-parity}
The connection \eqref{e:Bargmann estimate} between torus quasimodes 
$|\Phi^{\rm cont}_{\phi}\rangle$ and generalized eigenstates 
$|\Psi^{(even)}_{\phi}\rangle$
hints at a property we have not used much, namely parity. 
We have already mentioned that for each energy $E=-\hbar\phi$, 
$\hat H$ admits two independent generalized eigenstates,
 $|\Psi^{(even)}_\phi\rangle$ of even parity, and a second one of odd parity,
which we denote by $|\Psi^{(odd)}_{\phi}\rangle$. On the one hand, the Bargmann
function of the latter can be
expressed similarly as in Eq.~\eqref{e:parabolic cylinder functions}:
$$
\langle x,\tilde c_0|\Psi^{(odd)}_{\phi}\rangle
=C'_\phi \e^{-|Z|^2}\: \left\{ D_{-1+2\mu}(2Z) 
- D_{-1+2\mu}(-2Z)\right\}.
$$
On the other hand, as we did for $|\Psi^{(even)}_\phi\rangle$, 
we can build this odd eigenstate 
by propagating
an ``odd'' coherent state at the origin, \emph{i.e.} replacing the initial
$|\tilde c_0\rangle$ in Eq.~\eqref{e:plane projector} 
by the first excited squeezed state 
$$|\tilde c_0\rangle_1\defi \hat M_{(\tilde c_0,0)}a^\dagger|0\rangle.
$$
The Bargmann function of the corresponding quasimode $|\Psi_{\phi,T}\rangle_1$
is given by an integral similar to \eqref{e:Bargmann for quasimode}, 
with the integrand multiplied by the factor 
$\frac{Q'-\mi P'}{\sqrt{2}\cosh \lambda s}$: this is therefore an odd
function of $x$, semiclassically close to 
to $\langle x,\tilde c_0|\Psi^{(odd)}_{\phi}\rangle$

\begin{figure}[htbp]
{\par\centering
\resizebox*{0.45\columnwidth}{!}{\includegraphics{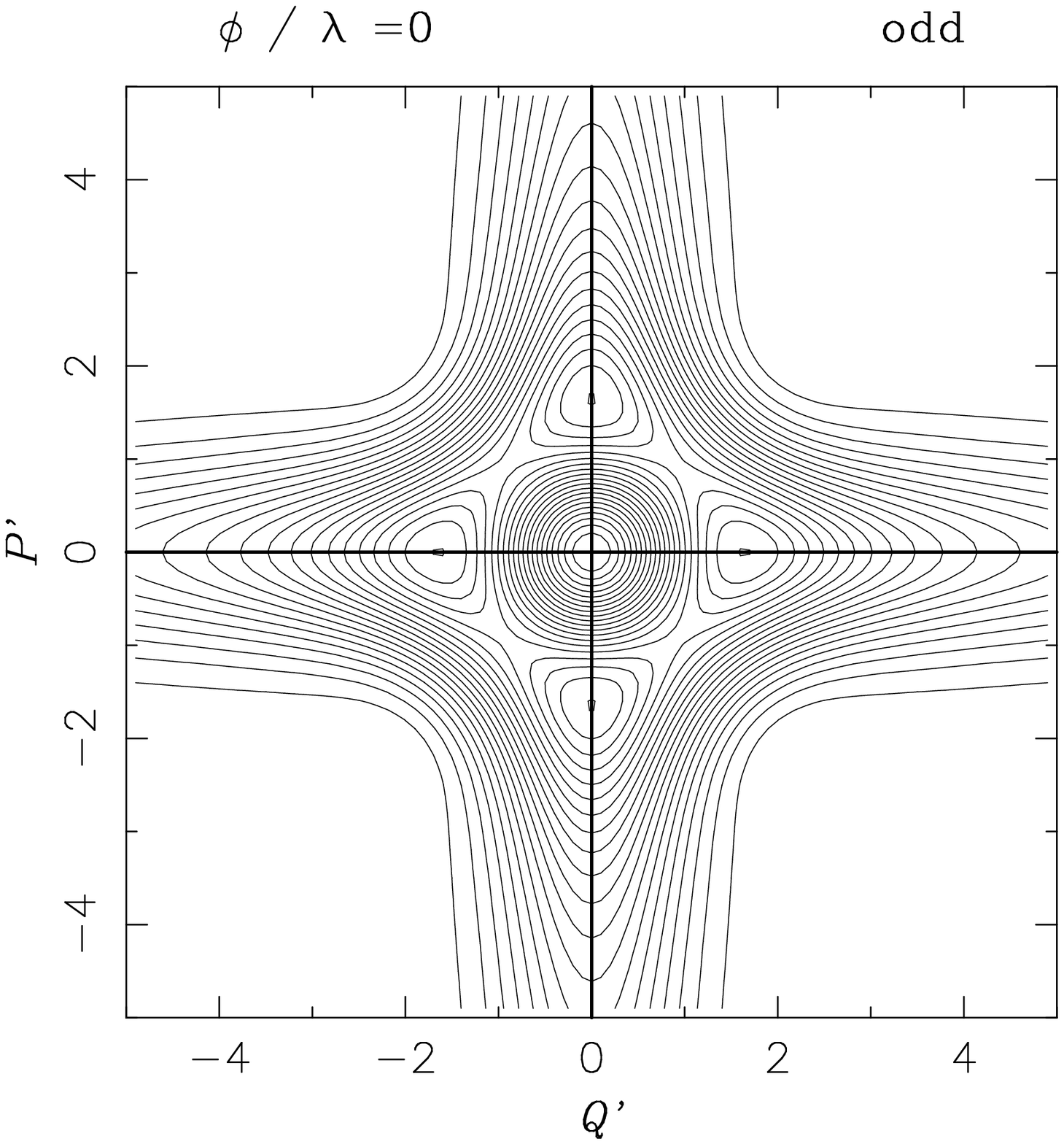}}
\resizebox*{0.50\columnwidth}{!}{\includegraphics{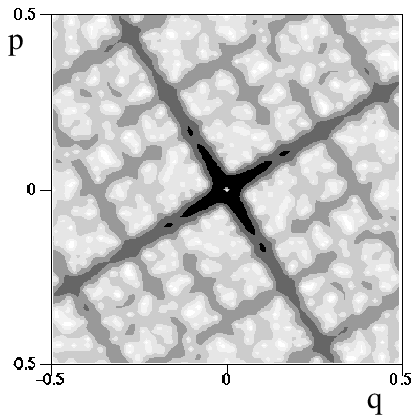}}
\par}
\caption{\label{fig:odd quasimode} Husimi functions of the odd eigenstate
$|\Psi^{(odd)}_{\phi=0}\rangle$ (linear scale) and the torus quasimode 
$|\Phi^{(odd)}_{\phi=0}\rangle$ (logarithmic scale) for $N=500$. 
Notice the zero at the origin.} 
\end{figure}

Projecting this plane odd quasimode to the torus through $\hat P_\theta$, 
one obtains a quasimode $|\Phi^{(odd)}_\phi\rangle$ of $\hat M$ with 
quasiangle $\phi$. 
Provided one has selected periodicity conditions $\theta\equiv (0,0)\bmod \pi$,
parity is conserved by $\hat P_\theta$, so that
the Bargmann function $\langle x,\tilde c_0,\theta|\Phi_\phi\rangle$ 
(resp. $\langle x,\tilde c_0,\theta|\Phi^{(odd)}_\phi\rangle$) is an even 
(resp. an odd) function of $x$. As a result, these two quasimodes are 
mutually orthogonal. 
The Bargmann and Husimi functions of $|\Phi^{(odd)}_\phi\rangle$ can
be described as precisely as for its even counterpart, in particular its
normalized Husimi and Wigner functions converge as well to the measure 
$\frac{1+\delta_0}{2}$, with a remainder $\cO(T^{-1/2})$.

\subsection{On the ``robustness'' of continuous quasimodes}\label{s:robustness}

We want to show that the continuous quasimodes $|\Phi^{\rm cont}_\phi\rangle$, 
$|\Phi^{odd}_\phi\rangle$ are ``stable'' with respect to a change of 
the initial state ($|\tilde c_0\rangle$ and $|\tilde c_0\rangle_1$, respectively).
One can indeed obtain an even quasimode very close to $|\Phi^{\rm cont}_\phi\rangle$ 
by propagating a different
initial state $|\psi_0\rangle$: this state needs to be of even parity,
sufficiently localized ({\it e.g.} a
finite combination of excited coherent states), and taken away from
a subspace of ``bad'' initial states. These remarks will be made more quantitative in 
Appendix~\ref{s:changesqueeze}, which treats the case where $|\psi_0\rangle$ is a squeezed
coherent state of arbitrary squeezing. 

To explain this ``robustness'', we notice that the operator
$$
\hat \P^{\rm cont}_{-\infty,\infty,\phi}\defi\int_{-\infty}^{\infty}ds\, \e^{-\mi \phi s} 
\e ^{-\mi \hat{H}s/\hbar}
$$
projects $L^2(\R)$ onto the 2-dimensional space spanned by 
$|\Psi^{(even)}_\phi\rangle$ and $|\Psi^{(odd)}_\phi\rangle$.
Any even state $|\psi_0\rangle\in L^2(\R)$
will thus be projected onto $C_\phi(\psi_0)|\Psi^{(even)}_\phi\rangle$, with the prefactor
$$
C_\phi(\psi_0)=\frac{\langle\Psi^{(even)}_\phi|\psi_0\rangle}
{\langle\Psi^{(even)}_\phi|\tilde c_0\rangle}.
$$
This prefactor vanishes iff there exists a state
$|\varphi_0\rangle\in L^2(\R)$ 
such that $|\psi_0\rangle=(\hat H+\nolinebreak\hbar\phi)|\varphi_0\rangle$; 
such $|\psi_0\rangle$ form a ``bad'' subspace of codimension $1$ inside the space of even states.

If $|\psi_0\rangle$ is localized inside a disk of radius $C\hbar^{1/2+\eps}$ 
at the origin, 
one can describe the plane quasimode $\hat\P^{\rm cont}_{-T,T,\phi}|\psi_0\rangle$ 
as in \eqref{closeness}:
\begin{equation}\label{e:general_close}
\langle x,\tilde c_0|\hat\P^{\rm cont}_{-T,T,\phi}|\psi_0\rangle=
C_\phi(\psi_0)\langle x,\tilde c_0|\Psi^{(even)}_\phi\rangle+\cO(\hbar^{1/2-\eps})\quad
\mbox{uniformly in $x\in\R^2$}.
\end{equation}
If $C_\phi(\psi_0)$ is of order unity, this estimate
shows that $\hat\P^{\rm cont}_{-T,T,\phi}|\psi_0\rangle$ resembles
the quasimode $|\Psi_{\phi,T}\rangle$. 
One can then show (as in Section~\ref{s:pointwise description}) that 
the torus state $\hat P_\theta\hat\P^{\rm cont}_{-T,T,\phi}|\psi_0\rangle$ is
close to the quasimode $C_\phi(\psi_0)|\Phi^{\rm cont}_\phi\rangle$. 

As an example, consider the case $\phi=0$: one can
start from any (finitely) excited coherent state of the form
$|\tilde c_0\rangle_{4n}\propto \hat M_{(\tilde c_0,0)}\big(a^\dagger\big)^{4n}|0\rangle$ to
obtain a quasimode asymptotically close to 
$|\Phi^{\rm cont}_0\rangle$. On the opposite, the states $|\tilde c_0\rangle_{4n+2}$ are
``bad'' initial states, because they are in the range of $\hat H$.

This discussion straightforwardly transposes to the construction of the odd quasimodes
$|\Phi^{odd}_\phi\rangle$ starting from odd localized states.


\section{Quasimodes on a general periodic orbit}\label{s:other_orbit}

We have so far described the construction of quasimodes localized on the fixed
point $0$ of the classical map $M$. We will now generalize this construction
to a general periodic orbit of $M$. The associated Husimi densities will be
shown to be (semiclassically) partly localized on the orbit and partly
equidistributed. The proofs require some minor changes with respect to the
previous case, but no fundamentally new ingredients.

We consider a fixed periodic orbit $\mathcal{P}=\left\{ x_{\ell}\in {\mathcal
F}\right\}^{\tau}_{\ell=0}$ of (primitive) period $\tau$, in other words, for
$0\leq\ell<\tau$, $M x_\ell = x_{\ell+1}$ mod $\Z^2$ and $x_{\tau}=x_0$.  Note
that $M^\tau x_\ell = x_\ell \ \mathrm{mod}\ \Z^2$, so that all $x_\ell$, when
viewed as points on the torus, are fixed points of $M'\defi M^\tau$.
Furthermore, for all $0\leq \ell \leq \tau$, there exist $m_\ell\in\Z^2$ so
that
$$
x_\ell = M^\ell x_0 - m_\ell.
$$
We will first introduce the discrete time quasimode defined in
(\ref{eq:quasimode}) and will consider its continuous time analog below:
$$
|\Phi^{\rm disc}_\phi\rangle = \sum_{t=-T}^{T-1} \e^{-\mi\phi t} \hat M^t \hat
P_\theta\hat T_{x_0}|\tilde c_0\rangle.
$$

Letting $T$ be the integer multiple of $\tau$ that is closest to
$|\ln\hbar|/\lambda$, and setting $T'=T/\tau$, a  simple computation yields
\begin{equation}\label{e:orbitdecompo2}
|\Phi_\phi^{\rm disc}\rangle = \sum_{\ell=0}^{\tau-1}
\e^{-\mi\phi\ell}|\Phi^{\rm disc}_\ell\rangle\quad\mathrm{where}\quad
|\Phi_\ell^{\rm disc}\rangle=\hat M^{\ell}
\left(\sum_{k=-T'}^{T'-1}\e^{-\mi\phi\tau k}\hat {M'}^k\right)\hat P_\theta\hat
T_{x_0}|\tilde c_0\rangle.
\end{equation}
It is  easy to see that
\begin{equation}\label{e:help7}
\hat M^\ell \hat P_{\theta} \hat T_{x_0}= \hat P_\theta \hat T_{x_\ell +
m_\ell} \hat M^\ell = \e^{\mi S_\ell}\hat P_\theta \hat T_{x_\ell} \hat M^\ell,
\end{equation}
where $S_\ell= \theta \cdot m_l+\mi \delta_{m_l}+\mi \pi N m_l\wedge x_{l}$ (see
(\ref{e:Heisenberg})). This phase can partly be interpreted in terms of the
\emph{action} along the classical orbit; however, the $\theta$-term is
non-classical, akin to the quantum phase due to a pointwise magnetic flux tube
on a charged particle (Aharonov-Bohm effect) \cite{mezzadri}. Hence
\begin{equation}\label{e:qmelldef}
|\Phi_\ell^{\rm disc}\rangle=\e^{\mi
S_\ell}\left(\sum_{k=-T'}^{T'-1}\e^{-\mi\phi\tau k}\hat {M'}^k\right)\hat
P_\theta\hat T_{x_\ell}\hat M^{\ell}|\tilde c_0\rangle.
\end{equation}
This suggests that $|\Phi_\ell^{\rm disc}\rangle$ is a quasimode of quasiangle
$\phi\tau$ for $\hat M'$, associated to the fixed point $x_\ell$ of
$M'$. This is basically the content of Proposition~\ref{prop:qmell}. There
is another instructive way of rewriting $|\Phi_\ell^{\rm disc}\rangle$ which
corroborates this idea. For that purpose, we first draw from
Eq.~(\ref{e:help7})
\begin{equation}\label{e:help10}
\hat M^{\tau k}\hat P_\theta \hat T_{x_0} =
\e^{\mi k S_\tau}\hat P_\theta \hat T_{x_0}\hat M^{\tau
k}\quad \mathrm{and} \quad \hat M^{\tau k+\ell}\hat P_\theta \hat T_{x_0} =
\e^{\mi (k S_\tau+S_\ell)}\hat P_\theta\hat T_{x_\ell}\hat M^{\tau k +\ell}.
\end{equation}
Using this, one can write
\begin{equation}\label{e:help20}
|\Phi_\ell^{\rm disc}\rangle=\e^{\mi S_\ell}\hat T_{x_\ell}\hat P_{\tilde
\theta_\ell} \left(\sum_{k=-T'}^{T'-1}\e^{-\mi(\phi\tau-S_\tau) k} \hat
{M'}^k\right)\hat M^{\ell}|\tilde c_0\rangle,
\end{equation}
where we used $\hat P_\theta \hat T_{x_{\ell}} = \hat T_{x_{\ell}}\hat
P_{\tilde \theta_{\ell}}$ with $\tilde \theta_{\ell} = \theta + 2\pi
N(p_{\ell}, - q_{\ell})$. A simple computation shows that, because
$x_\ell$ is a fixed point for $M^\tau$,  $\tilde\theta_\ell$ is a fixed
point for the map $\theta\to\theta'$ defined in (\ref{e:dual map}), with $M$
replaced by $M'$. Consequently, $|\Phi^{\rm disc}_\phi\rangle$ is the
$x_\ell$ translate of a quasimode for $\hat M'$ at the origin with
quasiangle $\phi\tau-S_\tau$, of the type studied in the previous sections.

\medskip

To build continuous time quasimodes, we replace in all the above formulas
$|\tilde c_0\rangle$ by
\begin{equation}
\label{e:generalized initial state} \frac{1}{\tau}\int_{0}^{\tau}dt\, \e^{-\mi t\tilde\phi}\,
\e^{-\frac{\mi}{\hbar}\hat H t}|\tilde c_0\rangle,
\end{equation}
where the ``quasienergy'' $\tilde\phi\in\R$ is chosen so that
\begin{equation}\label{tilde phi}
\tau\tilde\phi\equiv \tau\phi -S_\tau \bmod{2\pi}.
\end{equation}
Whereas the quasiangle $\phi$ is defined modulo $2\pi$, the
quasienergy $\tilde\phi$ is chosen in $\R$. The continuous quasimode reads:
\begin{equation}\label{eq:qmellcontdef}
|\Phi_\ell^{\rm cont}\rangle = \e^{\mi S_\ell}\hat T_{x_\ell}\hat P_{\tilde
\theta_\ell}\ \frac{1}{\tau}\int_{-T}^{T} dt\, \e^{-\mi\tilde \phi
t}\e^{-\frac{\mi}{\hbar}\hat H t} \hat M^\ell|\tilde c_0\rangle.
\end{equation}
All the above quasimodes can of course in obvious ways be split into a
localized and an equidistributing part, as before.
For both the discrete and continuous time quasimodes we
have the following estimates:
\begin{prop} \label{prop:qmell} For all $0\leq \ell'<\ell\leq \tau-1$,
for all $f\in C_0^\infty(\T^2)$, for all $k\in\Z$,
\begin{eqnarray}
\langle\Phi_\ell|\Phi_\ell\rangle &=&
2T'S_1(\phi\tau -S_\tau, \tau \lambda) + \mathcal O(1)\label{e:qmell1}\\
\lim_{\hbar \to0}\ _n\!\langle\Phi_\ell|\hat f|\Phi_\ell\rangle_n
&=&\frac{1}{2}f(x_\ell) + \frac{1}{2}\int_{\T^2} f(x) dx.\label{e:qmell3}\\
\lim_{\hbar\to0}\ _n\!\langle \Phi_{\ell'}|\hat
T_{k/N}|\Phi_{\ell}\rangle_n&=&0\label{e:qmell4}
\end{eqnarray}
The quasimodes $|\Phi_\phi\rangle$ satisfy (\ref{e:equidistributeall}), the limit being
uniform for $\phi$, $\tilde\phi$ in a bounded interval.
\end{prop}

Starting from (\ref{eq:qmellcontdef}) a pointwise analysis of 
the continuous time quasimode
$|\Phi^{\rm cont}_{\mathcal{P},\phi}\rangle$ can be performed as well, 
along the lines of Section \ref{s:pointwise description}.
One should notice that the Husimi function of
$|\Phi^{\rm cont}_{\mathcal{P},\phi}\rangle$ in the $\sqrt{\hbar}$-vicinity 
of a periodic point $x_l$ is dominated by the
contribution of $|\Phi_{l}^{\rm cont}\rangle$; it is
concentrated on a hyperbola which depends on the quasienergy
$\tilde\phi$ rather than on the quasiangle $\phi$.

\begin{proof}[Proof of the proposition] 
We write the proof for the discrete time quasimodes only. 
(\ref{e:help20}) immediately implies (\ref{e:qmell1}) and
(\ref{e:qmell3}) as a
consequence of the results of Section~\ref{s:quasimodes}. To prove
(\ref{e:qmell4}) when $k=0$, {\em i.e.} the asymptotic orthogonality of the 
$|\Phi_\ell\rangle$,
we write, using (\ref{e:orbitdecompo2}) and (\ref{e:help10})
\begin{equation*}
\!\langle \Phi_{\ell'}^{\rm disc}|\Phi_{\ell}^{\rm disc}\rangle=
\sum_{k'=-T'}^{T'-1}\sum_{k=-T'}^{T'-1} \e^{-\mi(\phi\tau-S_\tau)(k-k')+\mi
S_{\ell-\ell'}} 
\langle \tilde c_0| \hat T_{-x_0} \hat P_\theta\hat T_{x_{\ell-\ell'}}
\hat M^{(\ell-\ell')+\tau(k-k')}|\tilde c_0\rangle,
\end{equation*}
so that 
\begin{equation}
\begin{split}\label{e:ellellprime}
\big|\langle \Phi_{\ell'}^{\rm disc}
|\Phi_{\ell}^{\rm disc}\rangle \big|&\leq
\sum_{k'=-T'}^{T'-1}\sum_{k=-T'}^{T'-1} \sum_{m\in\Z^2}
|\langle \tilde c_0| \hat T_{-x_0} \hat T_m\hat T_{x_{\ell-\ell'}}
\hat M^{(\ell-\ell')+\tau(k-k')}|\tilde c_0\rangle|\\
&\leq\sum_{k'=-T'}^{T'-1}\sum_{k=-T'}^{T'-1} J_r(\tau (k-k')+\ell-\ell',0)
\leq C,
\end{split}
\end{equation}
where
$r=x_0-x_{\ell-\ell'}$, and
where we used the estimate $J_r(t,0)\leq C\hbar \e^{\lambda |t|/2}$ extracted
from Appendix~\ref{a:interference term}. 
To prove (\ref{e:qmell4}) when $k\not=0$, one repeats the arguments of Section
\ref{s:quasimodes}: we omit the details. For continuous quasimodes, 
the proofs are
analogous, using this time the same estimate on $J_r(t,s)$. 
The proof of (\ref{e:equidistributeall}) 
follows immediately.
\end{proof}

\paragraph{Convex combinations of limit measures}

We can further enlarge the set of semiclassical limit measures by taking
finite convex combinations of the previous ones. Consider a finite set of 
periodic orbits
$\{\mathcal{P}_1,\ldots,\mathcal{P}_f\}$, and complex coefficients
$\{\alpha_1,\ldots,\alpha_f\}$ satisfying $\sum_{i=1}^f|\alpha_i|^2=1$. Let
$|\Phi_{\mathcal{P}_i,\phi}\rangle$ be quasimodes (discrete or continuous time
) associated to $\mathcal{P}_i$, as defined above, with  the same quasiangle
$\phi$. We can then combine them into the quasimode
$$
|\Phi\rangle\defi\sum_{i=1}^f \alpha_i\, |\Phi_{\mathcal{P}_i,\phi}\rangle_n.
$$
One readily shows along the lines of the proof  of Proposition \ref{prop:qmell} that
for $i\not=j$, and for all $k\in\Z^2$, one has
$$
\lim_{\hbar\to0}\ _n\langle \Phi_{\mathcal{P}_i}|\hat T_{k/N}|\Phi_{\mathcal{P}_j}\rangle_n=0.
$$
This together with (\ref{e:equidistributeall}) shows that the
Husimi and Wigner functions of $|\Phi\rangle_n$  converge to the limit measure
$\frac{1}{2}\left(dx +\sum_{i=1}^f
|\alpha_i|^2\,\delta_{\mathcal{P}_i}\right)$.


\section{Scarred eigenstates for quantum cat maps of short quantum periods}
\label{s:min quantum period}

We will now slightly extend an argument from \cite{bondb1} in order to show 
that the quasimodes we have built and studied in the previous sections are
\emph{exact eigenstates} of the quantum map $\hat M$ for certain special
values of $\hbar$ and we will prove Theorem \ref{th:betascar}.

For that purpose, we first recall a few facts about quantum cat maps
\cite{hb}.
For a given value of $N =(2\pi \hbar)^{-1}$, every quantum map 
$\hat M$ has a ``quantum period''
$P(N )$  defined to be the smallest
nonnegative integer such that
\begin{equation}
\label{e:min quantum period} \hat{M}^{P(N )}=\e^{i\varphi (N)}
\hat I_{\hn} \quad \textrm{for a certain }\varphi (N
)\in [-\pi ,\pi [.
\end{equation}
It follows that, if $\phi $ is of the type $\phi =\phi_{j} =\frac{\varphi (N
)+2\pi j}{P(N)}$, then
$\frac{1}{P(N)}\hat{\mathcal{P}}_{t_1,t_1+P(N),\phi_j}$ is independent of
$t_{1}$, and is the spectral projector onto the eigenspace of $\hat{M}$ inside
$\hn$ associated to the eigenvalue $\e ^{\mi \phi _{j}}$
(the normalization factor $1/P(N)$ ensures that it is indeed a projector). All
eigenvalues of $\hat{M}$ on $\hn$ are necessarily of that form.

The function $P(N)$ depends on $N$ in an erratic way, and no closed formula
exists for it \cite{ke}. It satisfies the general bounds
\begin{equation}\label{e:bounds for period}
\exists C>0,\quad\forall N\in \N^*,\quad 
\frac{2}{\lambda}\log N -C
\leq P(N)\leq C\; N\log\log N.
\end{equation}
It is moreover known that, for ``almost all'' integers,  $P(N)\geq \sqrt{N}$
\cite{kuru2}. 
We will now give an elementary argument to show that, 
given any hyperbolic matrix in SL$(2,\Z)$, there
exists an infinite sequence of integers $N_k$ for which the quantum period 
is very
short in the sense that it saturates the above lower bound:
\begin{equation}
\label{e:short quantum period} P(N_k)=2\frac{\log N_k}{\lambda
} +\cO (1)=2T_k+\cO (1),
\end{equation}
where the Ehrenfest time $T$ was defined in (\ref{e:Ehrenfest time}).

Let us first recall that, for all $k\in \N^*$, one has
$$
M^k = p_k M -p_{k-1},\quad\mathrm{where}
\quad p_k=\frac{\e^{\lambda k}-\e^{-\lambda k}}{\e^\lambda -\e^{-\lambda}},\ p_0=0.
$$
It was proven in \cite{bondb1} that, for all $k\geq 1$, the integer 
$\tilde N_k=\mathrm{GCD}(p_k, p_{k-1}+1)$ satisfies 
\begin{equation}\label{estimate N_k}
\frac{2}{\lambda}\log\tilde N_k =k+\cO(1),
\end{equation}
and that
\begin{equation}\label{eq:classasymp}
 M^k=I + \tilde N_k M_k,\quad \mbox{with $M_k$ an integer matrix}.
\end{equation}
We now set $N_k=\tilde N_k$ if $\tilde N_k$ is odd, $N_k=\tilde N_k/2$
if $\tilde N_k$ is even. Choosing the periodicity angle $\theta=(0,0)$ 
when $N_k$ is even and $\theta=(\pi,\pi)$
when $N_k$ is odd 
(which makes sense, cf. the end of Section~\ref{s:quantum torus}),  
we prove below the following lemma:
\begin{lem}
With $N_k$, $\theta$ given as above, $\hat M^k=\e^{i\varphi}
\hat I_{\mathcal{H}_{N_k,\theta}} \quad$ for a certain 
$\varphi\in [-\pi ,\pi[$.
\end{lem}
This means that the quantum period $P(N_k)$ of $\hat M$ on
$\mathcal{H}_{N_k,\theta}$ divides $k$. Comparing 
\eqref{e:bounds for period} with \eqref{estimate N_k} entails that for 
$k$ large enough, $P(N_k)=k$ and 
\eqref{e:short quantum period} holds.

\begin{proof}[Proof of the lemma]
The case $\tilde N_k=2N_k$, $\theta=(0,0)$ was treated in \cite{hb}. We
give a different proof, which works for both cases.

From Schur's Lemma and the irreducibility of the  $\hat T_{n/N_k}$,
it suffices to show that $\left[\hat M^k, \hat T_{n/N_k}\right]=0$ 
on $\mathcal{H}_{N_k,\theta}$,
for all $n\in\Z^2$. Setting $\tilde N_k=\eps N_k$,  
$\theta=\eps'(\pi,\pi)$ and using the definition of 
$\hat P_\theta$, Eqs.~\eqref{e:Heisenberg}
and \eqref{e: M and T}, one readily computes
\begin{align*}
\hat M^k \hat T_{n/N_k} \hat M^{-k}\hat P_\theta&=
\e^{i\pi\eps(n\wedge M_kn)}\hat T_{n/N_k} \hat T_{M_kn}\hat P_\theta\\
&=(-1)^{\eps(n\wedge M_kn)+\eps\eps'[(M_kn)_1+(M_kn)_2+(M_kn)_1(M_kn)_2]}\hat T_{n/N_k}\hat P_\theta.
\end{align*}
This phase is trivial if $\eps=2$. In the case $\eps=\eps'=1$ 
(that is, $\tilde N_k$ odd), one must  consider
the $6$ possible values of $M$ modulo $2$: in all cases, the
phase is trivial. \end{proof}

If we now consider such a value $N_k$ together with an admissible eigenangle
$\phi_{j_k}$, the eigenstates
$$|\Phi_k\rangle=\sum_{t=-P(N_k)/2}^{P(N_k)/2-1} \e^{-\mi\phi_{j_k} t}\hat M^t
|x_0,\tilde c_0\rangle
$$
are (discrete time) quasimodes of the quantum map as studied in the previous
sections. Indeed, as discussed at the end of Section \ref{s:quasimodes}, since
$T$ and $P(N_k)/2$ differ by a bounded number of terms in the semiclassical
limit, we can replace one by the other in (\ref{eq:quasimode}), without
affecting any of the semiclassical properties of the quasimodes. One can 
similarly construct
eigenfunctions that are continuous time quasimodes.

\begin{proof}[Proof of Theorem \ref{th:betascar}] 
The previous arguments settle the case $\beta=
1/2$. To treat the general case, we recall that the Schnirelman theorem 
implies the existence of  a sequence of eigenfunctions $|\varphi_k\rangle_n$ of
$\hat M$ on $\mathcal{H}_{N_k,\theta}$
(with corresponding eigenvalues $(\phi_{j_k})_{k\in\N}$)  that equidistribute as $k\to\infty$.  We then construct, for $0\leq \alpha \leq
1$:
$$
|\psi_k\rangle = \alpha|\Phi_k\rangle_n + \sqrt{1-\alpha^2}|\varphi_k\rangle_n
$$
If we show that, for all $n\in\Z^2$,
\begin{equation*}
\lim_{\hbar\to0}\ _n\!\langle \varphi_k |\hat T_{n/N_k}|\Phi_k\rangle_n=0,
\end{equation*}
a simple computation implies that the $|\psi_k\rangle_n$ satisfy 
(\ref{eq:scarbeta})
with $\beta=\alpha^2/2$. We have
$$
\lim_{\hbar\to0}\ _n\!\langle \varphi_k |\hat T_{n/N_k}|\Phi_k\rangle_n =
\lim_{\hbar\to0}\left(\ _n\!\langle \varphi_k |\hat T_{n/N_k}|
\Phi_{k,\mathrm{erg}}\rangle_n
+\ _n\!\langle \varphi_k |\hat T_{n/N_k}
|\Phi_{k, \mathrm{loc}}\rangle_n\right).
$$
The second limit vanishes with an argument as  in (\ref{eq:decoupe}),
whereas for the first, we use the further decomposition
$|\Phi_{k, \mathrm{erg}}\rangle =|\Phi_{k, 1}\rangle +|\Phi_{k, 4}\rangle$ with
$|\Phi_{k, 1}\rangle=\e^{\mi \phi_{j_k} T/2}\hat M^{-T/2}|\Phi_{k,2}\rangle$,
$|\Phi_{k, 4}\rangle=\e^{-\mi \phi_{j_k} T/2}\hat M^{T/2}|\Phi_{k,3}\rangle$ (see (\ref{e:definition |Phi_j>})). Now, since
$|\varphi_{j_k}\rangle_n$ is an eigenfunction, we have
$$
\big|\, _n\!\langle\varphi_k|\hat T_{n/N_k}| \Phi_{k,4}\rangle_n\big| =
\big|\, _n\!\langle\varphi_k|\hat M^{-T/2}\hat T_{n/N_k}\hat M^{T/2}
| \Phi_{k,3}\rangle_n\big|.
$$
As in the proof of Proposition \ref{prop:ergodicity of |Phi_erg>}, and more 
specifically Eq.~(\ref{eq:decoupe}), this tends to $0$ with $\hbar$.
\end{proof}

For matrices $M$ of ``checkerboard structure'',  
the results of \cite{kuru1,mezz2} imply that, given 
an {\em arbitrary sequence} of eigenvalues $(\phi_{j_k})_{k\in\N}$,
there exists a corresponding sequence of eigenvectors 
$|\varphi_k\rangle \in\mathcal{H}_{N_k,\theta}$ that semiclassically
equidistribute. 
One can then construct for the same eigenvalues eigenstates
$|\psi_k\rangle$ satisfying Eq.~\eqref{eq:scarbeta}.

The $P(N_k)$ eigenstates with distinct eigenvalues constructed above are of
course \emph{exactly} orthogonal to each other, and not just asymptotically as
proven in Section~\ref{s:Delta phi-orthogonality}. On the other hand, two
continuous time eigenstates of identical eigenangle $\phi_j$ but different
quasienergies $\tilde\phi-\tilde\phi'=s\pi/T$, $s\neq 0$ become orthogonal in
the semiclassical limit. This is also the case for two eigenstates with the
same eigenangle supported on different periodic orbits $\mathcal P\neq\mathcal
P'$.


\section{Conclusion}
In this article we have constructed and analyzed a certain class of
``quasimodes'' of hyperbolic quantized torus isomorphisms, which for certain
values of $\hbar$ become exact eigenstates. The characteristic property of
these quasimodes is that their ``quantum limit'', that is the weak limit of
their Husimi densities, does not yield the Liouville measure, but contains a
singular component supported on a (finite union of) periodic orbit(s). In our
case, this singular component has a relative weight $\beta\leq 1/2$, less than
or equal to the
weight of the Liouville part. As explained in the introduction, no limit
measure of eigenstates can have a ``larger'' singular component. We
further conjecture that no sequence of quasimodes ({\it i.e.} images of the
operators $\mathcal{\hat P}_{-T,T,\phi}$) can have a more singular limit
measure either.

The strong scarring of eigenstates exhibited in this paper  is directly linked
to the very large degeneracies of the eigenvalues of $\hat M$ for certain
special values of Planck's constant. Therefore, such sequences of eigenstates
are very probably absent as soon as one considers nonlinear perturbations of
the dynamics, for instance $\hat M_\epsilon=\e^{-\mi\eps \hat H_1/\hbar}\hat M$, for
any periodic Hamiltonian $H_1(x)$ and $\eps>0$ small enough. Such a
perturbation of the classical map is known to conserve the uniform
hyperbolicity, but destroys the ``action degeneracies'' characteristic of the
(linear) cat map. As a consequence the spectrum of the perturbed map exhibits
Random Matrix statistics, in particular ``repulsion'' between eigenangles
\cite{mezzadri}, which forbids degeneracies.

The precise characterization of some weaker form of scarring for individual
eigenstates that would remain valid for $\hat M_\eps$ remains therefore an open
problem. Nevertheless, it might be interesting to study the phase space
distribution of the ``nonlinear`` quasimodes of the type
$\sum_{t=-T}^T\e^{-\mi\phi t}\hat M^{t}_\epsilon|x_0,\tilde c\rangle$, for $x_0$
a periodic point of $M_\epsilon$, which may not be as simple to describe as 
for the
linear map.

\paragraph{Acknowledgments:} We have benefited from useful conversations
with Y.~Colin de Verdi\`ere, C.~G\'erard, E.~Vergini, A.~Voros, D.~Wisniacki.
F.~F. acknowledges the kind hospitality of the Service de Physique
Th\'eorique, CEA Saclay where part of this work was accomplished.


\section{Appendices}

\subsection{Estimate of the interference term $I(t,s)$}
\label{a:interference term} In this appendix we prove Proposition
\ref{prop:crucial}. For the purpose of Section \ref{s:other_orbit}  we will at
the same time give a bound for the more general overlap ($t,s\in\R$)
\begin{equation}\label{e:Bargmann bound}
\left| \langle r,\tilde c_s|\hat{P}_\theta \, \e^{-\mi \hat{H}t/\hbar}
|\tilde{c}_{s}\rangle \right| \leq \sum_{n\in \Z ^{2}}\left| \langle
r+n,\tilde c_s|\e^{-\frac{i}{\hbar}\hat H t}|\tilde c_s\rangle \right|,
\end{equation}
where $r\in\mathcal{F}$ (the fundamental domain) belongs to the lattice
$\left(\frac{1}{D}\Z \right) ^{2}$, with $D\in \N ^{*}$ and where $\theta\in
[0,2\pi[\times[0,2\pi[$ is {\em arbitrary} (in other words, $\theta$ need not
be equal to the fixed point of the map (\ref{e:dual map}). We define
\begin{equation}\label{eq:defJr}
J_r(t,s)\defi\sum_{n\in \Z ^{2},r+n\not=0}\left| \langle r+n,\tilde
c_s|\e^{-\frac{i}{\hbar}\hat H t}|\tilde c_s\rangle \right|.
\end{equation}
We first consider the case $s=0$, $t\geq 0$. Since $\shbar\leq \Delta p'\leq
\sqrt{2 \hbar }$ for all positive times, only the points $r+n$ near 
the unstable axis
can significantly contribute. Therefore, we subdivide the plane into strips
parallel to this axis: the ``outer'' strips
$$
\forall l\geq 1,\quad S_{\pm l}= \left\{ x\mid a_{l}\leq \pm
p'(x)<a_{l+1}\right\} ,
$$
 with $a_{l}=W_{0}+(l-1)W$, and the central strip $S_{0}=\left\{x\neq 0 \mid
p'(x)|<W_{0}\right\}$. The widths $W_{0},\: W$ will be explicitly set below.

We start by estimating the contribution of the points $r+n\in S_{l}$ with
$l\geq 1$. Due to the diophantine condition (\ref{eq:dioph3}), as long as $W$
is small enough, two points in this strip satisfy the property
$|q'(r+n)-q'(r+m)|>C_{o}/W$. Ordering these points according to their
abscissas: $q'(r+n_{j})<q'(r+n_{j+1})<q'(r+n_{j+2})$, we have for any $\alpha
>0$ :
\begin{equation}
\label{e:bound to theta} \sum _{j\in \Z }\exp \left\{ -\alpha
q'(r+n_{j})^{2}\right\} \leq \sum _{j\in \Z }\exp \left\{ -\alpha \left(
\frac{jC_{o}}{W}\right) ^{2}\right\} .
\end{equation}
The sum on the RHS is a one-dimensional theta function, which has the upper
bound (optimal for $0<\alpha$ small enough):
\begin{equation}
\label{e:theta estimate} \sum _{j\in \Z }\e ^{-\alpha j^{2}}\leq
1+\sqrt{\frac{\pi }{\alpha }}.
\end{equation}
As a result, using (\ref{e:<R|TM|R>}) it becomes clear that the contribution
to $J_r(t)$ of the points $r+n\in S_{l}$ is bounded above by
\begin{multline}
\sum _{r+n_{j}\in S_{l}}\frac{1}{\sqrt{\cosh \lambda t}}\; \exp
\left\{-\frac{1}{2}\left[ \frac{p'(r+n_{j})^{2}}{\Delta p^{\prime 2}}
+\frac{q'(r+n_{j})^{2}}{\Delta q^{\prime 2}}\right] \right\} \\
 \leq\sqrt{2}\e
^{-\lambda t/2}\, \e^{-\frac{a_{l}^{2}}{2\Delta p^{\prime 2}}}\, \left[
1+\sqrt{2\pi }\frac{W\Delta q'}{C_{o}}\right] .
\end{multline}
The estimate (\ref{e:theta estimate}) can then be applied to the sum over the
strips $S_{l}$, $l\neq 0$, to obtain (remind $|t;\tilde c_0\rangle=\e^{-\mi\hat Ht/\hbar}|\tilde c_0\rangle$)
$$
\sum_{l\neq 0}\: \sum _{r+n\in S_{l}} \left|\langle
r+n,\tilde{c}_{0}|t;\tilde c_0\rangle\right| \leq
\sqrt{2}\e ^{-\lambda t/2}\, \e^{-\frac{W_0^2}{2\Delta p^{\prime 2}}}\, \left[
2+\frac{\sqrt{2\pi}\Delta p'}{W}\right] \, \left[ 1+\frac{\sqrt{2\pi }W\Delta
q'}{C_{o}}\right] .
$$
 For each time $t$, we can minimize the RHS with respect to
$W$ by taking $W^2=C_{o}\frac{\Delta p'}{2\Delta q'}=\frac{C_{o}}{2}\e
^{-\lambda t}$, which leads to the bound
\begin{equation}
\label{e:bound S_l} \sum_{l\neq 0}\: \sum _{r+n\in S_l} \left|\langle
r+n,\tilde{c}_{0}|t;\tilde c_0\rangle\right| \leq
2\sqrt{2}\e ^{-\frac{W_0^2}{2\Delta p^{\prime 2}}}\, \e ^{-\lambda t/2}\,
\left[1+\sqrt{\frac{\pi }{C_{o}}}\Delta p'\e ^{\lambda t/2}\right]^{2}.
\end{equation}
Notice that this upper bound is independent of the point $r$.

We now estimate the contribution of the strip $S_{0}$, which requires more
care, and will depend on $r$. For any point $r'\neq 0$ on the lattice
$\left(\frac{1}{D}\Z \right)^2$ sufficiently close to the unstable axis, the
diophantine property (\ref{eq:dioph3}) implies $|p'(r')|\geq
\frac{C_o}{D^2|q'(r')|}$. As a consequence, the quadratic form appearing in
\eqref{e:<R|TM|R>} may be bounded inside $S_0$ by
\begin{equation}
\label{e:definition f_t} \frac{q'(r+n)^{2}}{2\Delta q^{\prime 2}}
+\frac{p'(r+n)^{2}}{2\Delta p^{\prime 2}} \geq \frac{q'(r+n)^{2}}{2\Delta
q^{\prime 2}} +\frac{C_{o}^{2}}{2D^{4}\Delta p^{\prime 2}\, q'(r+n)^{2}}\defi
f_{t}(q'(r+n)).
\end{equation}
The function $f_t$ satisfies the scaling property $f_{t}(q)=\frac{C_{o}\e
^{-\lambda t}}{2D^{2}\Delta p^{\prime 2}}\, f\left( \frac{qD\e ^{-\lambda
t/2}}{\sqrt{C_{o}}}\right) $, with $f(q)\defi q^{2}+q^{-2}$. This function
$f(q)$ is bounded below for all positive $q$ by the parabola
$g(q)=2+(q-1)^{2}$, so after rescaling we get
$$
\forall q>0,\quad f_{t}(q)\geq g_{t}(q)\defi \frac{C_{o}\e ^{-\lambda
t}}{D^{2}\Delta p^{\prime 2}}+\frac{\e ^{-2\lambda t}}{2\Delta p^{\prime
2}}\left( q-\sqrt{C_{o}}\e ^{\lambda t/2}/D\right) ^{2}.
$$
We consider the contributions of the points $r+n$ in $S_{0}$ such that
$q'(r+n)>0$ (the points with negative $q'$ can be treated identically). We
order these points as $0<q'_{0}<q'_{1}<\ldots $: each contribution is bounded
above by the quantity $(\cosh \lambda t)^{-1/2}\e ^{-g_{t}(q'_{j})}$, which is
maximal for the $q'_{j}$ close to $\sqrt{C_{o}}\e ^{\lambda t/2}/D$. The
diophantine inequality $|q'_{j}-q'_{j+1}|\geq C_{o}/W_{0}$ together with the
estimates (\ref{e:bound to theta},\ref{e:theta estimate}) then yield
\begin{equation}\label{e:bound S_0}
\sum_{r+n\in S_{0}}\left|\langle r+n,\tilde{c}_0|t;\tilde{c}_0\rangle\right|
\leq 2\sqrt{2} \exp\left\{-\frac{C_{o}\e ^{-\lambda t}}{D^2\,\Delta p^{\prime
2}}\right\} \, \e^{-\lambda t/2}\, \left(1+\frac{\sqrt{2\pi}W_{0}\Delta p'\e
^{\lambda t}}{C_{o}}\right).
\end{equation}
This contribution now depends on the rational point $r$ through its
denominator $D$: the upper bound increases with $D$. The full sum $J_{r}(t)$
is bounded above by the sum of the RHS in (\ref{e:bound S_l})-(\ref{e:bound
S_0}). For each time $t$, we adjust the value of $W_0$ to minimize that sum.
We do not search the exact minimum, but only its order of magnitude. We have
to distinguish two time intervals:

\begin{itemize}
\item for short times ($t<< T$), the behaviour of \eqref{e:bound S_0} is
governed by the first exponential (since $\Delta p^{\prime 2}\leq 2\hbar $).
We take $W_0$ such that the first exponential in (\ref{e:bound S_l}) is much
smaller than that factor, for instance by taking $W_0=2\sqrt{C_o}\e ^{-\lambda
t/2}/D$. Being careful for times around $t\lesssim T$, we find
$$
0\leq t\leq T\Longrightarrow J_{r}(t,0)\leq 2\sqrt{2}\, \exp\left\{ -\frac{C_{o}\e
^{-\lambda t}}{D^{2}\, \Delta p^{\prime 2}(t)}\right\} \, \e ^{-\lambda
t/2}\, \left[ 1+C\e ^{\lambda (t-T)/2}\right] ,
$$
where the constant $C$ is independent on the denominator $D$. One may replace
$\Delta p^{\prime 2}(t)$ by its maximum $2\hbar$ for positive times. The
RHS increases with the denominator $D$.
\item for times $t\geq T,$ the RHS of (\ref{e:bound S_0}) is now governed
by the factor between brackets, and we want to make sure that \eqref{e:bound
S_l} is not larger than it. Still taking $W_0=\e ^{-\lambda t/2}$ leads to
the estimate:
$$
T\leq t\leq 2T\Longrightarrow J_{r}(t,0) \leq \frac{2\pi
\sqrt{2}}{C_{o}}\hbar\, \e ^{\lambda t/2}\, \left[ 1+C'\e ^{\lambda
(T-t)/2}\right] .
$$
 The constant $C'$ is independent of $r$, so this bound applies uniformly
to any point $x\in \T$: it yields a $L^\infty$-bound for the Bargmann (or the
Husimi) function of $\hat M^t|\tilde c_0, \theta\rangle$.
\end{itemize}
The same bounds apply as well to $J_r(t,s)$ with $s\not=0$.
Indeed, replacing the initial
squeezing $\tilde c_0$ by its $s$-evolved value amounts to dilating the
coordinates of the points as $q'(r+n)\mapsto \e ^{\lambda t_1}q'(r+n)$,
$p'(r+n)\mapsto \e ^{-\lambda t_1}p'(r+n)$. One easily checks that this
dilation does not modify the above bounds.

The negative times are treated thanks to the identity 
$J_r(t,s)=J_{-M^{-t}r}(-t,s)$, and noticing that the above bounds
only depend on the denominator $D$, common to $r$ and $M^{-t}r$.


\subsection{Changing the initial squeezing}\label{s:changesqueeze}
We chose from the beginning to construct quasimodes starting from the coherent
state $|\tilde c_0\rangle$ defined in Section~\ref{s:evolution}. The
definition was motivated by the positivity property \eqref{e:positivity} of
the overlap $\langle \tilde c_0|\hat M^t|\tilde c_0\rangle$, and by choosing
the ``smallest'' parameter $\tilde c$ sharing this property. The simple
expression \eqref{e:positivity} was then used to control the ``interferences''
$I(t,s)$ (cf. Appendix \ref{a:interference term}), and to obtain from there the
asymptotic norm of the quasimode (Section \ref{s:orthogonality}), a crucial
step for further estimates. Similarly, we also {\it chose} to analyze the
quasimodes using the $\tilde c_0$-Bargmann representation, because of the
relatively simple formulas for $\langle x,\tilde c_0|\hat M^t|\tilde
c_0\rangle$ (see (\ref{e:<R|TM|R>})).

We want to stress (as we did towards in Section~\ref{s:robustness} for the continuous quasimodes) 
that both these choices were made purely for
convenience, and are not crucial for the results of this paper. The construction
of quasimodes can be extended in many ways. In this appendix, 
we will consider discrete or continuous quasimodes starting from a squeezed state 
$|\tilde c_1\rangle$, with an arbitrary  (possibly $\hbar$-dependent)
squeezing $\tilde c_1$. We also want to analyze these quasimodes using the Bargmann
function $\langle x,\tilde c_2,\theta|\Phi\rangle$ for some $\tilde c_2\in
\C$ which could depend on $\hbar$ as well.

\begin{prop}
The convergence \eqref{e:equidistributeall} holds
the above quasimodes, as long as $\tilde c_1$ and
$\tilde c_2$ stay in a fixed compact set $K\subset \C$ for all $\hbar$. 
\end{prop}

\begin{proof}[Sketch of proof]
For an initial state $|\tilde c_1\rangle$, the overlap 
$\langle x,\tilde c_1|\e^{-\mi t\hat H/\hbar}|\tilde c_1\rangle$, crucial in the calculation of
$I(t,s)$, is still given by closed formulas. We only give it for the simpler case $x=0$:
$$\langle\tilde c_1|\e^{-\mi t\hat H/\hbar}|\tilde c_1\rangle =\langle \tilde 
c_1'|\hat D(\lambda t)|\tilde c_1'\rangle
=\big(\cosh(\lambda t)+\mi R(\tilde c'_1)\sinh(\lambda t)\big)^{-1/2},
$$
where $|\tilde c'_1\rangle\propto \hat Q^{-1}|\tilde c_1\rangle$ and 
$R(c)=-\Re(c)\frac{\sinh(2|c|)}{2|c|}$. In general, this overlap is therefore not real. However,
it still decreases exponentially fast with time, and its average 
$$
S_1(\tilde c_1,\lambda,\phi)=\int_{\R}dt\;\e^{-\mi\phi t} \langle \tilde c_1|
\e^{-\mi t\hat H/\hbar}|\tilde c_1\rangle
$$
can be easily related with $S_1(\lambda,\phi)$ through a change of variable. One gets
$S_1(\tilde c_1,\lambda,\phi)=\e^{-\phi \tau_1}\sqrt{\cos(\lambda \tau_1)}
\;S_1(\lambda,\phi)$ with the 'complex time' $\tau_1=\arctan\{R(\tilde c'_1)\}/\lambda$.

For $x\neq 0$, the expression for $\langle x,\tilde c_1|\hat M^t|\tilde c_1\rangle$ is
more cumbersome than \eqref{e:<R|TM|R>}. Yet, it is still a
Gaussian having an elliptic profile of width $\sim \sqrt{\hbar}$, length $\sim
\sqrt{\hbar}\e^{\lambda |t|}$ and height $\sim \e^{-\lambda |t|/2}$, and
its long axis is asymptotically lined up with the
unstable direction for $t\to\infty$.
As a result, the results of 
Sections~\ref{s:interference bound}--\ref{s:Delta phi-orthogonality} still hold 
(replacing $S_1(\lambda,\phi)$ by $S_1(\tilde c_1,\lambda,phi)$). 
The localization property \eqref{eq:help3} holds as well, even if one replaces in the bras
$\tilde c_0$ by $\tilde c_2$, 
as long as $\tilde c_2$ remains bounded. The
rest of the proof  to obtain \eqref{e:equidistributeall} 
(Sections~\ref{s:phi erg}--\ref{s:phi tot}) goes through unaltered.\end{proof}

Following the Section~\ref{s:robustness},  
the plane quasimode $\hat\P^{\rm cont}_{-T,T,\phi}|\tilde c_1\rangle$ 
can be analyzed pointwise through
the estimate \eqref{e:general_close}; one now has explicitly
$C_\phi(|\tilde c_1\rangle)=\e^{-\phi \tau_1}\sqrt{\cos(\lambda \tau_1)}$. One 
may replace $\tilde c_0$ by $\tilde c_2$ in that estimate.
As opposed to Eq.~\eqref{e:parabolic cylinder functions}, the Bargmann function 
$\langle x,\tilde c_2|\Psi^{(even)}_\phi\rangle$ is not given in terms of cylinder
parabolic functions. Yet, its
behaviour ``far'' from the origin will be similar to \eqref{inner sector}.
As a consequence, the pointwise estimate
\eqref{e:Bargmann estimate} (with $\tilde c_0\to\tilde c_2$ in the bras)
will apply to the torus quasimode 
$\hat P_\theta\hat\P^{\rm cont}_{-T,T,\phi}|\tilde c_1\rangle$ as well, upon taking the
prefactor $C_\phi(|\tilde c_1\rangle)$ into account and replacing in the bras $\tilde c_0\to
\tilde c_2$ on both sides. The estimates of Sections~\ref{s:controlling}--\ref{Ls norms}
may be generalized as well to the present case.


\end{document}